%% file: brane_mergers_FINAL.tex
\documentclass[a4paper]{article}
\usepackage{jheppub}
\usepackage{dsfont}
\usepackage{amsmath,amssymb,amscd,amsfonts,mathtools}
\DeclareMathOperator{\Tr}{Tr}

\DeclareMathOperator{\Tanh}{Tanh}

\usepackage[toc,page]{appendix}
\usepackage{epsfig}
\usepackage{epstopdf}
\usepackage{latexsym}
\usepackage{enumerate} 
\usepackage{graphicx}
\usepackage{placeins}
\usepackage{floatrow}
\usepackage{subfig}
\usepackage{caption}
\usepackage{booktabs}
\usepackage{bbm}
\usepackage{color}
\usepackage{physics}
\usepackage{tensor}
\usepackage{tikz}
\usetikzlibrary{matrix}
\usetikzlibrary{decorations.markings,calc,shapes,decorations.pathmorphing,patterns,decorations.pathreplacing,arrows.meta}
\usetikzlibrary{positioning}
\usepackage{hyperref}

\pdfoutput=1
\makeatletter
\def\@fpheader{\relax}
\makeatother

\def\deltabar{{\mathchar '26\mkern -10mu\delta}}

\newcommand{\sanjit}[1]{\textcolor{black!40!green}{\textbf{SS:} #1}}

\DeclareMathSizes{10}{10}{7}{7}

\title{Holographic BCFT Spectra from Brane Mergers}

\author[a]{Shovon Biswas,}
\author[b,c]{Jani Kastikainen,}
\author[d]{Sanjit Shashi,}
\author[a]{James Sully}
\affiliation[a]{Department of Physics and Astronomy, University of British Columbia, 6224 Agricultural Road, Vancouver, British Columbia V6T 1Z1, Canada.}
\affiliation[b]{Department of Physics, P.O. Box 64, FIN-00014 University of Helsinki, Finland.}
\affiliation[c]{Université Paris Cité, CNRS, Astroparticule et Cosmologie, F-75013 Paris, France.}
\affiliation[d]{Theory Group, Weinberg Institute, Department of Physics, University of Texas, 2515 Speedway, Austin, Texas 78712, USA.}
\emailAdd{shovon432@gmail.com}
\emailAdd{jani.kastikainen@helsinki.fi}
\emailAdd{sshashi@utexas.edu}
\emailAdd{jamie.sully@gmail.com}

\abstract{We use holography to study the spectra of boundary conformal field theories (BCFTs). To do so, we consider a 2-dimensional Euclidean BCFT with two circular boundaries that correspond to dynamical end-of-the-world branes in 3-dimensional gravity. Interactions between these branes inform the operator content and the energy spectrum of the dual BCFT. As a proof of concept, we first consider two highly separated branes whose only interaction is taken to be mediated by a scalar field. The holographic computation of the scalar-mediated exchange reproduces a light scalar primary and its global descendants in the closed-string channel of the dual BCFT. We then consider a gravity model with point particles. Here, the interaction of two separated branes corresponds to a heavy operator which lies below the black hole threshold. However, we may also consider branes at finite separation that ``merge" non-smoothly. Such brane mergers can be used to describe unitary sub-threshold boundary-condition-changing operators in the open-string spectrum of the BCFT. We also find a new class of sub-threshold Euclidean bra-ket wormhole saddles with a factorization puzzle for closed-string amplitudes.}

\begin{document}
\maketitle
\flushbottom

\input{figs/cylinderDuals}

\section{Introduction} 

Traditionally, boundary conformal field theory (BCFT) \cite{Cardy:1984bb,McAvity:1995zd,Cardy:2004hm} refers to a quantum field theory on a manifold with a boundary that has ``boundary" conformal symmetry. On a flat manifold, boundary conformal symmetries are contained in the subgroup of the global conformal group that preserves the location of the boundary. Hence a BCFT comes equipped with ``conformal" boundary conditions \cite{Cardy:1984bb} that respect this subgroup. BCFT has seen various applications, from being a natural descriptor of critical phenomena in finite systems \cite{Affleck:1995ge,Affleck2000} to its role in the worldsheet description of D-branes in string theory \cite{Sagnotti:1987tw,Polchinski:1995mt,Gaberdiel:2002my,Recknagel:2013uja}. However, just like CFT, BCFT is difficult to study at strong coupling. This motivates the use of holography to better understand physics of boundaries.

The relevant extension of holography is known as the AdS/BCFT correspondence \cite{Karch:2000gx,Takayanagi:2011zk,Fujita:2011fp} in which a BCFT is dual to AdS gravity with a bulk codimension-1 end-of-the-world (EOW) brane. This correspondence provides a geometrization of strongly coupled BCFT \cite{Karch:2017fuh} and has been applied to a number of studies on entanglement entropy dynamics in toy models of black holes \cite{Rozali:2019day,Almheiri:2019hni,Geng:2020qvw,Chen:2020uac,Sully:2020pza,Chen:2020hmv,Geng:2021iyq,Ageev:2021ipd,Hollowood:2021wkw,Geng:2021mic,Geng:2022dua}. The construction we employ is bottom-up in that we are starting with Einstein gravity, but there are also top-down constructions in the literature \cite{DHoker:2007zhm,DHoker:2007hhe,Gaiotto:2008sa,Gaiotto:2008ak,Aharony:2011yc,Chiodaroli:2011fn,Chiodaroli:2012vc,Berdichevsky:2013ija,Bachas:2013vza,Uhlemann:2021nhu,Uhlemann:2021itz,Coccia:2021lpp,Martinec:2022ofs,Karch:2022rvr}. 

As is usual in holography, EOW branes are dynamical objects subject to backreaction and interactions with other fields of the bulk theory. For example, two separated branes can interact by exchanging light bulk fields which may drive them to merge with one another and give rise to novel brane-merging saddles in the gravitational path integral (Figure \ref{figs:cylinderDuals}). In this work, we are interested in how such brane dynamics are encoded in the dual field theory and what they tell us about data of a 2-dimensional BCFT.

In a minimal model consisting of just Einstein gravity and branes of fixed tension \cite{Takayanagi:2011zk}, it has been shown that the merging of two branes of equal tensions is dual to a transition between the closed-string and open-string channels of a BCFT with two circular boundaries. However, the main shortcoming of this model is that it does not include excited states in either channel. The spectrum is thus extremely simple. This is most exemplified by the available boundary-condition-changing (BCC) operators, i.e. the open-string ground states arising when the boundary conditions on the circular boundaries are different \cite{affleck_fermi_1994}. In the minimal model, \cite{Miyaji:2021ktr} only finds BCC operators with scaling dimension at the black hole threshold ($\Delta_{\text{bcc}} = \frac{c}{12}$).


We thus extend the holographic setup by introducing scalar fields, point particles, and non-smooth brane intersections to the model. The scalar fields and point particles lead to long-distance interactions between two highly separated EOW branes on the gravity side, and they can be used to describe excited closed-string states in the BCFT. For scalar-field interactions, we extend the single-brane model used in \cite{Kastikainen:2021ybu} to describe a probe scalar field coupled to two separate branes. The probe scalar field calculation produces the $\text{SL}(2,\mathbb{R})$-character of a light $O(c^0)$ closed-string operator whose dimension is related to the scalar mass. On the other hand, point particle interactions are dual to exchanges of heavy $O(c)$ closed-string and open-string operators whose scaling dimensions lie below the black hole threshold, i.e. $\Delta \in \left(0,\frac{c}{12}\right)$.

As the two branes get close, their interactions become strongly coupled, and the branes eventually merge. When the conformal boundary conditions and the brane tensions are different, these merging configurations cannot be smooth as in \cite{Miyaji:2021ktr}. In 3-dimensional gravity, one is thus forced to consider non-smooth brane intersections, and these allow for ``sub-threshold" BCC operators ($\Delta_{\text{bcc}} < \frac{c}{12}$) in the BCFT (see also \cite{Geng:2021iyq}). To support non-smooth configurations, the bulk theory requires matter content at the brane intersections. We show that it is exactly this matter content that determines the scaling dimension of the BCC operator. We find that the dimension falls into the range $\Delta_{\text{bcc}} \in \left(0,\frac{c}{12}\right)$ without an extra gap of the type argued in \cite{Geng:2021iyq}. In the limit $\Delta_{\text{bcc}} \rightarrow 0$, the intersection becomes smooth, thereby reproducing the configuration of \cite{Miyaji:2021ktr}.

In allowing for non-smooth brane intersections, we look for other types of intersecting configurations. As a result, we find a new Euclidean wormhole saddle (of the bra-ket type \cite{chen_bra-ket_2020}) whose throat is bounded by two non-smoothly intersecting EOW branes. We also find brane mergers that appear to belong to the closed-string sector of the BCFT as black hole states above the threshold. However, these states turn out to be physically problematic---their scaling dimensions depend on the modular parameter, even though such data should be input of a conformal theory---and so their interpretation is subtle.

\subsection*{Overview}

To keep this article self-contained, we first review some facts about BCFT in Section \ref{sec:BCFT}, discussing specifics about BCFT on a cylinder and duality between the open-string and closed-string channels. This machinery makes manifest the idea that conformal boundary conditions may be treated algebraically as ``boundary states" \cite{Ishibashi:1988kg,Onogi:1988qk,Cardy:1989ir}. Our BCFT conventions are defined here.

We then review the AdS/BCFT correspondence in Section \ref{sec:AdSBCFT}, laying out our bulk conventions in the process. Specifically, we consider Einstein gravity in Euclidean spaces with boundaries and corners, which involves the corner Einstein equation proven in Appendix \ref{app:cornereinstein}. We also describe our bulk brane constructions therein.

To illustrate long-distance interactions between branes, we first consider the infinite-width-cylinder limit in Section \ref{sec:scalarCase}. In this limit, one expects exchanges to be mediated by the lightest mode, which for simplicity we assume to be a scalar. We demonstrate that the scalar exchange in the bulk directly reflects the exchange of an SL$(2,\mathbb{R})$ representation between the boundary states.

In Section \ref{sec:cornerCase}, we take the finite-width cylinder. In pure gravity \cite{Takayanagi:2011zk,Fujita:2011fp,Miyaji:2021ktr}, one only considers the sort of saddles shown in Figures \ref{figs:disconnected} and \ref{figs:connectedSmooth}---a very strong constraint. However with gravity furnished by a corner term \cite{Hayward:1993my}, we find the existence of two novel brane-merging saddles of the gravitational path integral that both schematically look like Figure \ref{figs:connectedNonSmooth}. Although the corresponding exchanges in the dual BCFT are strongly coupled, holography gives us access to such physics.

In Section \ref{sec:discussion}, we reiterate our general findings. We also discuss some of the more mysterious configurations, expanding briefly on the relationship between our bra-ket-type wormholes and ensemble averaging, the analytic continuation of particular closed-string black hole states to the open-string channel, and the existence of multi-intersecting configurations.

\begin{quote}
\textit{Note: At the final stages of this work, another paper \cite{Miyaji:2022cma} appeared which also studied intersecting branes in AdS/BCFT. Their results have some overlap with ours on the intersecting annulus branes. In particular, they are also able to achieve BCC operators of any sub-threshold dimension.}
\end{quote}

\section{Review of 2-Dimensional BCFT}\label{sec:BCFT}

We first review the key basic facts about 2-dimensional Euclidean BCFT in a relatively self-contained manner.\footnote{See \cite{Cardy:2004hm} for a canonical review.} Because our goal is to study multi-brane interactions and overlaps of different boundary states, we start by discussing BCFT with two boundaries, each with their own boundary conditions. Specifically, we focus on BCFTs living on a finite cylinder with two circular boundaries.

\subsection{The Open-String and Closed-String Sectors}\label{sec:bcftRev}

When considering two boundaries, the BCFT partition function can be equivalently expanded in two different channels: an open-string channel and a closed-string channel. As indicated by the names, the former describes the worldsheet theory of an open-string running between the two boundaries (Figure \ref{figs:openChannel}) while the latter describes the worldsheet theory of a closed string homotopic to the boundaries (Figure \ref{figs:closedChannel}).\footnote{Regardless of naming conventions, note that the BCFTs considered in this work are not worldsheet theories of any string theory.} The statement that these two ways of computing the partition function give the same result is known as the open-closed-string duality.

\input{figs/openClosedDuality}

In the open-string channel, boundary conditions constrain the values of fundamental fields at the boundaries and appear as restrictions on the open-string spectrum. Meanwhile, in the closed-string channel the boundary conditions are encoded by associated boundary states \cite{Ishibashi:1988kg,Onogi:1988qk,Cardy:1989ir} representing ``initial" and ``final" states of some transition amplitude. The duality between these two channels imposes constraints on the energy spectrum and on the possible boundary conditions that can appear in the BCFT.

\subsubsection*{Open-Closed-String Duality}

Start with a Euclidean cylinder with circumference $\beta$ and width $W$. For now, we keep these parameters general. There is a single dimensionless ``shape modulus" $ \frac{W}{\beta} $ which characterizes this cylinder up to its conformal class. Note that conformal symmetry allows us to rescale $\beta$ and $W$ so long as $\frac{W}{\beta}$ is unchanged.

Consider now the Euclidean path integral $ Z_{AB} $ of a CFT over the cylinder with boundary conditions $A$ and $B$ at the boundaries. There are two ways to slice the path integral and interpret it from the point of view of the operator formalism. One is as a thermal partition function
\begin{equation}
Z^{\text{op}}_{AB} = \text{Tr}\left(e^{-\beta H^{\text{op}}}\right)
\label{openZ}
\end{equation}
of a theory quantized on the interval with boundary conditions $A$ and $B$ at the two ends---the open-string channel (Figure \ref{figs:openChannel}). The other is as a transition amplitude
\begin{equation}
Z^{\text{cl}}_{AB} = \bra{A} e^{-WH^{\text{cl}}} \ket{B}
\label{closedZ}
\end{equation}
in a theory quantized on the circle---the closed-string channel (Figure \ref{figs:closedChannel}). These two quantizations lead to different Hilbert spaces of states $ \mathcal{H}^{\text{op}}_{AB} $ and $ \mathcal{H}^{\text{cl}} $ on which the trace \eqref{openZ} and the matrix element \eqref{closedZ} are respectively evaluated. These two Hilbert spaces are called, respectively, the open-string and closed-string sectors of the CFT. The open-closed-string duality then states that
\begin{equation}
Z_{AB} = Z^{\text{op}}_{AB} = Z^{\text{cl}}_{AB},
\label{duality}
\end{equation}
which imposes constraints on allowed boundary conditions of the BCFT. In string theory, this duality relates closed-string (graviton) amplitudes to open-string excitations (gauge fields). Its classical version corresponds to the double-copy relations \cite{Kawai:1985xq}. Because of conformal invariance and $W$ and $\beta$ being the only scales in the setup, the partition function depends only on $ \frac{W}{\beta} $.\footnote{The open-string Hamiltonian depends on $ W $ and the closed-string Hamiltonian on $ \beta $ such that only the combination $ \frac{W}{\beta} $ appears, as we will see below.}

\subsubsection*{Ishibashi States}

For the boundary conditions to preserve restricted conformal symmetry, the corresponding boundary states $\ket{B}$ have to satisfy
\begin{equation}
\mathcal{L}_{n}\lvert B\rangle = \bar{\mathcal{L}}_{-n}\lvert B\rangle, \quad \forall n\in \mathbb{Z},
\label{conformalBC}
\end{equation}
where $\{\mathcal{L}_n\}$ and $\{\bar{\mathcal{L}}_n\}$ each generate the holomorphic and antiholomorphic copies of the Virasoro algebra. Recall the defining commutation relation of the Virasoro algebra:
\begin{equation}
[\mathcal{L}_n,\mathcal{L}_m] = (n-m)\mathcal{L}_{n+m} + \frac{c}{12} n(n^{2}-1) \delta_{n,-m},\ \ \mathcal{L}_{n}^{\dagger} = \mathcal{L}_{-n}.
\label{viralgebra}
\end{equation}
A general state in $ \mathcal{H}^{\text{cl}} $ which solves \eqref{conformalBC} is a linear combination of \textit{Ishibashi states} \cite{Ishibashi:1988kg,Onogi:1988qk},\footnote{See also \cite{behrend_boundary_2000} for a proof of this statement based on the Schur lemma.}
\begin{equation}
\ket{I_h} =\sum_{\mathfrak{m}}\lvert h,\mathfrak{m} \rangle \otimes \overline{\lvert h,\mathfrak{m}\rangle},\label{ishibashidef}
\end{equation}
where $\mathfrak{m}$ is any multiset of positive integers. We define the holomorphic states as
\begin{equation}
\lvert h,\{m_1,\ldots,m_n\} \rangle = \prod_{k=1}^{n}\mathcal{L}_{-m_k}\lvert h\rangle, \quad \mathcal{L}_0\lvert h\rangle = h \lvert h\rangle, \quad \mathcal{L}_{n>0}\lvert h \rangle = 0,
\end{equation}
with a similar definition for $ \overline{\lvert h,\mathfrak{m}\rangle} $ in terms of $ \bar{\mathcal{L}}_n $. Essentially, each Ishibashi state is obtained by taking a spinless\footnote{That an Ishibashi state can only be constructed from a spinless primary comes from the $n = 0$ case of \eqref{conformalBC}, which is the level-matching condition.} primary $(h,h)$ in the closed-string sector and summing over ``symmetric" descendants for which the holomorphic and antiholomorphic factors are described by the same $\mathfrak{m}$ (i.e. constructed from isomorphic Virasoro generators).

The inner products of Ishibashi states are infinite series that do not converge, so they are nonnormalizable states. However, for any $0 < \tilde{p} < 1$, they satisfy the equality
\begin{equation}
\langle I_h \lvert \tilde{p}^{\frac{1}{2}\left(\mathcal{L}_0 + \bar{\mathcal{L}}_0 - \frac{c}{12}\right)}\lvert I_{h'} \rangle = \delta_{hh'} \chi_{h}(\tilde{p}).\label{normp}
\end{equation}
$\chi_h(\tilde{p})$ is the Virasoro character of the weight $h$ irreducible representation $\mathcal{H}_h$. For $c > 1$ (meaning that there are no null states descended from the $h > 0$ primaries), this is 
\begin{equation}
\chi_h(\tilde{p}) = \Tr_{\mathcal{H}_h}\left(\tilde{p}^{\mathcal{L}_0-\frac{c}{24}}\right) = \begin{cases}
\dfrac{\tilde{p}^{h-\frac{c}{24}}}{\prod_{k=1}^\infty \left(1 - \tilde{p}^k\right)},&\text{if}\ h > 0,\vspace{0.1cm}\\
\dfrac{\tilde{p}^{h-\frac{c}{24}}}{\prod_{k=2}^\infty \left(1 - \tilde{p}^k\right)},&\text{if}\ h = 0.
\end{cases}\label{virasoroChar}
\end{equation}
\eqref{normp} can be seen by reorganizing the sum in \eqref{ishibashidef} to be over descendant level $N_d$ and noting that, for fixed level, each $\mathfrak{m}$ is a partition of $N_d$ (or the empty multiset for $N_d = 0$).

Furthermore, we note the linear transformation rule for Virasoro characters under modular $S$-transformations \cite{lacki_modular_1990,Cardy:2004hm},
\begin{equation}
\chi_h(\tilde{p}) = \sum_{h'}S_{hh'}\chi_{h'}(p),
\label{ishexp}
\end{equation} 
where $p = e^{4\pi^2/\log\tilde{p}}$ is the modular $S$-transform of $\tilde{p}$.

\subsubsection*{Decomposition into Virasoro Characters}

We can use the open-closed-string duality \eqref{duality} to expand the Euclidean path integral on the cylinder as a sum over states in the open-string and closed-string sectors respectively. We show this now.

The open-string sector consists of irreducible representations of the Virasoro algebra,
\begin{equation}
\mathcal{H}^{\text{op}}_{AB} = \bigoplus_{h}\mathcal{N}^{h}_{AB}\mathcal{H}_{h}.
\end{equation}
The coefficients $ \mathcal{N}^{h}_{AB}\geq 0 $ are degeneracy factors of the open-string spectrum. The open-string Hamiltonian $ H^{\text{op}} $ generates translation around the cylinder, which is equivalent to translation along a strip of width $W$. By using a conformal transformation from this strip to the upper half-plane and employing the conformal boundary condition \cite{Cardy:2004hm}, we can relate $H^{\text{op}}$ to the generator of dilatations on the upper half-plane $\mathcal{L}_0$ (Figure \ref{figs:hopenPlane}),
\input{figs/hOpenClosedPlane}
\begin{equation}
H^{\text{op}} = \frac{\pi}{W}\left(\mathcal{L}_0 - \frac{c}{24}\right).\label{openHam}
\end{equation}
Defining $ q = e^{-\pi\beta/W} $, it follows that
\begin{equation}
Z_{AB}^{\text{op}} = \text{Tr}_{\mathcal{H}^{\text{op}}_{AB}}\left({q^{\mathcal{L}_0-\frac{c}{24}}}\right) = \sum_{h}\mathcal{N}_{AB}^{h} \chi_{h}(q),
\label{openZfinal}
\end{equation}

When the path integral is sliced by circles, both copies of the Virasoro algebra are preserved. Thus the closed-string sector instead decomposes according to two copies of the Virasoro algebra \eqref{viralgebra} generated by $ \mathcal{L}_n $ and $ \bar{\mathcal{L}}_n $,
\begin{equation}
\mathcal{H}^{\text{cl}} = \bigoplus_{h,\bar{h}}d_{h\bar{h}} \mathcal{H}_{h}\otimes \bar{\mathcal{H}}_{\bar{h}},
\end{equation}
with a second set of degeneracy factors $ d_{h\bar{h}}\geq 0 $ depending on the circle CFT in question. The closed-string Hamiltonian $ H^{\text{cl}} $ generates longitudinal translation along the cylinder and so, after a conformal transformation, corresponds to the generator of dilatation on the plane $\mathcal{D} = \mathcal{L}_0 + \bar{\mathcal{L}}_0$ \cite{Cardy:2004hm} (Figure \ref{figs:hclosedPlane}),
\begin{equation}
H^{\text{cl}} = \frac{2\pi}{\beta}\left(\mathcal{L}_0+\bar{\mathcal{L}}_0 - \frac{c}{12}\right).\label{closedHam}
\end{equation}
It follows that
\begin{equation}
Z_{AB}^{\text{cl}} = \bra{A} \tilde{q}^{\frac{1}{2}(\mathcal{L}_0+\bar{\mathcal{L}}_0 - \frac{c}{12})}\ket{B},\label{closedstringamplitude}
\end{equation}
where $ \tilde{q} = e^{-4\pi W/\beta} $ is related to $q = e^{4\pi^2/\log\tilde{q}}$ by a modular $S$-transformation:
\begin{equation}
q = e^{2\pi i \omega},\quad \tilde{q} =e^{-\frac{2\pi i}{\omega}}, \quad \omega = \frac{i\beta}{2W}.\label{qqtildedef}
\end{equation}
To decompose this into characters, we use the fact that a general boundary state is a linear combination of Ishibashi states,
\begin{equation}
\lvert B \rangle = \sum_{h}\bra{I_h}\ket{B}\ket{I_h},
\end{equation}
and the orthogonality relation \cite{behrend_boundary_2000}
\begin{equation}
\bra{I_h} \tilde{q}^{\frac{1}{2}\left(\mathcal{L}_0+\bar{\mathcal{L}}_0 - \frac{c}{12}\right)} \ket{I_{h'}} = \delta_{hh'}\chi_{h}(\tilde{q}), \quad 0< \tilde{q} < 1,
\end{equation}
to write the transition amplitude \eqref{closedstringamplitude} as
\begin{equation}
Z_{AB}^{\text{cl}} = \sum_{h} \bra{A} \ket{I_{h}} \bra{I_{h}} \ket{B}\chi_{h}(\tilde{q}).\label{closedCharacterForm}
\end{equation}

\subsubsection*{Open- and Closed-String Limits}

The duality \eqref{duality} between the open-string and closed-string channels leads to constraints on the allowed boundary states---the so-called Cardy conditions. Specifically, recall the definition of the Virasoro character \eqref{virasoroChar} and its linear transformation rule under $S$ \eqref{ishexp}. In conjunction with open-closed-string duality \eqref{duality} and the linear independence of the Virasoro characters, we obtain the following two equivalent constraints:
\begin{align}
\mathcal{N}_{AB}^h &= \sum_{h'} S_{hh'} \bra{A}\ket{I_{h'}}\bra{I_{h'}}\ket{B},\\
\sum_{h} S_{hh'} \mathcal{N}_{AB}^{h} &= \bra{A}\ket{I_{h'}}\bra{I_{h'}}\ket{B}.  
\label{cardyconditions}
\end{align}
These are the Cardy conditions. Boundary states which satisfy these are called \textit{Cardy states}. The coefficients of Cardy states in the Ishibashi-state basis are specified by BCFT data---particularly $S$ and the irreducible representations.

In a BCFT whose data and boundary conditions that obey the Cardy conditions \eqref{cardyconditions}, the Euclidean path integral has the two equivalent series representations
\begin{equation}
    Z_{AB} = \sum_{h}\mathcal{N}_{AB}^{h}\, \chi_{h}(q)  =\sum_{h} \bra{A} \ket{I_{h}} \bra{I_{h}} \ket{B}\chi_{h}(\tilde{q}).
    \label{ZABexpansion}
\end{equation}
Now in the limit $\frac{W}{\beta} \rightarrow 0$ in which $q \rightarrow 0$ and $\tilde{q}\rightarrow 1$, the Virasoro characters behave as
\begin{equation}
    \chi_{h}(q) = q^{h-\frac{c}{24}}+\ldots,\quad \chi_{h}(\tilde{q}) = \sum_{h'}S_{hh'}q^{h'-\frac{c}{24}}+\ldots,\quad  \frac{W}{\beta} \rightarrow 0,
\end{equation}
where we have used the modular transformation law \eqref{ishexp}. We may either use the first of these expansions directly or the second of these expansions in conjunction with the Cardy conditions to write $Z_{AB}$ as
\begin{equation}
    Z_{AB} = \sum_{h}\mathcal{N}_{AB}^{h}\,e^{-\frac{\pi\beta}{W}\left(h-\frac{c}{24}\right)} + \ldots,\quad  \frac{W}{\beta} \rightarrow 0,
    \label{openstringlimit}
\end{equation}
where contributions from open-string sector representations with weight $h \in\left[0,\frac{c}{24}\right)$ are divergent while those with weight $h \in \left(\frac{c}{24},\infty\right)$ are exponentially suppressed. We refer to the $\frac{W}{\beta} \rightarrow 0$ limit as the open-string limit.

In the opposite limit $\frac{W}{\beta} \rightarrow \infty$ in which $q \rightarrow 1$ and $  \tilde{q} \rightarrow 0$, we may similarly and unambiguously expand the partition function as
\begin{equation}
    Z_{AB} = \sum_{h} \bra{A} \ket{I_{h}} \bra{I_{h}} \ket{B}\,e^{-\frac{4\pi W}{\beta}\left(h-\frac{c}{24}\right)}+\ldots,\quad \frac{W}{\beta}\rightarrow \infty,
    \label{closedstringlimit}
\end{equation}
where contributions from closed-string representations with weight $h \in\left[0,\frac{c}{24}\right)$ diverge while those with weight $h \in \left(\frac{c}{24},\infty\right)$ are suppressed. The limit $\frac{W}{\beta} \rightarrow\infty$ is called the closed-string limit.

We are mainly interested in holographic BCFTs with large central charge $c \rightarrow\infty$. In this limit, there are light and heavy states depending on whether their dimension is of the order $\mathcal{O}(c^0)$ or $\mathcal{O}(c)$, respectively. All light states in the open-string (or closed-string) sector give divergent contributions to the open-string (or closed-string) limit, but the same is not true for heavy states: only heavy states $h = \frac{c}{24}(1-\alpha^2) $ with $\alpha \in [0,1]$ give divergent contributions, while the rest of the heavy states are suppressed. The transition from divergent to suppressed contributions happens at $h = \frac{c}{24} $, which is known as the black hole threshold.

\subsection{Boundary Entropy}

The \textit{boundary entropy} is a quantity which represents a particularly basic piece of information about boundary states. Specifically, it is the temperature independent contribution to the thermal entropy of the open-string theory in the thermodynamic limit $W\rightarrow \infty$ \cite{Affleck:1991tk}. The thermal entropy $S_{\text{th}}$ is obtained from the thermal free energy $ F_{\text{th}} = -\beta^{-1}\log{Z_{AB}^{\text{op}}}$ (of the open-string theory) as
\begin{equation}
    S_{\text{th}} = \beta^2 \pdv{F_{\text{th}}}{\beta}.
    \label{Sth}
\end{equation}
We can expand \eqref{Sth} in the thermodynamic limit by noting that it is equivalent to the closed-string limit $\frac{W}{\beta} \rightarrow \infty$ with $\beta$ fixed. Using the duality \eqref{duality} and the expansion \eqref{closedstringlimit}, we get that
\begin{equation}
    Z_{AB}^{\text{op}} = g_Ag_B\,e^{\frac{\pi W}{\beta}\frac{c}{6}} + \ldots, \quad W\rightarrow\infty,
\end{equation}
where we have defined $g_A = \bra{A}\ket{0}$ and $g_B = \bra{0}\ket{B}$. Hence the thermal entropy \eqref{Sth} has the expansion
\begin{equation}
    S_{\text{th}} = \left(\frac{\pi c}{3}\right)\frac{W}{\beta} + s_A + s_B + \ldots, \quad W\rightarrow\infty,
\end{equation}
where $s_A = \log g_A$ and $s_B = \log g_B$ are the boundary entropies associated to the boundary conditions $A$ and $B$ respectively. Because $s_{A,B}$ are independent of the temperature, they give zero-temperature microcanonical entropy, so $g_{A,B}$ are also called ``ground state degeneracies" which can be non-integer in the thermodynamic limit \cite{Affleck:1991tk}.\footnote{The quantity $g_A$ is also called the $g$-function and is similar to the usual $c$-function in that it is monotonic under RG flow \cite{Affleck:1991tk,Friedan:2003yc}, making it a good count of the degrees of freedom associated to a single boundary.} Note that the phase of the closed-string vacuum $\lvert 0 \rangle$ can be chosen such that its overlaps with all boundary states are real and positive, i.e. $\bra{A}\ket{0} = \bra{0}\ket{A}>0$, $\bra{0}\ket{B} = \bra{B}\ket{0}>0$, and $s_{A,B}\in\mathbb{R}$.

\subsection{Boundary-Condition-Changing Operators}

An aspect of BCFT which requires at least two boundaries to describe is the notion of a \textit{boundary-condition-changing (BCC) operator}. Imprints of such operators are seen in correlation functions in a BCFT with different boundary conditions $A \neq B$ \cite{Cardy:2004hm}.

Formally, a BCC operator is defined as the primary of smallest dimension\footnote{Note that the open-string spectrum does not accommodate states with spin because there is only one Virasoro algebra. So, any primary of weight $h$ will have conformal dimension $\Delta = 2h$.} in the open-string spectrum given a pair of different boundary conditions $A,B$. Specifically, we generally require that $\mathcal{N}_{AB}^{0} = \delta_{AB}$. In other words, if the two boundary conditions are the same, then the lowest-dimension operator that may be inserted is the identity (which is unique in a unitary theory) acting on $\mathcal{H}_{AA}^{\text{op}}$. If the two boundary conditions are different, however, then the lowest-dimension operator cannot be the identity and by unitarity must have strictly positive dimension. The dimension $\Delta_{\text{bcc}}$ of the BCC operator can be extracted from the open-string limit $\frac{W}{\beta} \to 0$ \eqref{openstringlimit} of the Euclidean path integral
\begin{equation}
Z_{AB} = \mathcal{N}_{AB}^{\,\text{bcc}}\, e^{-\frac{\pi \beta}{2W} \left(\Delta_{\text{bcc}} - \frac{c}{12}\right)} + \cdots,\ \ \frac{W}{\beta} \to 0.\label{openLimit}
\end{equation}
where $\mathcal{N}_{AB}^{\,\text{bcc}}$ counts the degeneracy of this operator which can be any positive integer. This is the zero-temperature limit of the open-string theory so that $\Delta_{\text{bcc}}$ measures the increase in ground state energy due to different boundary conditions at the end-points \cite{affleck_fermi_1994}.

\section{Extending Bottom-Up AdS/BCFT}\label{sec:AdSBCFT} 

We now review the AdS/BCFT correspondence. We set the speed of light and $\hbar$ to $1$ and $\kappa = 8\pi G_N$, and we work in Euclidean signature. Additionally, while we present the actions in this section in general $d$, we will perform more concrete calculations specifically in $d = 2$. There we may recast gravitational couplings in terms of the boundary central charge through the Brown--Henneaux formula \cite{Brown:1986nw},
\begin{equation}
c = \frac{12\pi \ell}{\kappa},\label{brownHen}
\end{equation}
where $\ell$ is the AdS radius. Since we are concerned with studying the Euclidean BCFT partition function, we will work in Euclidean signature in this paper, using the same conventions for the overall signs of Euclidean actions as \cite{Takayanagi:2011zk,Fujita:2011fp,Miyaji:2021ktr}. However, much of the AdS/BCFT machinery presented here works in Lorentzian signature, as well.

Lastly, for convenience we only write the integration measures (or, more specifically, the differentials) when the coordinates are specified in the integrand.

\subsection{AdS Gravity with Intersecting Branes} 

A holographic $d$-dimensional BCFT is dual to AdS gravity on a $(d+1)$-dimensional manifold $\mathcal{M}$ containing a $d$-dimensional EOW brane $\mathcal{Q}$.\footnote{This discussion, and in particular the notation $\mathcal{Q}$, is schematic. $\mathcal{Q}$ may represent multiple branes which are either disconnected or have a non-smooth intersection in the bulk.} The brane is a boundary of the bulk geometry and thus necessitates the presence of a Gibbons--Hawking--York boundary term \cite{York:1972sj,Gibbons:1976ue}. We then take then this brane to satisfy a dynamical Neumann-type\footnote{We do this as opposed to taking a Dirichlet boundary condition.} boundary condition \cite{Takayanagi:2011zk,Fujita:2011fp} which determines the embedding of $\mathcal{Q}$ in $\mathcal{M}$. A simple toy model for AdS/BCFT is Einstein gravity with a negative cosmological constant on $\mathcal{M}$ and a constant brane tension $T$,
\begin{equation}
I_{\text{G}} = -\frac{1}{2\kappa}\int_{\mathcal{M}} \sqrt{g}\left(R + \frac{d(d-1)}{\ell^2}\right) - \frac{1}{\kappa}\int_{\mathcal{Q}} \sqrt{h}\,(K-T),\label{basicAdSBCFTGrav}
\end{equation}
where $\ell$ is the AdS radius, $h_{ab} = g_{ab} - n_an_b$ is the projector onto $\mathcal{Q}$, $n^a$ is the outward-directed unit normal of $\mathcal{Q}$, $K_{ab} = h^c_ah^d_b\,\nabla_c n_d $ is the extrinsic curvature of $\mathcal{Q}$, and $K = g^{ab}K_{ab}$. The tension term $T$ is sometimes called a Randall--Sundrum (RS) term \cite{Randall:1999vf}. We take the tension to be ``subcritical" ($|T|\ell < d-1$), in which case $\mathcal{Q}$ is a Karch--Randall (KR) brane \cite{Karch:2000ct,Karch:2000gx}. As $T$ is in a one-to-one relationship with possible holographic boundary entropies \cite{Takayanagi:2011zk}, we denote the corresponding boundary state in the dual BCFT as $\ket{T}$.\footnote{We emphasize that $\ket{T}$ is actually defined from bulk parameters. In principle, it may be any boundary state whose boundary entropy is computed by $T$, or it may even be an ensemble average of such boundary states.}

We are interested in studying more comprehensive setups which accommodate two-brane interactions and intersections at corners (represented as $\mathcal{C}$). The action is
\begin{equation}
I = -\frac{1}{2\kappa}\int_{\mathcal{M}} \sqrt{g}\,\left(R+ \frac{d(d-1)}{\ell^2} -\mathcal{L}_\mathcal{M}\right)- \frac{1}{\kappa}\int_{\mathcal{Q}} \sqrt{h}\,(K-\mathcal{L}_\mathcal{Q}) - \frac{1}{\kappa}\int_{\mathcal{C}} \sqrt{\sigma}\,(\Theta-\mathcal{L}_\mathcal{C}),
\label{gravityCorners}
\end{equation}
where $\sigma_{ab} = g_{ab} - n_an_b-t_at_b$ is the projector onto the corner $\mathcal{C}$, $t^a$ is the tangent vector of $\mathcal{Q}$ (so that $n_at^a = 0$), and $\Theta$ is the local intersection angle between the two branes comprising $\mathcal{C}$. The last term involving $\Theta$ is the Hayward corner term \cite{Hayward:1993my}. In addition, we have included arbitrary (for now) bulk $\mathcal{L}_\mathcal{M}$, brane $\mathcal{L}_\mathcal{Q}$ and corner $\mathcal{L}_\mathcal{C}$ matter Lagrangians to the action. 

The variational problem for the action \eqref{gravityCorners} is to keep embeddings of the branes $\mathcal{Q}$ (and hence those of the corners $\mathcal{C}$) fixed while varying the component functions $g^{ab}$ of the inverse metric in the region $\mathcal{M}$ bounded by the branes. Under $\delta g^{ab}$, the variation of the action \eqref{gravityCorners} is\footnote{Note the opposite overall sign multiplying the corner integral.} (see Appendix \ref{app:cornereinstein} for details)
\begin{equation}
\begin{split}
\delta I =\ &-\frac{1}{2\kappa}\int_{\mathcal{M}}\sqrt{g}\,\left(G_{ab} -\frac{d(d-1)}{2\ell^2} g_{ab}+\frac{1}{2} T^{\mathcal{M}}_{ab}\right)\delta g^{ab}\\
&\ \ -\frac{1}{2\kappa}\int_{\mathcal{Q}}\sqrt{h}\,\left(K_{ab}-K h_{ab}+ T^{\mathcal{Q}}_{ab}\right)\delta h^{ab}\\
&\ \ +\frac{1}{2\kappa}\int_{\mathcal{C}} \sqrt{\sigma}\,\left(\Theta\,\sigma_{ab}- T^{\mathcal{C}}_{ab}\right)\delta \sigma^{ab},
\end{split}
\label{fullvariationAbridged}
\end{equation}
where $T_{ab}^{\mathcal{M}}$, $T_{ab}^{\mathcal{Q}}$, and $T_{ab}^{\mathcal{C}}$ are respectively the bulk, boundary, and corner stress tensors:
\begin{equation}
T_{ab}^{\mathcal{M}} = -\frac{2}{\sqrt{g}}\frac{\partial(\sqrt{g}\,\mathcal{L}_\mathcal{M})}{\partial g^{ab}},\ \ T_{ab}^{\mathcal{Q}} = -\frac{2}{\sqrt{h}}\frac{\partial(\sqrt{h}\,\mathcal{L}_\mathcal{Q})}{\partial h^{ab}},\ \ T_{ab}^{\mathcal{C}} = -\frac{2}{\sqrt{\sigma}}\frac{\partial(\sqrt{\sigma}\,\mathcal{L}_\mathcal{C})}{\partial \sigma^{ab}}.
\end{equation}
The variational principle $\delta I = 0$ then produces the usual bulk Einstein equation coupled to matter,
\begin{equation}
G_{ab} - \frac{d(d-1)}{2\ell^2} g_{ab} + \frac{1}{2}T_{ab}^{\mathcal{M}} = 0,
\end{equation}
and the ``boundary" Einstein equation,
\begin{equation}
K_{ab} - K h_{ab} + T^{\mathcal{Q}}_{ab} = 0,
\end{equation}
since we will not impose Dirichlet boundary conditions for the induced metrics of the branes. As a result, the induced metric of the corner $\mathcal{C}$ is not fixed either, so we get the ``corner" Einstein equation
\begin{equation}
\Theta\, \sigma_{ab} - T_{ab}^{\mathcal{C}} = 0.
\end{equation}
An on-shell bulk metric $g_{ab}$ satisfies these three Einstein equations from which the on-shell induced metrics $h_{ab}, \sigma_{ab}$ and the extrinsic data $K_{ab}, \Theta$ are determined. 

In our first calculation of Section \ref{sec:scalarCase}, we will consider a scalar-field sector which couples to $\mathcal{Q}$,
\begin{equation}
\mathcal{L}_\mathcal{M} = \frac{\kappa}{\kappa_{\Phi}}\left(\nabla^a \Phi \nabla_a \Phi + m_{\Phi}^2 \Phi^2\right),\ \ \mathcal{L}_{\mathcal{Q}} = T - \frac{\kappa}{\kappa_{\Phi}} V(\Phi),
\end{equation}
where $T$ is the subcritical brane tension, $V(\Phi)$ is brane localized potential, and $\kappa/\kappa_\Phi$ is a dimensionless normalization factor. This action will provide the model for long range interactions between highly separated branes that do not intersect so that $\mathcal{C} = \varnothing$. Specifically, we will take a probe limit (see Section \ref{sec:eomLims} for details) in which the equations of motion become (with $\ell = 1$):
\begin{equation}
\begin{matrix*}[l]
R_{ab} - \dfrac{1}{2}Rg_{ab} - \dfrac{d(d-1)}{2} g_{ab} = 0,\quad & K_{ab} - (K-T)h_{ab} = 0,\vspace{0.1cm}\\
(\nabla_a \nabla^a - m_\Phi^2)\Phi = 0,\quad & \left[n^a \partial_a \Phi - V'(\Phi)\right]|_{\mathcal{Q}} = 0.
\end{matrix*}
\end{equation}
In our analysis of brane-merging saddles in Section \ref{sec:cornerCase}, we will only work in three bulk dimensions ($d+1 = 3$) and take the matter Lagrangians to be
\begin{equation}
    \mathcal{L}_{\mathcal{M}} = m\delta_{\mathcal{D}},\quad \mathcal{L}_{\mathcal{Q}} =T,\quad \mathcal{L}_{\mathcal{C}} = M,
\end{equation}
where $m$ is the mass of a point particle with worldline $\mathcal{D}$ ($\delta_{\mathcal{D}}$ being a localized Dirac delta function on $\mathcal{D}$), $T$ is again a subcritical brane tension, and $M$ is a tension parameter for the corner $\mathcal{C}$ at the intersection of two branes (and thus called the ``intersection mass"). The resulting stress tensors are given by
\begin{equation}
T_{ab}^{\mathcal{M}} = m g_{ab}\, \delta_{\mathcal{D}},\ \ T^{\mathcal{Q}}_{ab} = Th_{ab},\ \ T^{\mathcal{C}}_{ab} = M\sigma_{ab}.\label{defectStress}
\end{equation}
The corresponding Einstein equations for $g_{ab}$ become (again with $\ell = 1$)
\begin{equation}
R_{ab} - \frac{1}{2}Rg_{ab} -  g_{ab} +\frac{1}{2}\,m\,g_{ab}\, \delta_{\mathcal{D}} = 0, \quad K_{ab} - (K-T)\, h_{ab} = 0, \quad \Theta -M = 0\label{neumann}
\end{equation}
We will solve these equations in the next two sections.


\subsection{Brane Embeddings in Conical AdS$_3$}

We start by solving for geometries sourced by a single point particle and containing a single EOW brane. Intersecting brane configurations that also involve the corner Einstein equation are studied in the next section. First, consider the locally AdS$_3$ metric (with radius $\ell = 1$) given by
\begin{equation}
ds^{2} = f_{\alpha}(r)\,d\tau^{2} + \frac{dr^{2}}{f_{\alpha}(r)} + r^{2}d\phi^{2},
\label{conicalglobalads}
\end{equation}
where $ f_{\alpha}(r) = r^{2} + \alpha^2 $ with $\alpha > 0$ being a free parameter. The coordinates have ranges $\tau \in \mathbb{R}$, $r\geq 0$, and $\phi\in \mathbb{R}$ with $\phi \sim \phi + 2\pi$. Topologically, the space is an infinite solid cylinder, and the conformal boundary is an infinite cylinder equipped with the flat metric $ds^2 = d\tau^2 + d\phi^2$.

The metric is locally AdS$_3$, so it solves the bulk Einstein equation in the region $r>0$. For $\alpha \neq 1 $, the metric has a conical line defect at $r = 0$ with a deficit angle $2\pi\,(1-\alpha)$. The conical line defect has to be supported by a point particle whose mass $m$ is related to $\alpha$. Using the results of \cite{fursaev_description_1995}, we have for the $\tau\tau$-component that
\begin{equation}
    R_{\tau\tau} - \frac{1}{2}Rg_{\tau\tau}-g_{\tau\tau} = -2\pi\,(1-\alpha)\,g_{\tau\tau}\,\delta_{\mathcal{D}},
\end{equation}
so the bulk Einstein equation \eqref{neumann} implies
\begin{equation}
    m = 4\pi\,(1-\alpha).
    \label{pointparticlemass}
\end{equation}
To keep the mass positive, we only consider $\alpha \in [0,1]$. If $\alpha > 1$, then the dual operator breaks the CFT unitarity bound \cite{sarkar_first_2020} (see also Section \ref{sec:cornerCase}).

There are two types of brane embeddings that solve the boundary Einstein equation locally in the point mass background \eqref{conicalglobalads} (Figure \ref{figs:diskstripbranes}).\footnote{See Appendix \ref{app:foliations} to see how these embeddings are constructed as foliations of smooth AdS space. These can be translated into the coordinates \eqref{conicalglobalads}.} We write both in the coordinates of \eqref{conicalglobalads}. The first type consists of ``disk" branes (Figure \ref{figs:diskbrane}) given by
\begin{equation}
\tau = F(r;T,\tau_0) \equiv \tau_0+\frac{1}{\alpha}\text{Tanh}^{-1}{\left(\frac{T\alpha}{\sqrt{f_{\alpha}(r)-T^{2}\,r^{2}}}\right)},\ \ r \geq 0.
\label{diskbrane}
\end{equation}
$T$ denotes the tension,\footnote{In treating this as an EOW brane with tension $T$, we are assuming that the part of the geometry $\tau > F(r;T,\tau_0)$ is being excised. If we excise the complementary region instead, then the tension would be $-T$.} and $\tau_0 \equiv F(\infty;T,\tau_0) $ is a free constant denoting the value of $\tau $ at which the brane asymptotes to the conformal boundary. The disk branes are invariant under $\phi$-translation.

\input{figs/diskstripbranes}

\begin{figure}[t]
\centering
\includegraphics{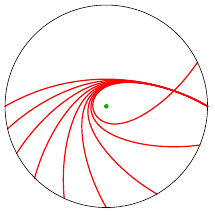}
\caption{Cross sections of $|T| = 1/2$ strip branes for decreasing values of $\alpha$. For $\alpha = 1$, the brane's endpoints are antipodal. For $\alpha < 1\slash 2$, the brane self-intersects in the bulk.} 
\label{figs:wrappingstrips}
\end{figure}

The second type consists of ``strip" branes. These may be written in a branched way as
\begin{equation}
\phi = P(r;T,\phi_0), \quad \phi = \frac{\pi}{\alpha} + 2\phi_0 - P(r;T,\phi_0),
\label{stripbrane}
\end{equation}
where we have defined
\begin{equation}
P(r;T,\phi_0) \equiv 
\phi_0 - \frac{1}{\alpha}\text{Tan}^{-1}{\left(\frac{T\alpha}{\sqrt{r^{2}-T^{2}f_{\alpha}(r)}}\right)},\ \ r \geq \frac{|T|\alpha}{\sqrt{1-T^2}}.
\end{equation}
Here, $\phi_0$ is a free parameter, and the brane intersects the conformal boundary at $\phi = \phi_0 $ and $\phi = \phi_0+\frac{\pi}{\alpha} $. Note that it is more convenient to parameterize the strip branes as
\begin{equation}
r = p(\phi;T,\phi_0),
\end{equation}
where $p(P(r;T,\phi_0);T,\phi_0) = r$. Then we may write each strip brane in terms of a single expression,
\begin{equation}
p(\phi;T,\phi_0) = -\frac{T\alpha}{\sqrt{1-T^{2}}}\csc{[\alpha\,(\phi-\phi_0)]},\ \ \phi\in \left(\phi_0,\phi_0+\frac{\pi}{\alpha}\right),\label{stripbranepfunction}
\end{equation}
where we use the standard convention of polar coordinates where $(r,\phi)$ for $r < 0$ corresponds to $(|r|,\phi+\pi)$. Strip branes project onto lines in the $ (r\cos{(\alpha \phi)},r\sin{(\alpha \phi)}) $-plane such that the zero-tension ($T = 0$) brane runs through the origin:
\begin{equation}
r\sin{(\alpha \phi)}  =  r\cos{(\alpha \phi)}\,\tan{(\alpha \phi_0)}-\frac{T\alpha}{\sqrt{1-T^{2}}}\sec(\alpha \phi_0).\label{rcosrsinline}
\end{equation}
It is enlightening to plot these lines on the Poincar\'e disk, which has the radial coordinate
\begin{equation}
\rho = \frac{-1 + \sqrt{1+r^2}}{r} \in [0,1).
\label{poincaredisk}
\end{equation}
We have done so in Figure \ref{figs:wrappingstrips}. When $\alpha < \frac{1}{2}$, the brane wraps around the defect and intersects with itself in the bulk \cite{Geng:2021iyq}.

We note that disk and strip brane embeddings are related to each other by analytic continuation and changes of variables. By first analytically continuing $\alpha \rightarrow i \alpha$ and then subsequently doing the coordinate transformation $r\rightarrow \sqrt{f_\alpha(r)}$, the disk brane embedding function transforms to a strip brane embedding function:
\begin{equation}
    F(r;T,\tau_0) \rightarrow P(r;T,\tau_0).
    \label{analyticontinuation}
\end{equation}
We will utilize this analytic continuation later to describe a mapping of ``bad" holographic closed-string states to ``good" open-string states.

Disk and strip branes can be used to construct bulk geometries that asymptote to a finite cylinder with modulus $\frac{W}{\beta}$. Depending on the brane tensions, there may be both disconnected and connected brane configurations. The connected configurations can also include non-smooth intersections between branes.

\subsection{Brane Configurations Dual to a Finite Cylinder}\label{subsec:braneconfs}

\begin{figure}[t]
\subfloat[\label{figs:posdisconnected}]
{\includegraphics{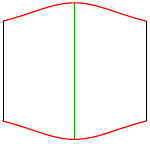}}
\hspace{1cm}
\subfloat[\label{figs:posintersecting}]
{\includegraphics{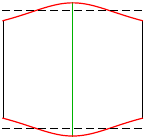}}
\hspace{1cm}
\subfloat[\label{figs:negdisconnected}]
{\includegraphics{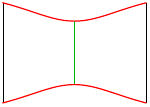}}
\hspace{1cm}
\subfloat[\label{figs:negintersecting}]
{\includegraphics{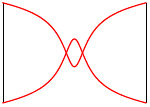}}
\caption{Four types of disk-brane configurations studied in this paper: (a) disconnected positive-tension branes, (b) intersecting positive-tension branes obtained by cutting and gluing along the dashed black lines, (c) disconnected negative-tension branes, and (d) intersecting negative-tension branes.}
\label{figs:diskbraneconfs}
\end{figure}

We now categorize Euclidean brane configurations which asymptote to a finite cylinder of fixed modulus. In particular, this lays the groundwork for our action calculations in Section \ref{sec:cornerCase}. A representative sample of the relevant configurations is shown in Figure \ref{figs:diskbraneconfs}.

\subsubsection*{Non-Intersecting Disk Branes}

Consider two non-intersecting disk branes with different parameters,
\begin{equation}
\tau = -F_1(r) \equiv -F(r;T_1,\tau_1),\quad \tau = F_2(r) \equiv F(r;T_2,\tau_2),
\label{twodisks}
\end{equation}
with $\tau_i > 0$ so that the branes asymptote to $\tau = -\tau_1$ and $\tau = \tau_2$ respectively. To keep two disk branes disconnected, the geodesic distance between these branes $\alpha\left[F_1(r) + F_2(r)\right]$ must be positive. This is equivalent to the following constraint on the brane tensions:
\begin{equation}
\left(\frac{1 + T_1}{1-T_1}\right)\left(\frac{1 + T_2}{1-T_2}\right) > e^{-2\alpha\left(\tau_1 + \tau_2\right)}.\label{tensionIneqAlpha}
\end{equation}
For now, we only focus on branes and values of $\alpha$ that satisfy \eqref{tensionIneqAlpha}. In this case, the bulk region bounded by two disk branes,
\begin{equation}
-F_1(r)<\tau <F_2(r),\ \ r>0,
\label{disconnecteddisksregion}
\end{equation}
asymptotes to a finite cylinder $ (\tau,\phi)\in [\tau_1,\tau_2]\times S^1$ of width $\tau_1+\tau_2$ and modulus
\begin{equation}
\frac{\tau_1 + \tau_2}{2\pi} = \frac{W}{\beta}.\label{twodiskmodulus}
\end{equation}
By setting $\tau_1+\tau_2 = \frac{2\pi W}{\beta}$, we obtain a cylinder of the required modulus on the conformal boundary. Doing so defines a one-parameter family of non-intersecting disk-brane configurations labeled by $\alpha$, with $\alpha = 1$ being the disconnected configuration of \cite{Takayanagi:2011zk}. Imposing \eqref{twodiskmodulus} also allows us to rewrite \eqref{tensionIneqAlpha} in terms of the boundary modulus,
\begin{equation}
\left(\frac{1 + T_1}{1-T_1}\right)\left(\frac{1 + T_2}{1-T_2}\right) > e^{-\frac{4\pi W}{\beta}\alpha} = \tilde{q}^{\alpha}.\label{tensionIneqAlpha2}
\end{equation}
Furthermore, requiring that the branes are non-intersecting for \textit{all} moduli (and in particular as $\frac{W}{\beta} \to 0$) gives the inequality
\begin{equation}
T_1+T_2 > 0.
    \label{tensionIneqStrong}
\end{equation}
In particular, this inequality implies that any two negative tension branes $T_i < 0$ \textit{will} intersect for some finite value of the modulus.

\subsubsection*{Intersecting Disk Branes}

We now assume that the two disk branes intersect and that their intersection is supported by a corner stress tensor. There are two ways to realize such solutions. \textit{A priori}, at a fixed modulus, the disk branes either intersect automatically if they violate the bound \eqref{tensionIneqAlpha2} or may be made to intersect by a cutting-and-gluing procedure otherwise. In the latter class of construction, the conical line defect is exposed to the conformal boundary, whereas it is hidden in the former.

More specifically, keeping $T_1$, $T_2$, and $\tilde{q}$ (or, equivalently, the boundary modulus $\frac{W}{\beta}$) fixed, we may consider three possible scenarios based on \eqref{tensionIneqAlpha2} and the stronger constraint \eqref{tensionIneqStrong}:
\begin{enumerate}[(i)]
\item When the brane tensions satisfy the inequality \eqref{tensionIneqStrong}, they can \textit{only} be made to intersect by cutting and gluing. This also leaves the conical line defect exposed. In this case, we get a two-parameter family of disk-brane configurations labeled by the defect parameter $\alpha$ and the intersection depth of the identified branes $r_*$.

\item When the brane tensions violate \eqref{tensionIneqStrong}, we may still have that \eqref{tensionIneqAlpha2} holds. In this case, in principle we can make the branes intersect by a cutting-and-gluing procedure, leading to the same sort of two-parameter family of configurations as in case (i). However, we can actually prove that this case leads to a contradiction, which implies that cutting and gluing is only mathematically consistent in case (i) if \eqref{tensionIneqStrong} holds---see Appendix \ref{app:proofNoCaseii}.

\item We may consider branes which violate not just \eqref{tensionIneqStrong} but also \eqref{tensionIneqAlpha2}, i.e. branes corresponding to finite boundary modulus but with negative geodesic distance. This time, the conical line defect is behind the branes and thus ``hidden" from the boundary. The configurations in this case comprise a one-parameter family of solutions labeled by $\alpha$.
\end{enumerate}
Let us first consider the cutting-and-gluing constructions of cases (i) and (ii) in detail. We assume that we cut the solid cylinder at $\tau = -\tau_{1*}$ and $\tau = \tau_{2*}$ such that both of the branes will be cut along a circle of radius $r = r_*$. The branes can be glued together at these circles by periodically identifying in the $\tau$-direction.

\input{figs/diskBraneCutGlueTorus}

Another way to think about this construction is as the embedding of two disk branes in a solid torus in a way that makes the branes intersect---see Figure \ref{figs:diskBraneCutGlueTorus}. From this perspective, the freedom associated with the $r_*$ parameter is captured by the circumference of the torus. Additionally, for the cutting-and-gluing to be a well-defined procedure, we must have that
\begin{equation}
\tau_{1*} + \tau_{2*} > \tau_1 + \tau_2.\label{tauconstraint}
\end{equation}
This is just the condition that the width of the boundary cylinder must be shorter than the circumference of the torus obtained from the $\tau$-identification. We can use this to show that case (i) is consistent with a well-defined cutting-and-gluing procedure, whereas case (ii) is not and so need not be considered. See Appendix \ref{app:proofNoCaseii} for details.

Because cutting and gluing requires that \eqref{tensionIneqStrong} be satisfied, we need at least one positive tension brane. Upon following through with the procedure, we have the following bulk regions:
\begin{equation}
\begin{matrix*}[l]
-F_1(r) < \tau < F_2(r),\ \ &r > r_*,\\
-F_1(r_*) < \tau < F_2(r_*),\ \ &r_* \geq r > 0,
\end{matrix*}\label{cutglueBound}
\end{equation}
Additionally, the periodic identification of $\tau$ turns the $r=0$ central line of the cylinder into a circle. 

This construction gives us a two-parameter family of intersecting disk-brane configurations parametrized by $(\alpha,r_*)$. The intersection depth $r_*$ can be traded for the intersection angle $\Theta$, which is given by\footnote{This is found by computing the inner product of the outward-pointing unit normal vectors (given in \eqref{normsBranesDisks}) of the two branes at the intersection depth $r = r_*$.}
\begin{equation}
    \cos{\Theta} = \frac{1}{f_\alpha(r_*)}\biggl(T_1T_2\,r^2_*-\sqrt{f_\alpha(r_*)-T_1^2\,r^2_*}\sqrt{f_\alpha(r_*)-T_2^2\,r^2_*}\biggr).
    \label{intangle}
\end{equation}
Observe that for $0\leq r_*<\infty$, $\Theta$ is bounded to the interval
\begin{equation}
-1\leq\cos{\Theta} <  T_1T_2-\sqrt{1-T_1^2}\sqrt{1-T_2^2}<1.
\end{equation}
The lower bound corresponds to $\Theta = \pi$ where the tips of the two branes barely intersect and the normal vectors of the branes point in opposite directions.

A single solution from the family $(\alpha,\Theta)$ is chosen by the bulk matter content via the Einstein equations: the value of the point particle mass $m$ fixes the deficit angle via \eqref{pointparticlemass} while the intersection mass $M = \Theta$ fixes the intersection angle.

Now, let us then consider the case (iii) which corresponds to the only intersecting construction consistent with $T_1+T_2 \leq 0$. In this case, the disk branes will always intersect for $\tilde{q} \leq \tilde{q}_c$ where
\begin{equation}
\tilde{q}_c^{\alpha} = \left(\frac{1 + T_1}{1-T_1}\right)\left(\frac{1 + T_2}{1-T_2}\right).\label{diskMergeIntBad}
\end{equation}
The intersection angle is given by \eqref{intangle}. However, the intersection depth $r_* = r_*(\tilde{q})$ is no longer a free parameter. This leads us to being able to write $\alpha$ as a function of the boundary modulus. While this dependence is mathematically sensible, we find it to be physically inconsistent because it implies that the dual operator dimension may be treated as a function of the modulus, despite such a quantity being data of the BCFT. Nonetheless, seeing this inconsistency requires performing the bulk analysis, as we do in Section \ref{sec:actiondiskconfs}.

\subsubsection*{Intersecting Annulus Branes}

An alternative construction of the cylinder is realized when quotienting the region bounded by a single infinite strip brane \eqref{stripbrane}. Specifically, we periodically identify $\tau \sim \tau + \tau_0$, where $\tau_0$ corresponds a circumference on the boundary. After cutting and gluing in this manner, an infinite strip brane inherits the topology of an annulus, and so we will call it an annulus brane.\footnote{These are precisely the same annulus branes as those constructed natively in BTZ coordinates. See Appendix \ref{app:stripAnnEq} for details.}

On the conformal boundary, a single annulus brane with tension $T$ has domain $\phi\in\left(\phi_0,\phi_0+\frac{\pi}{\alpha}\right)$. To prevent this brane from self-intersecting, we must take $\alpha > \frac{1}{2}$ as in \cite{Geng:2021iyq}. The size of this interval and the sign of $T$ determine the angular width $\Delta\phi$ of the region included in the bulk. Specifically, a negative-tension brane bounds the interval $\left(\phi_0,\phi_0+\frac{\pi}{\alpha}\right)$, while a positive-tension brane bounds the complementary interval:\footnote{Observe that a $T = 0$ brane will always have a kink unless $\alpha = 1$, and so we cannot have a conical defect for a smooth $T = 0$ annulus brane.}
\begin{equation}
\begin{split}
T > 0 &\implies \Delta\phi = 2\pi - \frac{\pi}{\alpha},\\
T < 0 &\implies \Delta\phi = \frac{\pi}{\alpha}.\label{Twidthsingle}
\end{split}
\end{equation}
Meanwhile, the circumference of the cylinder is $\tau_0$ after quotienting, so taking
\begin{equation}
\tau_0 = \frac{\beta}{W}\Delta\phi\label{tauStrip1}
\end{equation}
gives a cylinder of the required modulus $\frac{W}{\beta}$. The result is a bulk configuration with a single connected brane of fixed tension $T$ interpolating between the two ends of the cylinder. For $\alpha = 1$, this is the connected configuration of \cite{Takayanagi:2011zk}.

\begin{figure}[t]
\subfloat[\label{figs:rotstripbranes}]
{\includegraphics{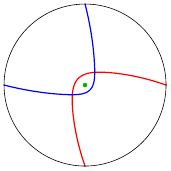}}\qquad
\subfloat[\label{figs:intstripbranes}]
{\includegraphics{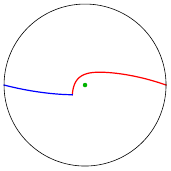}}
\qquad
\subfloat[\label{figs:rotstripbranesneg}]
{\includegraphics{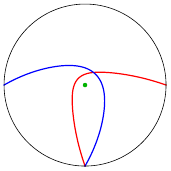}}
\qquad
\subfloat[\label{figs:intstripbranesneg}]
{\includegraphics{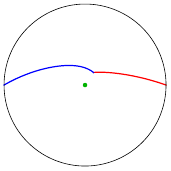}}
\caption{Intersecting annulus brane configurations obtained from taking two different strip branes at $\alpha = \frac{2}{3} < 1$ and rotating them with respect to each other. By taking particular legs of the individual branes that meet at a single intersection point, we get (b) from (a) and (d) from (c). The Lorentzian version of this picture was also considered by \cite{Geng:2021iyq}.}
\label{figs:intstripbraneconf1}
\end{figure}

However, a single annulus brane is not enough if the two brane tensions, and thus the boundary conditions in the dual BCFT, are different, and so configurations with just one annulus brane cannot describe BCC operators in the open-string spectrum \cite{Miyaji:2021ktr}. This problem is remedied by allowing two annulus branes with parameters $(T_1,\phi_{1})$ and $(T_2,\phi_{2})$ to intersect non-smoothly. Assuming that
\begin{equation}
0 \leq \phi_1 < \phi_2 < 2\pi,\ \ \alpha \leq 1,
\end{equation}
the branes intersect non-trivially since the second brane has been rotated counter-clockwise with respect to the first. We may then construct a single, non-smooth brane by keeping particular ``legs" of the two initial annulus branes that meet at an intersection point---see Figure \ref{figs:intstripbraneconf1}. The resulting configuration is an annulus of the form Figure \ref{figs:intersectAnnulus}.

If the circumference $\tau_0$ of the dual cylinder is \eqref{tauStrip1}, then the intersecting configuration will correspond to a different modulus than those of either of the individual branes. Specifically, the width $\Delta \phi$ of this new boundary cylinder is a function of $\phi_1$ and $\phi_2$ depending on which legs and which side of the bulk we keep. Furthermore, the choice of legs determines the relative sign of the two brane tensions, and the absolute signs are set by which side of the bulk we then keep.

For example, assume that we keep the leg of $\mathcal{Q}_1$ that intersects the boundary at $\phi = \phi_1$. If we also keep the leg of $\mathcal{Q}_2$ that intersects the conformal boundary at $\phi=\phi_2$ (e.g. Figure \ref{figs:intstripbranes}), then the two brane tensions will have opposite signs---this is the ``positive-negative" case. We may then either keep the side of the bulk with the interval $\phi \in (\phi_1,\phi_2)$ or its complement. In the former case, we have that $T_1 < 0$, $T_2 > 0$, and $\Delta\phi = \phi_2 - \phi_1$. In the latter case, the tensions have the opposite signs, and $\Delta\phi = 2\pi - (\phi_2 - \phi_1)$.

In contrast, if we keep the $\phi = \phi_1$ leg of $\mathcal{Q}_1$ and the $\phi = \phi_2 + \frac{\pi}{\alpha}$ (modulo $2\pi$) leg of $\mathcal{Q}_2$, then the two brane tensions will have the same sign. In Figure \ref{figs:intstripbranesneg} for example, we may then keep the side of the bulk containing either the interval $\phi\in\left(\phi_1,\phi_2-2\pi+ \frac{\pi}{\alpha}\right)$ or its complement. In the former case, we have $T_{1,2} < 0$ (the ``negative-negative" case) and $\Delta\phi = \phi_2 - \phi_1 - 2\pi +  \frac{\pi}{\alpha}$. In the latter, we have $T_{1,2} > 0$ (the ``positive-positive" case) and $\Delta \phi = 4\pi - \left(\phi_2 - \phi_1 +  \frac{\pi}{\alpha}\right)$.

\input{figs/intersectAnnulus}

To summarize, it is possible to use these intersecting-annulus-brane configurations with \textit{any} combination of two brane tensions to realize BCFT on a cylinder of width $\Delta\phi$. To achieve a modulus $\frac{W}{\beta}$, we must identify $\tau$ such that
\begin{equation}
\tau_0 = \frac{\beta}{W}\Delta\phi.\label{tau0WidthVar}
\end{equation}
For simplicity, from here on out we focus on the positive-negative case; we will see that this is sufficient to show that intersecting-annulus configurations realize BCC operators of any sub-threshold scaling dimension.

So far, for fixed tensions, we have a two-parameter family of intersecting brane configurations $(\alpha,\Theta)$ which are dual to states on a cylinder with fixed modulus $\frac{W}{\beta}$. However, observe that the intersection angle $\Theta$ of the branes is
\begin{equation}
\cos\Theta = \frac{1}{r_*^2}\left(T_1 T_2 f_\alpha(r_*) + \sqrt{r_*^2 - T_1^2 f_\alpha(r_*)} \sqrt{r_*^2 - T_2^2 f_\alpha(r_*)}\right),\label{stripTheta}
\end{equation}
where the intersection depth $r_*$ may be found by writing the branes in the form \eqref{rcosrsinline}:\footnote{Note that we may take $\phi_1 = 0$ and $\phi_2 = \Delta\phi$ without loss of generality. Only the difference in the anchoring points of the legs on the conformal boundary is important here.}
\begin{equation}
r_*^2 = \alpha^2\csc^2{(\alpha\Delta\phi)}\,\biggl(\frac{T_1^2}{1-T_1^2}+\frac{T_2^2}{1-T_2^2}+\frac{2\,T_1T_2\cos{(\alpha\Delta\phi)}}{\sqrt{1-T_1^2}\sqrt{1-T_2^2}}\biggr).\label{striprsGen}
\end{equation}
By plugging \eqref{striprsGen} into \eqref{stripTheta}, we can see that the individual factors of $\alpha$ cancel and $\Theta = \Theta(\alpha\Delta\phi)$. Thus, $\Theta$ may be traded for $\alpha\Delta\phi$. Recalling the equations of motion \eqref{neumann}, we then note that the intersection mass $M$ picks out a particular value for $\alpha\Delta\phi$.

At this stage, it is convenient to restrict ourselves to a particular interval size. When doing so, we are also restricting to a one-parameter family of intersecting-annulus-brane configurations labeled by $\alpha$. The natural value to consider is $\Delta\phi = \pi$, because then the boundary cylinder's width then does not depend on which side of the bulk we keep. In this case, the modulus is
\begin{equation}
\frac{\beta}{W} = \frac{\tau_0}{\pi},
   \label{annulusbranetau0}
\end{equation}
and the intersection depth is
\begin{equation}
r_*^2 = \alpha^2\csc^2{(\pi\alpha)}\,\biggl(\frac{T_1^2}{1-T_1^2}+\frac{T_2^2}{1-T_2^2}+\frac{2\,T_1T_2\cos{(\pi \alpha)}}{\sqrt{1-T_1^2}\sqrt{1-T_2^2}}\biggr).\label{striprs}
\end{equation}
Furthermore, because $M$ picks out some value for $\alpha$, we have a constraint on the point particle mass $m$ (when the conical line defect is included in the geometry):
\begin{equation}
m = 4\pi\left[1-\alpha(M)\right].\label{stabCond}
\end{equation}
One can understand this as a type of stability condition for the interaction between the point particle and the intersection. Indeed, without a point particle (i.e. if $m = 0$), we cannot have a non-smooth intersection of two different-tension branes because $\alpha = 1$ implies that the intersection point \eqref{striprs} runs to the conformal boundary $r_* = \infty$.\footnote{The exception is when $T_1 = T_2$, but for $\Delta\phi = \pi$ this configuration is really just that of a single brane with no conical defect.}

\begin{figure}[t]
\centering
\includegraphics{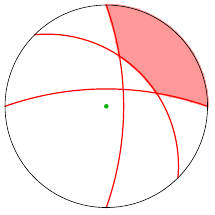}
\caption{A cross-section of an intersecting annulus brane configuration involving three branes. The red shaded region is taken to be the bulk region dual to the BCFT. There are two corners, both of which involve a brane not directly attached to the conformal boundary.} 
\label{figs:doubleintersection}
\end{figure}

Note that this construction from two strips does not exhaustively lead to all possible configurations with a cylindrical boundary. In principle, we may also consider geometries comprised of multiple EOW branes, with only two being anchored to the conformal boundary and the rest ``floating" in the bulk---see Figure \ref{figs:doubleintersection}. This would require multiple corner terms and could accommodate multiple independent intersection masses. We speculate on these configurations in Section \ref{sec:discussion}, but we leave further analysis to future work.

\section{Scalar Exchanges in the Closed-String Limit} \label{sec:scalarCase}

As a proof of concept, we first explore the long-range interaction between two highly separated disk branes. By considering the branes to be far apart, the interaction is mediated by a light bulk field, which we model as being a probe scalar field with mass $m_{\Phi}$ propagating on the disk brane background. We will work in the closed-string limit of the dual BCFT,
\begin{equation}
\frac{W}{\beta} \to \infty,
\end{equation}
which ensures that the disk branes are far apart and in the disconnected phase. Furthermore, we turn off sources at the conformal boundary.



Let us be clear about what exactly we are computing here. Schematically, we will consider a total action $I = I_{\text{G}} + I_{\text{S}}$, where
\begin{align}
I_{\text{G}} &= -\frac{1}{2\kappa}\int_{\mathcal{M}} \sqrt{g}\left[R + \frac{d(d-1)}{\ell^2} \right] - \frac{1}{\kappa}\int_{\mathcal{Q}} \sqrt{h}\,(K-T),\label{scalarGravact}\\
I_{\text{S}} &= \frac{1}{2\kappa_\Phi} \int_{\mathcal{M}} \sqrt{g}\,\left(\nabla^a \Phi \nabla_a \Phi + m_{\Phi}^2 \Phi^2\right) - \frac{1}{\kappa_\Phi} \int_{\mathcal{Q}} \sqrt{h}\, V(\Phi),\label{scalarAct}
\end{align}
as a minimal toy model for interactions between AdS branes. $\kappa^{-1}$ measures the number of dual CFT degrees of freedom and in $d = 2$ is related to the dual CFT central charge $c$ by the Brown--Henneaux formula \eqref{brownHen}. Meanwhile, $\kappa_\Phi^{-1}$ measures the number of bulk scalar degrees of freedom.

As there is no conical deficit angle in the space, the bulk geometry will be smooth. Additionally, we will work in a probe limit $\kappa/\kappa_\Phi \ll 1$. It turns out that $I_\text{S} \ll I_{\text{G}}$ in our regime of interest, and so we will approximate the semiclassical partition function of the Einstein + scalar theory as
\begin{equation}
Z \approx e^{-I_{\text{G}}^{\text{on-shell}}}\left(1 - I_{\text{S}}^{\text{on-shell}}\right),\label{pfDecoupling}
\end{equation}
where we have taken precisely the minimal saddles in the saddle-point approximation and kept the first subleading term in small $I_{\text{S}}$. We will split this subleading term into ``self-energy" and ``exchange" terms. Focusing on the latter, we will then find that it reproduces an SL$(2,\mathbb{R})$ character corresponding to a scalar primary operator exchange in the closed-string channel expansion \eqref{closedCharacterForm} of the BCFT. In other words, we will deduce that the exchange of a scalar between the bulk branes is accounted for by a dual scalar operator and its SL$(2,\mathbb{R})$ descendants in the closed-string sector of the BCFT.



\subsection{Equations of Motion and Convenient Limits}\label{sec:eomLims}

While $\Phi$ is a Klein-Gordon field in the bulk, in principle it may have any potential $V$ on the brane.\footnote{Without loss of generality, we may assume that $V$ does not have a constant term. This is because any such term may be absorbed into the RS tension term.} For now, we will keep $V$ generic. The classical dynamics are then governed by the following four equations of motion (defining $\square = -\nabla_a \nabla^a$):
\begin{align}
G_{ab} - \frac{d(d-1)}{2\ell^2}g_{ab} &= \frac{\kappa}{\kappa_\Phi}\left[\nabla_a\Phi \nabla_b \Phi - \frac{g_{ab}}{2}\left(\nabla_c \Phi \nabla^c \Phi + m_{\Phi}^2 \Phi^2\right)\right],\label{eq1}\\
K_{ab} - \left[K - T + \frac{\kappa}{\kappa_\Phi} V(\Phi)\right]h_{ab} &= 0,\label{eq2}\\
\left(\square + m_{\Phi}^2\right)\Phi &= 0,\label{eq3}\\
\left[n^a \partial_a \Phi - V'(\Phi)\right]|_{\mathcal{Q}} &= 0.\label{eq4}
\end{align}
\eqref{eq1} and \eqref{eq2} are respectively the Einstein equation with scalar matter and the boundary condition dictating the dynamics of the brane, all obtained from varying the action with respect to the metric. \eqref{eq3} and \eqref{eq4} are the Klein-Gordon equations and the boundary conditions on $\Phi$, all from varying the action with respect to $\Phi$. Note that $n^a$ denotes the outward-pointing unit normal vector at the brane $\mathcal{Q}$.

Equations \eqref{eq1}--\eqref{eq4} are necessary to fully analyze backreaction on the metric due to the presence of the scalar field. However, we aim to use the scalar field as a toy model for long-range ``light" interactions between two infinitely-separated branes. To this end, it is convenient to work in particular limits of the theory in which the physics simplifies.

One limit that we will make extensive use of is the \textit{probe limit}. This is described by the following double scaling limit (in units of $\ell=1$):
\begin{equation}
\kappa^{-1} ,\kappa_\Phi^{-1} \to \infty,\ \ \frac{\kappa}{\kappa_\Phi} = \text{fixed} \ll 1.
\label{probelimit}
\end{equation}
Physically, the limit $\kappa_{\Phi}^{-1}\rightarrow \infty$ corresponds to taking the total number of scalar fields to be large. The probe limit \eqref{probelimit} is the statement that the number $\kappa^{-1}\rightarrow\infty$ of D-branes sourcing the AdS geometry is much larger so that the gravitational backreaction of the scalars is negligible. In a top-down context where scalar-field actions can be uplifted to flavor-brane actions \cite{Karch:2002sh}, the limit \eqref{probelimit} is dual to the quenched approximation of the CFT \cite{nunez_unquenched_2010}.

In the probe limit, \eqref{eq1}--\eqref{eq2} reduce to the vacuum equations. Thus, the metric and branes are solutions to pure gravity, and solving \eqref{eq3}--\eqref{eq4} becomes a matter of doing field theory on a curved background and with a field-independent embedding of the boundary. In this regime, obtaining analytic results becomes more feasible.\footnote{It would still be interesting to consider backreaction effects which are neglected in the probe limit. This would likely require numerics. We leave such an analysis to future work.}

Another simple regime is the \textit{large-mass limit}, $m_{\Phi}\ell \to \infty$. Recall that the bulk scalar field is dual to a boundary scalar operator $\mathcal{O}_{\Phi}$ with conformal dimension $\Delta_{\Phi}$ such that \cite{Aharony:1999ti}
\begin{equation}
(m_{\Phi}\ell)^2 = \Delta_{\Phi}(\Delta_{\Phi} - d).\label{massDim}
\end{equation}
Working in standard quantization $\Delta_{\Phi}>\frac{d}{2}$, the large $m_{\Phi}\ell$ corresponds to large $\Delta_{\Phi}$, so the mass-dimension relation \eqref{massDim} simplifies to
\begin{equation}
m_{\Phi}\ell \approx \Delta_{\Phi}.
\end{equation}
It is in this regime that correlation functions of $\mathcal{O}_{\Phi}$ in the BCFT are encoded by geodesics probing classical geometry in the bulk \cite{Balasubramanian:1999zv,Balasubramanian:2012tu,Kastikainen:2021ybu}. Furthermore, the classical bulk fields are easier to compute analytically in this limit, as we will see later in this section.

\subsection{Computing Scalar Actions with Branes}

To solve the boundary condition for the scalar field, we split it as
\begin{equation}
\Phi = \varphi_{\text{B}} + \varphi.
\end{equation}
$\varphi_{\text{B}}$ is the \textit{background field}, which we take as always on-shell. By expanding the brane potential around $\varphi = 0$,
\begin{equation}
V(\Phi) = V(\varphi_{\text{B}}) + V'(\varphi_{\text{B}}) \,\varphi + \frac{V''(\varphi_{\text{B}})}{2}\, \varphi^2 + O(\varphi^3),
\end{equation}
the boundary condition \eqref{eq4} becomes
\begin{equation}
\left.\left[n^a \partial_a \varphi_{\text{B}} + n^a \partial_a \varphi - \left( V'(\varphi_{\text{B}}) + V''(\varphi_{\text{B}})\,\varphi + O(\varphi^2)\right)\right]\right|_{\mathcal{Q}} = 0.\label{orgexpanded}
\end{equation}
Imposing separate boundary conditions on $\varphi_{\text{B}}$ and $\varphi$,
\begin{equation}
\left.\left[n^a \partial_a \varphi_{\text{B}} - V'(\varphi_{\text{B}})\right]\right|_{\mathcal{Q}} =0, \quad  \left.\left[n^a \partial_a \varphi - \left(V''(\varphi_{\text{B}})\,\varphi + O(\varphi^2)\right)\right]\right|_{\mathcal{Q}} = 0,
\label{phiBphiBCs}
\end{equation}
is consistent with \eqref{orgexpanded}, and so the variational principle for $ \Phi $ is well defined.

The purpose of the on-shell background field is to absorb the linear source term $ V'(0) $ of the brane-localized potential. Furthermore, near the conformal boundary we require $ \varphi_{\text{B}} $ to contain the nonnormalizable mode determining the source $ J $ in the dual field theory. In that case, $ \varphi $ is a normalizable fluctuation with a Neumann-type boundary condition \eqref{phiBphiBCs} at the brane, and this fluctuation is integrated over in the bulk path integral.

After integration by parts, the scalar action \eqref{scalarAct} can be written as (defining $\square = -\nabla_a \nabla^a$)
\begin{align}
I_{\text{S}}
&= \frac{1}{2\kappa_\Phi}\int_{\mathcal{M}} \sqrt{g} \left[\varphi_{\text{B}}\left(\square + m_{\Phi}^2\right)\varphi_{\text{B}} + 2\varphi\left(\square + m_{\Phi}^2\right)\varphi_{\text{B}} + \varphi\left(\square + m_{\Phi}^2\right)\varphi\right]\nonumber\\
&\quad - \frac{1}{\kappa_\Phi}\int_{\mathcal{Q}} \sqrt{h}\left[
\biggl(V(\varphi_{\text{B}}) - \dfrac{1}{2}\varphi_{\text{B}} n^a \partial_a \varphi_{\text{B}}\biggl) - \biggl(n^a \partial_a \varphi_{\text{B}} - V'(\varphi_{\text{B}})\biggl)\varphi\right]\nonumber\\
&\quad + \dfrac{1}{2\kappa_\Phi} \int_{\mathcal{Q}} \sqrt{h} \left[n^a \partial_a \varphi -\biggl( V''(\varphi_{\text{B}})\varphi + O(\varphi^2) \biggl)\right]\varphi\nonumber\\
&\quad +\frac{1}{2\kappa_\Phi}\int_{\mathcal{B}} \sqrt{\gamma} \left(\varphi_{\text{B}}r^a \partial_a \varphi_{\text{B}} + 2\varphi\, r^a \partial_a \varphi_{\text{B}} + \varphi\, r^a \partial_a \varphi\right).
\end{align}
We have two boundary terms---one supported on $\mathcal{Q}$ (whose outward-directed unit normal vector is $n^a$) and another on an asymptotic cutoff surface $\mathcal{B}$ (whose outward-directed unit normal vector is $r^a$). With $\varphi_{\text{B}}$ on-shell, the only remaining bulk term is the quadratic term $\varphi(\square + m_{\Phi}^2)\varphi$ which can be integrated over in the bulk path integral in the usual way (with appropriate boundary conditions at the brane). However, in approximating such a path integral by a saddle-point approximation, we also put $\varphi$ on-shell in the bulk. The resulting action consists purely of boundary terms,
\begin{align}
I_{\text{S}}^{\text{on-shell}}
=\ &-\frac{1}{\kappa_\Phi}\int_{\mathcal{Q}} \sqrt{h}
\left[\biggl(V(\varphi_{\text{B}}) - \dfrac{1}{2} \varphi_{\text{B}} n^a \partial_a \varphi_{\text{B}}\biggl) - \biggl(n^a \partial_a \varphi_{\text{B}} - V'(\varphi_{\text{B}})\biggl)\varphi\right]\nonumber\\
&\ \ + \dfrac{1}{2\kappa_\Phi}\int_{\mathcal{Q}}  \sqrt{h} \left[n^a \partial_a \varphi - \biggl(V''(\varphi_{\text{B}})\, \varphi + O(\varphi^2)\biggl)\right]\varphi\nonumber\\
&\ \  +\frac{1}{\kappa_\Phi}\int_{\mathcal{B}} \sqrt{\gamma} \left(\frac{1}{2}\varphi_{\text{B}} r^a \partial_a \varphi_{\text{B}} + \varphi\, r^a \partial_a \varphi_{\text{B}} + \frac{1}{2} \varphi\, r^a\partial_a \varphi\right).\label{onshellscalartext}
\end{align}
We will now take the source to be turned off in the CFT. The $\mathcal{B}$ term in \eqref{onshellscalartext} then vanishes both because $\varphi_{\text{B}}$ and $\varphi$ are normalizable and we are in the standard quantization scheme $\Delta_{\Phi} > \frac{d}{2}$. Thus,
\begin{equation}
\begin{split}
\left.I_{\text{S}}^{\text{on-shell}}\right|_{J = 0}
=\ &-\frac{1}{\kappa_\Phi}\int_{\mathcal{Q}} \sqrt{h} \left[\left(V(\varphi_{\text{B}}) - \frac{1}{2}\varphi_{\text{B}} n^a \partial_a \varphi_{\text{B}}\right) - \biggl(n^a \partial_a \varphi_{\text{B}} - V'(\varphi_{\text{B}})\biggl)\,\varphi\right]\\
&\ \ + \frac{1}{2\kappa_\Phi}\int_{\mathcal{Q}} \sqrt{h}\left[n^a \partial_a \varphi - \biggl(V''(\varphi_{\text{B}})\varphi + O(\varphi^2)\biggl)\right]\varphi,
\end{split}\label{onshellNoSource}
\end{equation}
which after imposing the boundary conditions \eqref{phiBphiBCs} takes the form
\begin{equation}
I_{\text{S}}^{\text{on-shell}} = -\frac{1}{\kappa_\Phi}\int_{\mathcal{Q}} \sqrt{h}\,\biggl[V(\varphi_{\text{B}})-\frac{1}{2}\,\varphi_{\text{B}} V'(\varphi_{\text{B}}) + O(\varphi^3)\biggr].
\label{IQ}
\end{equation}
The $O(\varphi^3)$ terms remain because the boundary condition for $\varphi$ does not generally cancel the $O(\varphi^2)$ term in \eqref{onshellNoSource}. However, the coefficients of these terms are all proportional to three-point or higher-point couplings of the scalar to the brane,\footnote{By keeping careful track of the higher-order terms, we find that they contribute to the action \eqref{IQ} as
\begin{equation}
\left.I_{\text{S}}^{\text{on-shell}}\right|_{O(\varphi^3)} = -\frac{1}{2\kappa_\Phi}\int_{\mathcal{Q}} \sqrt{h} \sum_{k=3}^{\infty} \frac{(2-k)}{k!}V^{(k)}(\varphi_{\text{B}})\varphi^k.
\end{equation}} so in the discussion that follows they identically vanish.

In the following subsection, we will focus on the minimal case of a linear potential
\begin{equation}
V(\Phi) = \lambda \Phi.
\end{equation}
For this potential, the boundary condition \eqref{phiBphiBCs} is
\begin{equation}
\left.\left(n^a \partial_a \varphi_{\text{B}} - \lambda\right)\right|_{\mathcal{Q}} =0, \quad  \left.n^a \partial_a \varphi\right|_{\mathcal{Q}} = 0,
\end{equation}
and we can solve for the background field with this boundary condition in terms of a bulk-to-bulk propagator satisfying a Neumann condition at the brane. Furthermore, the on-shell action \eqref{IQ} is simply
\begin{equation}
I_{\text{S}}^{\text{on-shell}} = -\frac{1}{2\kappa_\Phi}\int_{\mathcal{Q}}\sqrt{h}\,  \lambda \varphi_{\text{B}},\label{linearIQ}
\end{equation}
and so can be evaluated once we solve for $\varphi_{\text{B}}$.

This procedure can also be applied to the case when $ V(\Phi) $ contains higher-order polynomial interactions, but we reiterate that such terms generally modify both the boundary conditions \eqref{phiBphiBCs} and the on-shell action \eqref{IQ}. In fact, the on-shell action with higher-order couplings generally depends on the fluctuation $\varphi$. However, there is one unusual exception---when the potential is quadratic. In this case, just as for the linear potential the action \eqref{IQ} only depends on the background field, but we get Robin boundary conditions on $\varphi_{\text{B}}$ and $\varphi$:
\begin{equation}
V(\Phi) = \lambda \Phi + \frac{1}{2}\eta \Phi^2 \implies \begin{cases}
\left.\left(n^a \partial_a \varphi_{\text{B}} - \lambda - \eta\varphi_{\text{B}}\right)\right|_{\mathcal{Q}} =0,\\
\left.\left(n^a \partial_a \varphi - \eta \varphi\right)\right|_{\mathcal{Q}} = 0.
\end{cases}
\end{equation}
Hence the bulk-to-bulk propagator must satisfy a Robin condition instead of a Neumann one, even though the on-shell action is independent of the fluctuation.

\subsection{Scalar Exchange Between Two Branes}\label{sec:scalarEx}

The discussion thus far has been schematic, with $\mathcal{Q}$ representing some configuration of EOW branes. We are now ready to consider the case where we have two disconnected branes, i.e. $ \mathcal{Q} = \mathcal{Q}_1 \cup\mathcal{Q}_2 $. We respectively furnish these branes with distinct linear potentials:
\begin{equation}
V_{\mathcal{Q}_1}(\Phi) = \lambda_1 \Phi,\ \ V_{\mathcal{Q}_2}(\Phi) = \lambda_2 \Phi.
\end{equation}
Thus, from \eqref{linearIQ}, the on-shell action is
\begin{equation}
I_{\text{S}}^{\text{on-shell}} = -\frac{1}{2\kappa_\Phi}\left(\lambda_1\int_{\mathcal{Q}_1} \sqrt{h_1}\,  \varphi_{\text{B}} + \lambda_2 \int_{\mathcal{Q}_2} \sqrt{h_2} \,\varphi_{\text{B}}\right).\label{onshellscalar}
\end{equation}
To compute this, we have to solve the boundary value problem,
\begin{equation}
(\square + m_{\Phi}^{2})\,\varphi_{\text{B}} = 0,\ \ \left.\left(n_i^a \partial_a \varphi_{\text{B}} - \lambda_i\right)\right|_{\mathcal{Q}_i} = 0,\ \ (i = 1,2).
\label{backgroundfieldproblem}
\end{equation}
To do so, we will use an approach involving the propagator that yields a controlled expansion when the branes are far away from each other.

First, note that the boundary value problem \eqref{backgroundfieldproblem} can be solved using the bulk-to-bulk propagator $ G_{\text{N}}(X,X') $ ($X,X' \in \mathcal{M}$) that obeys Neumann boundary conditions at the two branes,
\begin{equation}
\left(\square + m_{\Phi}^2\right)G_{\text{N}}(X,X') = \delta_{\mathcal{M}}(X,X'),\ \ \left.n_i^a \partial_a G_{\text{N}}(X,X')\right|_{X \to \mathcal{Q}_i} = 0,\ \ (i = 1,2),
\end{equation}
where the derivative in the boundary condition is on $X$. In terms of this propagator, the solution of \eqref{backgroundfieldproblem} is given by \cite{mcavity_heat_1992,McAvity:1992ig}
\begin{equation}
\varphi_{\text{B}}(X) = \lambda_1\int_{\mathcal{Q}_{1}}d^{d}\hat{x}_1\sqrt{h_1}\,G_{\text{N}}(X,\hat{x}_1)+\lambda_2\int_{\mathcal{Q}_{2}}d^{d}\hat{x}_2\sqrt{h_2}\,G_{\text{N}}(X,\hat{x}_2),
\label{phiBtwobranes}
\end{equation}
where $\hat{x}_i$ denotes worldvolume coordinates of $\mathcal{Q}_i$ and we have used the short-hand $ G_{\text{N}}(X,\hat{x}) \equiv G_{\text{N}}(X,E_i(\hat{x})) $, with $ E_i(\hat{x})\in \mathcal{M} $ being the embedding of the point $\hat{x} \in \mathcal{Q}_i$ into the bulk. This expression satisfies the required boundary condition \eqref{backgroundfieldproblem} due to the identity \cite{mcavity_heat_1992}\footnote{Compared to \cite{mcavity_heat_1992}, our $n^a$ is outward-pointing so there is no minus sign on the right-hand side of \eqref{boundarydeltaG}.}
\begin{equation}
\left.n^a_i \partial_a G_{\text{N}}(X,\hat{x}')\right|_{X \to E(\hat{x})} = \delta_{\mathcal{Q}_{i}}(\hat{x},\hat{x}'),\ \ (i=1,2),
\label{boundarydeltaG}
\end{equation}
where the limit is taken from the interior and $\delta_{\mathcal{Q}_{i}}$ is the Dirac delta function on the worldvolume of the brane $\mathcal{Q}_i$. Substituting \eqref{phiBtwobranes} to \eqref{onshellscalar} gives
\begin{equation}
I^{\text{on-shell}}_{\text{S}} =I^{\text{self}}_{\text{S}} + I^{\text{exchange}}_{\text{S}},
\end{equation}
where the scalar self-interaction term is
\begin{equation}
I_{\text{S}}^{\text{self}}
= -\frac{1}{2\kappa_\Phi}\sum_{i=1}^{2} \lambda_i^2 \int_{\mathcal{Q}_i} d^d\hat{x}_i \sqrt{h_i}\int_{\mathcal{Q}_i} d^d\hat{x}'_i \sqrt{h_i'}\, G_{\text{N}}(\hat{x}_i,\hat{x}_i'),\label{braneselfgen}
\end{equation}
and the scalar exchange term is
\begin{equation}
I_{\text{S}}^{\text{exchange}} = -\frac{\lambda_1\lambda_2}{\kappa_\Phi} \int_{\mathcal{Q}_1} d^d\hat{x}_1 \sqrt{h_1} \int_{\mathcal{Q}_2} d^d\hat{x}_2 \sqrt{h_2}\, G_{\text{N}}(\hat{x}_1,\hat{x}_2).\label{braneintgen}
\end{equation}
For our purposes, we focus on the exchange contribution \eqref{braneintgen}. The self-interaction term corresponds to a Feynman diagram with a line whose both points end on the same brane, while the exchange term corresponds to a diagram with a line running between the two branes.

For general brane configurations, solving for the Neumann propagator $ G_{\text{N}} $ is very difficult. However, it can be written formally as an infinite series expansion of nested integrals of the pure AdS$_{d+1}$ bulk-to-bulk propagator $ G(X,X') $, which satisfies Green's equation without any boundary conditions,
\begin{equation}
(\square+m_{\Phi}^{2})\,G(X,X') = \delta_{\mathcal{M}}(X,X').
\end{equation}
The first two terms in the expansion are given by \cite{balian_distribution_1970,balian_distribution_1971,duplantier_exact_1990} as
\begin{equation}
G_{\text{N}}(X,X') = G(X,X') - 2\sum_{i=1}^2 \int_{\mathcal{Q}_{i}}d^{d}\hat{x}_i\sqrt{h_i}\, G(X,\hat{x}_i)\,n_i^a \partial_a G(\hat{x}_i,X') + \cdots.
\end{equation}
In the literature, this is called the \textit{multiple-reflection expansion} for the Neumann propagator, and it also exists for the Dirichlet propagator.

The multiple-reflection expansion truncates at the first term when the proper distance between the two branes is very large. Thus in this limit, we can replace $ G_{\text{N}}(X,X') $ in \eqref{braneintgen} by $ G(X,X') $,
\begin{equation}
I_{\text{S}}^{\text{exchange}} = -\frac{\lambda_1 \lambda_2}{\kappa_\Phi} \int_{\mathcal{Q}_1}d^{d}\hat{x}_1\sqrt{h_1}\int_{\mathcal{Q}_2}d^{d}\hat{x}_2\sqrt{h_2}\,G(\hat{x}_1,\hat{x}_2) + \cdots.
\label{braneintgenapp}
\end{equation}
In the following discussion, we compute this term explicitly for $ d=2 $ and show that it produces the  SL$(2,\mathbb{R}) $ character of a scalar operator in the closed-string channel.

\subsection{Reproducing the Scalar Character}\label{sec:scalarChar}

\input{figs/disconnectedAnnulusDual}

We now focus on the case of a scalar field propagating between two disconnected disk branes in AdS$_3$. The region bounded by the two branes is given in equation \eqref{disconnecteddisksregion} in global AdS$_3$ coordinates and we set the conical defect parameter $\alpha = 1$. Equivalently, we can work in Poincaré AdS coordinates in which the disk branes are hemispheres bounding an annulus with radii $\mathcal{R}_1 < \mathcal{R}_2$ on the conformal boundary (see Figure \ref{figs:disconnectedAnnulusDual}). The ratio $\mathcal{R}_2\slash \mathcal{R}_1 = e^{2\pi W/\beta} $ is related to the modular parameter $\frac{W}{\beta}$ of the cylinder of the CFT. Without loss of generality, we set $\mathcal{R}_1 = 1$ and $\mathcal{R}_2 = \tilde{q}^{-1/2}$ where $\tilde{q} = e^{-4\pi W/\beta}$ is defined in \eqref{qqtildedef}.

In Poincar\'e coordinates, the metric of AdS$_3$ is (setting $\ell = 1$)
\begin{equation}
ds^2 = \frac{dz^2 + dy^2 + dx^2}{z^2},\ \ z > 0,\ \ (y,x) \in \mathbb{R}^2.\label{poincareMetMain}
\end{equation}
In these coordinates, the embeddings \eqref{twodisks} of the two disk branes $\mathcal{Q}_1$ and $\mathcal{Q}_2$ are explicitly (see also Appendix \ref{app:foliations})
\begin{align}
\text{Brane 1:}&\ \ (z - \cot \theta_1)^2 + y^2 + x^2 =  \csc^2 \theta_1,\label{brane1}\\
\text{Brane 2:}&\ \ (z + \tilde{q}^{-1/2} \cot\theta_2)^2 + y^2 + x^2 = \tilde{q}^{-1} \csc^2\theta_2,\label{brane2}
\end{align}
and they intersect the conformal boundary at angles $\theta_1$ and $\theta_2$, respectively (see Figure \ref{figs:disconnectedAnnulusDual}).\footnote{Note that the way in which the angle for $\mathcal{Q}_1$ is defined here is the opposite from how the angles are defined in Appendix \ref{app:foliations}. Thus, we must plug $\pi-\theta_1$ into the expression \eqref{sphereBrane}, which is why we have a minus sign below.} The tensions of the branes are given by
\begin{equation}
T_1 = -\cos\theta_1,\ \ T_2 = -\cos\theta_2.
\label{tensionangle}
\end{equation}
We parametrize the embeddings in terms of the Cartesian coordinates $(y,x)$ so that $ z = z_1(y,x) $ solves \eqref{brane1} and $z = z_2(y,x)$ solves \eqref{brane2}. In terms of the embedding maps, we write that $E_i(y,x) = (z_i(y,x),y,x) \in \mathcal{M}$ (where $i = 1,2$) and the worldvolume coordinates of both branes are $\hat{x} = (y,x)$.

Lastly, we may write the AdS$_3$ bulk-to-bulk propagator in these coordinates. Generally the AdS$_3$ propagator is given by (see for example \cite{hijano_witten_2015})
\begin{equation}
G(X,X') = \frac{1}{2\pi}\frac{e^{-\Delta_{\Phi} L}}{1-e^{-2L}},\label{adsProp}
\end{equation}
where $L = L(X,X') $ is the geodesic distance between two bulk points and $\Delta_{\Phi} >\frac{d}{2}$ is given by the mass $m_{\Phi}$ through \eqref{massDim}. In Poincar\'e coordinates with $X = (z,y,x)$ and $X' = (z',y',x')$, this distance is
\begin{equation}
L = \log{\biggl(\frac{1 + \sqrt{1-\xi^{2}}}{\xi}\biggr)}, \quad \xi = \frac{2zz'}{z^2 + z'^2 + (y-y')^2 + (x-x')^2}.
\end{equation}
Thus, to leading order in the multiple-reflection expansion, the scalar exchange term \eqref{braneintgenapp} becomes
\begin{equation}
I^{\text{exchange}}_{\text{S}} = -\frac{\lambda_1\lambda_2}{2\pi\kappa_\Phi}\int_{\mathcal{Q}_1}d^{2}\hat{x}_1\int_{\mathcal{Q}_2}d^{2}\hat{x}_2\sqrt{h_1}\sqrt{h_2}\,\frac{e^{-\Delta_{\Phi} L}}{1-e^{-2L}},\label{ourintegral}
\end{equation}
where the geodesics being integrated over are restricted to those connecting $\mathcal{Q}_1$ and $\mathcal{Q}_2$.\footnote{Note that $L$ in this integral should be read as a function of two points with one on each brane, i.e. as a function of four free variables, rather than as a function of two bulk points consisting of six free variables.} The product of induced metric determinants for disk branes \eqref{brane1}--\eqref{brane2} is
\begin{equation}
\sqrt{h_1}\sqrt{h_2} = \frac{\tilde{q}^{-1/2} \csc{\theta_1}\csc{\theta_2}}{z_1(y_1,x_1)^{2}z_2(y_2,x_2)^{2}\sqrt{(\csc^{2}{\theta_1}-y_1^{2}-x_2^2)(\tilde{q}^{-1}\csc^{2}{\theta_2}-y_2^{2}-x_2^2)}}.
\end{equation}
Analytically performing this integral for general $ \Delta_{\Phi} $ is unfeasible. Hence we consider the large-mass limit $\Delta_{\Phi} \to \infty$ where the integral can be computed using the saddle-point approximation. To this end, we introduce the vector $ \vec{v} = (x_1,y_1,x_2,y_2) $ so that the integral is computed in $ \mathbb{R}^{4} $, and we denote the bulk geodesic anchored to the branes at $E_1(y_1,x_1)$ and $E_2(y_2,x_2)$ as $ L(\vec{v}) $. For a general integral over any open subset $\Omega$ of $ \mathbb{R}^{n} $,\footnote{The shape of the integration region is not relevant in the saddle-point approximation due to the rapid exponential damping of the integral away from the saddle-point.} the saddle-point approximation states that
\begin{equation}
\int_{\Omega} \prod_{i=1}^{n}dv_i\,g(\vec{v}) e^{-\Delta_{\Phi} L(\vec{v})} = \left(\frac{2\pi}{\Delta_{\Phi}}\right)^{n/2}\frac{g(\vec{v}_*)\, e^{-\Delta_{\Phi} L(\vec{v}_*)}}{\sqrt{\det\left(L_{ij}\right)}}\left[1+\mathcal{O}(\Delta_{\Phi}^{-1})\right], \label{saddlegeneral}
\end{equation}
where $ \vec{v}_{*}\in \Omega $ is a minimum of $ L(\vec{v}) $ and $L_{ij}$ is the Hessian of $L(\vec{v})$ at the saddle-point,
\begin{equation}
L_{ij} = \biggl(\frac{\partial^{2}L}{\partial v_{i}\partial v_{j}}\biggr)\biggr\lvert_{\vec{v} = \vec{v}_{*}}.
\end{equation}
The minimal geodesic runs between the inner-most points, i.e. the turning points, of the two branes which are located at the origin of the transverse plane $ \vec{v} = 0 $. The radial depths $z_{1*},z_{2*}$ of these two points are
\begin{equation}
z_{1*} = \cot{\frac{\theta_1}{2}},\ \  z_{2*} = \tilde{q}^{-1/2}\tan{\frac{\theta_2}{2}},\label{zstars}
\end{equation}
such that $ z_{2*} > z_{1*} $. Thus this geodesic has length
\begin{equation}
L_{*} = \log{\frac{z_{2*}}{z_{1*}}} = \log{\left(\tilde{q}^{-1/2}\tan{\frac{\theta_1}{2}}\tan{\frac{\theta_2}{2}}\right)}.
\label{lengthLs}
\end{equation}
Applying \eqref{saddlegeneral} with $ n=4 $ to our integral \eqref{ourintegral}, we get
\begin{equation}
I^{\text{exchange}}_{\text{S}}
= -\frac{2\pi}{\kappa_\Phi} \frac{\lambda_1\lambda_2}{\Delta_{\Phi}^{2}}\frac{1}{z_{1*}^{2}z_{2*}^{2}\sqrt{\det\left(L_{ij}\right)}}\frac{e^{-\Delta_{\Phi} L_{*}}}{1-e^{-2L_{*}}},\ \ \Delta_{\Phi} \to \infty.\label{saddleint}
\end{equation}
which is valid up to $ \mathcal{O}(\Delta_{\Phi}^{-3}) $ corrections. Furthermore, we may compute the saddle-point values of the components of the Hessian explicitly.\footnote{We do so by differentiating \eqref{brane1} and \eqref{brane2} to solve for the first and second derivatives of $z_i(y_i,x_i)$, then setting $y_i = x_i = 0$ and using \eqref{zstars}.} Because the Hessian is symmetric, we only need to evaluate 10 of its components:
\begin{align}
L_{11} &= L_{22} = \frac{z_{1*}^{2}+z_{2*}^{2}+(z_{2*}^{2}-z_{1*}^{2})\cos{\theta_1}}{z_{1*}^{2}(z_{2*}^{2}-z_{1*}^{2})},\\
L_{33} &= L_{44} = \frac{z_{1*}^{2}+z_{2*}^{2}+(z_{2*}^{2}-z_{1*}^{2})\cos{\theta_2}}{z_{2*}^{2}(z_{2*}^{2}-z_{1*}^{2})},\\
L_{13} &= L_{24} = \frac{2}{z_{1*}^{2}-z_{2*}^{2}},\\
L_{12} &= L_{14} = L_{23} = L_{34} = 0.
\end{align}
We then find that
\begin{equation}
\frac{1}{\sqrt{\det(L_{ij})}} = \frac{(z_{2*}^{2}-z_{1*}^{2})z_{1*}^{2}z_{2*}^{2}}{(1+\cos{\theta_1})(1+\cos{\theta_2})z_{2*}^{2}-(1-\cos{\theta_1})(1-\cos{\theta_2})z_{1*}^{2}}.\label{hessianDet1}
\end{equation}
Using \eqref{zstars} and \eqref{lengthLs}, we can write \eqref{hessianDet1} as
\begin{equation}
\frac{1}{\sqrt{\det(L_{ij})}} = \frac{\csc{\theta_1}\csc{\theta_2}}{1-\tilde{q}}\frac{1-e^{-2L_*}}{\cot{\frac{\theta_1}{2}}\cot{\frac{\theta_2}{2}}}\,z_{1*}^{2}z_{2*}^{2}.
\label{geodesic1loop}
\end{equation}
Substituting into \eqref{saddleint}, we get
\begin{equation}
I^{\text{exchange}}_{\text{S}} = -\frac{2\pi}{\kappa_\Phi}\frac{\lambda_1\lambda_2}{\Delta_{\Phi}^{2}}\frac{\csc{\theta_1}\csc{\theta_2}}{\cot\frac{\theta_1}{2}\cot\frac{\theta_2}{2}}\frac{e^{-\Delta_{\Phi} L_{*}}}{1-\tilde{q}}, \quad \Delta_{\Phi} \rightarrow \infty.
\end{equation}
Since $ e^{-\Delta_{\Phi} L_{*}}=\tilde{q}^{\Delta_{\Phi}/2}\left(\cot{\frac{\theta_1}{2}}\cot{\frac{\theta_2}{2}}\right)^{\Delta_{\Phi}}$, we finally write the scalar exchange term explicitly as
\begin{equation}
I_{\text{S}}^{\text{exchange}} = -\frac{2\pi}{\kappa_\Phi}\,\biggl(\frac{\lambda_1\csc{\theta_1}}{\Delta_{\Phi}}\biggr)\biggl(\frac{\lambda_2\csc{\theta_2}}{\Delta_{\Phi}}\biggr)\biggl(\cot{\frac{\theta_1}{2}}\cot{\frac{\theta_2}{2}}\biggr)^{\Delta_{\Phi}-1}\frac{\tilde{q}^{\Delta_{\Phi}/2}}{1-\tilde{q}}, \quad \Delta_{\Phi} \rightarrow \infty.
\end{equation}
Lastly, recall that the integrand of the path integral also incorporates the pure gravitational action with two disconnected branes. We will compute its on-shell value $I_\text{G}^{\text{on-shell}}$ with a defect present in Section \ref{sec:actionADM}. For now, we take the result \eqref{actionnonintdiskbranes} and set $\alpha = 1$ to write
\begin{equation}
\begin{split}
&I_{\text{G}}^{\text{on-shell}} = -\frac{c}{6}\left(\frac{\pi W}{\beta} + \text{Tanh}^{-1} T_1 + \text{Tanh}^{-1} T_2\right)\\
&\implies e^{-I_{\text{G}}^{\text{on-shell}}} = \tilde{q}^{-\frac{c}{24}} \left(\frac{1 + T_1}{1-T_1}\right)^{\frac{c}{12}} \left(\frac{1 + T_2}{1-T_2}\right)^{\frac{c}{12}}.
\end{split}
\end{equation}
We note that $I_{\text{S}}^{\text{on-shell}} \ll I_{\text{G}}^{\text{on-shell}}$,\footnote{It is reasonable to think that the $\lambda$ couplings may compensate for the fact that $\kappa/\kappa_\Phi \ll 1$. However, as we are working in the closed-string limit, $I_\text{S}^{\text{on-shell}}$ will still be subleading to $I_{\text{G}}^{\text{on-shell}}$ with respect to $\tilde{q}$.} and so we may use the approximation \eqref{pfDecoupling} to write the contribution of the exchange term to the partition function as
\begin{equation}
Z_{\text{exchange}} = -e^{-I_{\text{G}}^{\text{on-shell}}} I_{\text{S}}^{\text{exchange}},
\end{equation}
Putting everything together, we get
\begin{equation}
Z_{\text{exchange}} = \frac{2\pi}{\kappa_\Phi}\,\prod_{i=1}^2 \frac{\lambda_i}{\Delta_{\Phi}(1-T_i)}\left(\frac{1+T_i}{1-T_i}\right)^{\frac{c}{12}-h_{\Phi}} \chi_{h_{\Phi}}^{\text{SL}(2,\mathbb{R})}(\tilde{q}),
\label{finalZexchange}
\end{equation}
where $h_{\Phi} = \frac{\Delta_{\Phi}}{2}$ is the weight of the dual scalar operator (which is spinless). Additionally, we have used \eqref{tensionangle} and defined
\begin{equation}
    \chi_{h}^{\text{SL}(2,\mathbb{R})} = \frac{\tilde{q}^{h-\frac{c}{24}}}{1-\tilde{q}}.
\end{equation}
This is the character of an SL$(2,\mathbb{R})$ irreducible representation of weight $h$.\footnote{Recall that this character is computed as the trace of $\tilde{q}^{\mathcal{L}_0 - c/24}$ over the SL$(2,\mathbb{R})$ subalgebra (generated by $\{\mathcal{L}_{-1},\mathcal{L}_0,\mathcal{L}_1\}$) of the Virasoro algebra,
\begin{equation}
\chi_{h}^{\text{SL}(2,\mathbb{R})}(\tilde{q}) = \Tr_{\text{SL}(2,\mathbb{R})}\left(\tilde{q}^{\mathcal{L}_0 - c/24}\right) = \sum_{N=0}^\infty \tilde{q}^{h + N - c/24} = \frac{\tilde{q}^{h-c/24}}{1-\tilde{q}}.
\end{equation}}
We can see that $Z_{\text{exchange}}$ gives exactly the contribution of a scalar primary state $\lvert \Delta_{\Phi}\rangle$ and its SL$(2,\mathbb{R})$ descendants to the closed-string limit \eqref{closedstringlimit} of the Euclidean path integral of the BCFT. From \eqref{finalZexchange} we can then identify the overlaps
\begin{equation}
    \bra{T_i}\ket{\Delta_{\Phi}} = \sqrt{\frac{2\pi}{\kappa_\Phi}}\frac{\lambda_i}{\Delta_{\Phi}(1-T_i)}\left(\frac{1+T_i}{1-T_i}\right)^{\frac{c}{12}-h_{\Phi}}.
    \label{scalaroverlaps}
\end{equation}
which are valid at leading order in the large-$\Delta_{\Phi}$ limit. Subleading corrections to the geodesic approximation \eqref{saddlegeneral} only modify the coefficient of the character by terms of order $ \mathcal{O}(\Delta_{\Phi}^{-3}) $. Hence the $ \tilde{q} $-dependence remains the same at finite $ \Delta_{\Phi} $, which is expected from the CFT side.

We conclude that a scalar field in the disk-brane background describes a light closed-string state whose dimension is fixed by the scalar mass and whose overlaps with boundary states are given by \eqref{scalaroverlaps}. The reason why we obtained contributions from global descendants is that we took into account fluctuations around the minimal geodesic running between the tips of two the branes; the 1-loop correction \eqref{geodesic1loop} to the geodesic approximation contains the factor $(1-\tilde{q})^{-1}$ encoding the descendants. These fluctuations in the geodesic position can be equivalently described as small fluctuations of the classical background metric that correspond to tree level graviton--scalar interactions. The appearance of the full Virasoro character would require taking into account 1-loop interactions between gravitons and the scalar field which goes beyond our large-$c$ classical bulk computation above. We leave exploration of such subleading effects to future work.

\section{Conical Defect Exchanges at Finite Modulus} \label{sec:cornerCase} 

We now depart from the $\frac{W}{\beta} \to \infty$ regime dominated by light brane interactions in the bulk and focus on finite moduli. Geometrically, this corresponds to the branes being close so that merging configurations become possible.

In the bulk, we schematically have three possible EOW brane configurations corresponding to a cylinder of fixed modulus on the boundary (as shown in Section \ref{subsec:braneconfs}): two disconnected ``disk" branes (Figure \ref{figs:disconnected}), one smooth and connected ``annular" brane (Figure \ref{figs:connectedSmooth}), and one non-smooth and connected brane (Figure \ref{figs:connectedNonSmooth}). In addition to branes and their intersections, each configuration also contains a conical line defect. 

In this section, we compute Euclidean on-shell actions of these brane configurations and extract dual BCFT data. We compute the actions by writing them as boundary integrals consisting of the ADM mass and the Wald entropy. We find that disconnected-disk-brane configurations are dual to heavy closed-string states whose dimensions are of order $O(c)$ and below the black hole threshold, i.e. $\Delta^{\text{cl}} < \frac{c}{12}$. Similarly, intersecting annulus brane configurations are dual to heavy open-string states whose dimensions are also below the threshold. We identify these open-string states with BCC operators, and we demonstrate that the gap $\Delta_{\text{bcc}} \in \left(0,\frac{c}{12}\right)$ can be filled.

\subsection{Euclidean On-Shell Action and the ADM Mass}\label{sec:adm}

We first discuss how to compute Euclidean on-shell actions of our bulk brane configurations as boundary integrals. We can do so because all of our bulk geometries have a global Killing vector that can be used to foliate the spacetime. As a result, the gravitational on-shell action reduces to a difference between the Arnowitt--Deser--Misner (ADM) mass and the Wald entropy, each of which are codimension-2 integrals.

Consider a Euclidean manifold that has a Killing vector field $ \xi^{a} $ obeying $\nabla^{(a}\xi^{b)} = 0$. The vector $ \xi^{a} $ generates a flow, and we foliate the space by Cauchy slices $ \Sigma_{u} $ labeled by a parameter $ u\in (0,u_0) $ (which we will suppress for convenience)---see Figure \ref{figs:admSlices} for a pair of examples in conical AdS$_3$. These Cauchy slices are orthogonal to the flow lines of $ \xi^{a} $. This is essentially a general Euclidean version of the ADM decomposition \cite{Arnowitt:1959ah} in which the metric takes the form
\begin{equation}
ds^2 = N^2 du^2 + ds^2_{\Sigma}.\label{admMet}
\end{equation}

\input{figs/admSlices}

\noindent The Killing vector satisfies $ \xi_a\xi^{a} = N^{2} $, and so the unit normal vector field associated with it is $ u^{a} = N^{-1}\,\xi^{a} $ such that $ u_a u^{a} = 1$. Since $ \xi^a$ is a Killing vector, the lapse does not depend on $u$, i.e. $ \partial_u N = 0 $. We will assume that the $u$-direction is also periodic with period $u_0$ so that there are no future or past boundaries in the $u$-direction. This turns out to be the case for our brane configurations.

We start by considering the BCFT Hamiltonian which is an integral of the stress tensor. On the gravity side, it corresponds to the on-shell value of the gravitational Hamiltonian. This is a pure boundary term in the presence of a Killing vector, and it is identified as the \textit{ADM mass} \cite{wald_black_1993,iyer_properties_1994},\footnote{There is a choice of orientation for the surface $\mathcal{B}$ which shows up as a sign in the first term of \eqref{ADMmass}. We have chosen the orientation such that $r^a$ is outward pointing and $u^a$ points towards the direction of increasing $u$. If we were foliating a Lorentzian spacetime with $u^a$ being timelike and future directed, then we would get an additional minus sign in the first term of \eqref{ADMmass}.}
\begin{equation}
M_{\text{ADM}} = -\frac{1}{\kappa}\int_{\mathcal{B}\cap \Sigma}\sqrt{\hat{\gamma}}\,\bigl(u_{a}r_{b}\, \nabla^{[a}\xi^{b]} + N\mathcal{K}\bigr),
\label{ADMmass}
\end{equation}
where $ \mathcal{B}\cap\Sigma $ is a slice of the cutoff surface $\mathcal{B}$ near infinity, $r^a$ is its the outward-pointing unit normal vector (which is orthogonal to $u^a$), and
\begin{equation}
\gamma_{ab} = g_{ab} - r_ar_b, \quad \hat{\gamma}_{ab} = g_{ab} -r_ar_b - u_au_b,\quad \mathcal{K}_{ab} = \gamma^{c}_{a}\gamma^{d}_{b}\,\nabla_{c}r_{d}.
\end{equation}
The projector $\gamma$ gives the induced metric on $\mathcal{B}$, and the projector $\hat{\gamma}$ gives the induced metric on $\mathcal{B} \cap \Sigma$. The first term in \eqref{ADMmass} is proportional to the Komar mass, which is the Noether charge of the Killing symmetry at infinity, and the second term arises from the Gibbons--Hawking--York boundary term of the gravitational action \cite{wald_black_1993,iyer_properties_1994}.

In asymptotically locally AdS$ _{d+1} $ spaces, the expression \eqref{ADMmass} for the ADM mass is divergent. It can be renormalized by adding a counterterm $ \mathcal{L}_{\mathcal{B}} $ that only depends on the induced metric $\gamma$ of the cutoff surface $ \mathcal{B} $:
\begin{equation}
M_{\text{ADM}}^{\text{ren}} = -\frac{1}{\kappa}\int_{\mathcal{B}\cap \Sigma}\sqrt{\hat{\gamma}}\,\bigl(u_{a}r_{b}\, \nabla^{[a}\xi^{b]} + N\,(\mathcal{K}-\mathcal{L}_{\mathcal{B}})\bigr).
\label{renadm}
\end{equation}
This (renormalized) ADM mass satisfies the equation \cite{wald_black_1993}
\begin{equation}
    I^{\text{ren}}_{\text{on-shell}} = u_0\,\biggl(M_{\text{ADM}}^{\text{ren}} - \frac{\kappa_{\text{s}}}{2\pi}\,S_{\text{W}}\biggr), \quad S_{\text{W}} = \frac{2\pi}{\kappa}\,A_{\mathcal{H}},
\label{definingeq}
\end{equation}
where $I^{\text{ren}}_{\text{on-shell}}$ is the renormalized Euclidean on-shell action, $A_{\mathcal{H}}$ is the area of a horizon acting as a boundary in the Cauchy slice (with $S_{\text{W}}$ being the associated \textit{Wald entropy functional} \cite{wald_black_1993} in Einstein gravity), and $\kappa_{\text{s}}$ is the surface gravity of the horizon. In Euclidean signature, a horizon corresponds to a surface where the $u$-circle shrinks to zero size.\footnote{If the $u$-direction is non-compact, there are no horizons and the temperature is zero.} For a given a gravitational theory on the left-hand side, we emphasize this equation can be seen as the \textit{definition} of the ADM mass (and Wald entropy) in that theory.

The punchline is that we can calculate the Euclidean on-shell action from the ADM mass and Wald entropy, and so this is how we perform our later calculations. Prior to doing so, let us discuss how the expression for the ADM mass \eqref{renadm} can be simplified. We define
\begin{equation}
    \hat{g}_{ab} = g_{ab} - u_au_b,\quad \widehat{\mathcal{K}}_{ab} = \hat{g}^{c}_a\,\hat{g}^{d}_b\,\mathcal{K}_{cd}
\end{equation}
which are respectively the projector onto the slice $\Sigma$ and the extrinsic curvature of the cutoff surface $\mathcal{B}\cap \Sigma$ as embedded in the slice. We can then write the trace of the extrinsic curvature as
\begin{equation}
\mathcal{K} = \widehat{\mathcal{K}} + u^{a}u^{b}\mathcal{K}_{ab} = \widehat{\mathcal{K}}- N^{-1}\,u_{a}r_b\,\nabla^{[a}\xi^{b]},
\label{LtoLhat}
\end{equation}
where we have used $ r_au^{a} = 0 $, $u^a = N^{-1}\,\xi^a$, and $\nabla^{(a}\xi^{b)} = 0$. Substituting to the formula \eqref{renadm} for the ADM mass, we get\footnote{This can be seen as the ``absolute" ADM mass. The ``relative" ADM mass is the difference
\begin{equation}
M_{\text{ADM}}^{\text{rel}} \equiv M_{\text{ADM}}^{\text{ren}}-M_{\text{ADM}}^{\text{ren}}\lvert_{(0)}  = -\frac{1}{\kappa}\int_{\mathcal{B}\cap \Sigma}N\sqrt{\hat{\gamma}}\, \bigl(\widehat{\mathcal{K}}-\widehat{\mathcal{K}}\lvert_{(0)}\bigr),
\end{equation}
and is the formula originally presented in \cite{hawking_gravitational_1996} in asymptotically flat spaces.} 
\begin{equation}
	M_{\text{ADM}}^{\text{ren}} = -\frac{1}{\kappa}\int_{\mathcal{B}\cap \Sigma}N\sqrt{\hat{\gamma}}\, (\widehat{\mathcal{K}}-\mathcal{L}_{\mathcal{B}}),
	\label{absoluteADM}
\end{equation}
where the contribution from the Noether charge has cancelled. This form of the mass allows for easier computation.

The formula \eqref{renadm} (and \eqref{absoluteADM}) is valid when there are no additional boundaries such as EOW branes present in the geometry. In the presence of intersecting branes with matter content $\mathcal{L}_{\mathcal{Q}}$ and $\mathcal{L}_{\mathcal{C}}$, we claim that the formula \eqref{renadm} is modified to
\begin{align}
M_{\text{ADM}}^{\text{ren}} =\ &-\frac{1}{\kappa}\int_{\mathcal{B}\cap \Sigma}\sqrt{\hat{\gamma}}\,\bigl(u_{a}r_{b}\, \nabla^{[a}\xi^{b]}+ N\,(\mathcal{K}-\mathcal{L}_\mathcal{B})\bigr)\nonumber\\
&\ \ -\frac{1}{\kappa}\int_{\mathcal{Q}\cap \Sigma}\sqrt{\hat{h}}\,\bigl(u_{a}n_{b}\, \nabla^{[a}\xi^{b]} + N\,(K-\mathcal{L}_{\mathcal{Q}})\bigr)-\frac{1}{\kappa}\int_{\mathcal{C}\cap \Sigma}N\sqrt{\hat{\sigma}}\,(\Theta-\mathcal{L}_{\mathcal{C}}),
	\label{ADMmasswithbranes}
\end{align}
where we now have extra terms localized on the branes and on their intersections.\footnote{There are also Hayward terms and counterterms at the corners $\mathcal{Q}\cap \mathcal{B}$ shared by the branes and the conformal boundary. However, they are both purely divergent and will cancel each other without leaving any finite contributions as shown in our explicit disk-brane action calculation in Appendix \ref{app:gravActionexplicit}. Hence, we will neglect such divergences in the formula \eqref{ADMmasswithbranes} and later computations.} This new formula for the ADM mass is proven in Appendix \ref{app:admmassproof} by showing that it reproduces the defining equation \eqref{definingeq}. In addition, the result of the calculation in Appendix \ref{app:admmassproof} shows that an action $\mathcal{L}_{\mathcal{M}} \propto\delta_{\mathcal{D}}$ of a codimension-2 conical defect $\mathcal{D}$ does not directly contribute to the on-shell action appearing \eqref{definingeq} when the worldsheet of $\mathcal{D}$ is tangent to the Killing vector $\xi^a$.

Applying the identity \eqref{LtoLhat} to the extrinsic curvatures in \eqref{ADMmasswithbranes}, we can write it as
\begin{equation}
	M_{\text{ADM}}^{\text{ren}} = -\frac{1}{\kappa}\int_{\mathcal{B}\cap \Sigma}N\sqrt{\hat{\gamma}}\, (\widehat{\mathcal{K}}-\mathcal{L}_\mathcal{B})-\frac{1}{\kappa}\int_{\mathcal{Q}\cap \Sigma}N\sqrt{\hat{h}}\, (\widehat{K}-\mathcal{L}_{\mathcal{Q}})-\frac{1}{\kappa}\int_{\mathcal{C}\cap \Sigma}N\sqrt{\hat{\sigma}}\,(\Theta-\mathcal{L}_{\mathcal{C}}).
	\label{ADMbranes}
\end{equation}
with extra terms compared to \eqref{absoluteADM}. With this machinery, we can compute Euclidean on-shell actions of our bulk brane configurations simply by computing the ADM mass and areas of horizons. Both of these quantities are simple surface integrals that allow us to sidestep the volume integration appearing in the Euclidean action. This is rather convenient because such volume integrals can be complicated for multi-intersecting brane configurations.

With three bulk dimensions $d=2$, only the area counterterm is needed to renormalize the ADM mass (and the on-shell action---see Appendix \ref{app:gravActionexplicit}), and the coefficient is unity---$ \mathcal{L}_{\mathcal{B}} = 1 $. For constant tension branes with $\mathcal{L}_{\mathcal{Q}} = T$, the boundary Einstein equation implies that $ K_{ab}\lvert_{\mathcal{Q}}\, = T\,h_{ab}\lvert_{\mathcal{Q}} $, and when projected onto $ \mathcal{Q}\cap \Sigma $ this further implies $ \widehat{K}_{ab}\lvert_{\mathcal{Q}\cap \Sigma}\, = T\,\hat{h}_{ab}\lvert_{\mathcal{Q}\cap \Sigma} $. Taking the trace gives
\begin{equation}
\widehat{K}-T = 0.
\end{equation}
Similarly, for a corner mass term $\mathcal{L}_{\mathcal{C}} = M$, the corner Einstein equation states that
\begin{equation}
\Theta-M = 0.
\end{equation}
Hence the last two terms in the formula \eqref{ADMbranes} for the ADM mass vanish and we simply get
\begin{equation}
M_{\text{ADM}}^{\text{ren}} = -\frac{1}{\kappa}\int_{\mathcal{B}\cap \Sigma}N\sqrt{\hat{\gamma}}\, (\widehat{\mathcal{K}}-1).
\label{ADMmassonshell3d}
\end{equation}
Combining this with the defining expression \eqref{definingeq} we can compute Euclidean on-shell actions with ease. As a result in three bulk dimensions, brane matter content does not directly contribute to the on-shell action, but they can contribute indirectly by determining locations of the branes: the Wald entropy term in \eqref{definingeq} can be sensitive to how the branes are embedded deep in the bulk. The same holds true for the point particle action $\mathcal{L}_{\mathcal{M}} = m\delta_{\mathcal{D}}$, which does not directly contribute to the on-shell action either (when the worldline $\mathcal{D}$ is tangent to $\xi^a$). However, the presence of the particle can show up indirectly through the deficit angle created by its backreaction.\footnote{Including the point particle action $\mathcal{L}_{\mathcal{M}} = m\delta_{\mathcal{D}}$ to the on-shell action would show up as an extra bulk contribution to the ADM mass \eqref{ADMmassonshell3d}. This is unphysical from the CFT perspective, because the ADM mass is dual to an integral of the CFT stress tensor expectation value which is a boundary object. Indeed, it can be shown \cite{sarkar_first_2020} (which we also see below) that the mass \eqref{ADMmassonshell3d} without the point particle action computes correctly the scaling dimension of a heavy sub-threshold operator.}


\subsection{On-Shell Actions of Disk-Brane Configurations}\label{sec:actiondiskconfs}

We first compute on-shell actions of disk-brane configurations presented in Section \ref{subsec:braneconfs}. These configurations enjoy translation symmetry in the $\phi$-direction. Hence we foliate the conical global AdS$_3$ geometry \eqref{conicalglobalads} by constant-$\phi$ slices so that
\begin{equation}
    N = r,\quad u^a = \frac{1}{r}\,\delta^a_\phi,\quad \xi^a = \delta^a_\phi.
\end{equation}
See Figure \ref{figs:diskslices} for visual representations of this foliation. Note that the central $r = 0$ line acts as a horizon in these Cauchy slices, and so the Euclidean on-shell actions of the disk-brane configurations may see contributions from Wald entropy terms.

\subsubsection*{Non-Intersecting Disk Branes}

\begin{figure}[t]
\subfloat[\label{figs:diskslice}]
{\includegraphics{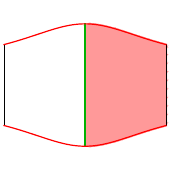}}
\hspace{1.5cm}
\subfloat[\label{figs:disksliceintpos}]
{\includegraphics{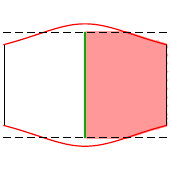}}
\hspace{1.5cm}
\subfloat[\label{figs:disksliceintneg}]
{\includegraphics{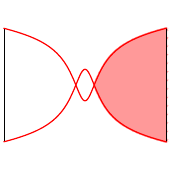}}
\caption{Slicings of disk brane configurations used in the computation of their Euclidean on-shell actions. We identify $u$ with the angular ($\phi$) direction in \eqref{conicalglobalads}. Note that the central $r = 0$ line (which is the position of the conical defect) is a horizon in this slicing.}
\label{figs:diskslices}
\end{figure}

Consider now two disconnected disk branes in the defect AdS$ _3 $ geometry such that there is a line defect exposed running between the two tips of the branes. As shown in Figure \ref{figs:diskslice}, we slice the geometry by constant-$ \phi $ slices so that the representative Cauchy slice is
\begin{equation}
\Sigma = \{\phi = 0,\ \ 0<r<\Lambda,\ \ -F_1(r) < \tau < F_2(r)\}
\end{equation}
with $F_i(r) \equiv F(r;T_i,\tau_i)$. The foliation is parameterized by $ \phi \in (0,2\pi) $, so $ u_0 = 2\pi $. The Cauchy slice has three boundary components--- $\mathcal{Q}_{1}\cap \Sigma$, $\mathcal{Q}_2 \cap \Sigma$, and $ \mathcal{D} $---with the conical line defect at $r=0$ being a horizon where the $\phi$-circle shrinks to zero. The Euclidean on-shell action is thus given by,
\begin{equation}
	I^{\text{ren}}_{\text{on-shell}} = 2\pi\,\biggl(M_{\text{ADM}}^{\text{ren}} - \frac{\kappa_{\text{s}}}{2\pi}\,S_{\text{W}}\biggr),
	\label{onshelldiskbranes}
\end{equation}
where the ADM mass is
\begin{equation}
M_{\text{ADM}}^{\text{ren}} = -\lim_{\Lambda\rightarrow \infty}\frac{1}{\kappa}\int_{-F_1(\Lambda)}^{F_2(\Lambda)} d\tau\, \Lambda\sqrt{f_{\alpha}(\Lambda)}\, \left[\frac{f'_{\alpha}(\Lambda)}{2\sqrt{f_\alpha(\Lambda)}}-1\right] = \frac{1}{2\kappa}\,\alpha^{2}\,(\tau_1+\tau_2),
\label{nonintdiskbraneadm}
\end{equation}
and the Wald entropy term is proportional to the proper length of the conical line defect,
\begin{equation}
\frac{\kappa_{\text{s}}}{2\pi}\,S_{\text{W}} = \frac{\kappa_{\text{s}}}{\kappa}\int_{-F_1(0)}^{F_2(0)}d\tau\,\alpha = \frac{1}{\kappa}\,\alpha^{2}\,(\tau_1+\tau_2) + \frac{1}{\kappa}\,\alpha\,(\Tanh^{-1}{T_1}+\Tanh^{-1}{T_2}).
\label{waldentropycalculation}
\end{equation}
Note that we have used $ F_i(0) = \tau_i+\frac{1}{\alpha}\Tanh^{-1}{T_i} $ and that the surface gravity in is given by $ \kappa_{\text{s}} = \alpha $.\footnote{One can compute Wald entropy explicitly (see \eqref{waldentropyexplicitly}) using $\phi_an_b^{\mathcal{D}}\,\nabla^{[a}\xi^{b]}\lvert_{r=0} \,= -\sqrt{f_\alpha(r)}\,\lvert_{r=0}\, = -\alpha $, where $n_a^{\mathcal{D}} = -r_a$ is the outward-pointing unit normal of $\mathcal{D}$. The minus sign is cancelled, because the integration direction over $\mathcal{D}$ is in the negative $\tau$-direction (opposite to the integration direction of $\mathcal{B}\cap \Sigma$ which is in the positive $\tau$-direction) leading to \eqref{waldentropycalculation}.} By plugging these expressions into the Euclidean on-shell action \eqref{onshelldiskbranes}, we find that
\begin{equation}
I^{\text{ren}}_{\text{on-shell}} = -\frac{2\pi}{\kappa}\,\biggl[\frac{1}{2}\,\alpha^{2}\,(\tau_1+\tau_2)+\alpha\,(\Tanh^{-1}{T_1}+\Tanh^{-1}{T_2})\biggr],
\label{nonintdiskbraneaction}
\end{equation}
which we have reproduced by an explicit calculation of the renormalized action in Appendix \ref{app:gravActionexplicit}.

Now, we write this in terms of boundary data. Recall that the modulus of the boundary cylinder is given by
\begin{equation}
\frac{\tau_1+\tau_2}{2\pi} = \frac{W}{\beta},
\label{diskbranesmodulus}
\end{equation}
which gives $ \tau_1+\tau_2 = \frac{2\pi W}{\beta} $. By plugging this into \eqref{nonintdiskbraneaction}, we write
\begin{equation}
I^{\text{ren}}_{\text{on-shell}} = -\frac{2\pi}{\kappa}\,\biggl[\alpha^{2}\,\frac{\pi W}{\beta}+\alpha\,\left(\Tanh^{-1}{T_1}+\Tanh^{-1}{T_2}\right)\biggr].
\label{actionnonintdiskbranes}
\end{equation}
The corresponding contribution to the BCFT partition function is
\begin{equation}
  e^{-I^{\text{ren}}_{\text{on-shell}}} = \left(\frac{1+T_1}{1-T_1}\right)^{\frac{c}{12}\alpha}\left(\frac{1+T_2}{1-T_2}\right)^{\frac{c}{12}\alpha}\,\tilde{q}^{-\frac{c}{24}\alpha^{2}}
\end{equation}
where we have used $\frac{2\pi}{\kappa} = \frac{c}{6}$ and $\tilde{q} = e^{-4\pi W\slash \beta}$. From the dependence on $\tilde{q}$, it follows that this geometry is dual to an eigenstate $\lvert \Delta^{\text{cl}}\rangle$ of the closed-string Hamiltonian $H^{\text{cl}}$ with dimension (noting that the spin vanishes)
\begin{equation}
    \Delta^{\text{cl}} = \frac{c}{12}\,(1-\alpha^2)
    \label{disconnecteddiskdimension}
\end{equation}
and with overlaps
\begin{equation}
    \langle T_i \lvert \Delta^{\text{cl}} \rangle =\biggl(\frac{1+T_i}{1-T_i}\biggr)^{\frac{c}{12}\alpha}.
\end{equation}
For $\alpha = 1$, the dual state is the closed-string vacuum as in \cite{Takayanagi:2011zk}. Meanwhile, for $\alpha < 1$, the dual states are excited and comprise a heavy but sub-threshold spectrum, since $\Delta^{\text{cl}} = O(c)$ and $\Delta^{\text{cl}} < \frac{c}{12}$.

\subsubsection*{Intersecting Disk Branes}

We now consider two intersecting disk branes. Recall that there are two ways in which such branes may intersect:
\begin{equation}
\begin{split}
T_1 + T_2 > 0 &\implies \text{cutting and gluing},\\
T_1 + T_2 \leq 0 &\implies \text{decreasing the modulus}.
\end{split}\label{intdiskcases}
\end{equation}
Let us start with the action of the first type of configuration for which the conical line defect is contained in the bulk. We assume that the branes are cut and glued at radial depth $ r = r_* $, which is a free parameter. Any representative Cauchy slice consists of two pieces \eqref{cutglueBound}. The outer $r \in (r_*,\Lambda)$ part of the $\phi = 0$ Cauchy slice is
\begin{equation}
\Sigma_{r > r_*} = \{\phi = 0,\ \ r_* < r < \Lambda,\ \ -F_1(r) < \tau < F_2(r)\},
\end{equation}
and the inner $r \in [0,r_*]$ part is
\begin{align}
\Sigma_{r < r_*} = \{\phi = 0,\ \ 0 \leq r\leq r_*,\ \ -F_1(r_*) < \tau < F_2(r_*)\},\label{casesTensCauchy}
\end{align}
As an example, the configuration with two positive-tension branes is shown in Figure \ref{figs:disksliceintpos}.

Just as before, the on-shell action is given by \eqref{onshelldiskbranes}. The ADM mass is the same as in the non-intersecting case \eqref{nonintdiskbraneadm}. The only difference this time is the calculation of the Wald entropy. Specifically, by integrating over the conical line defect in the domain \eqref{casesTensCauchy}, we have that
\begin{equation}
\frac{\kappa_{\text{s}}}{2\pi}\,S_{\text{W}} = \frac{\kappa_{\text{s}}}{\kappa}\int_{-F_1(r_*)}^{F_2(r_*)}d\tau\,\alpha = \frac{1}{\kappa} \alpha^2\,[F_1(r_*)+F_2(r_*)].
\end{equation}
Because $r_* > 0$, this term takes the form
\begin{equation}
\frac{\kappa_{\text{s}}}{2\pi}\,S_{\text{W}} = \frac{1}{\kappa}\alpha^2 (\tau_1 + \tau_2) + \frac{1}{\kappa}\alpha\left[\Tanh^{-1} T_1^{\text{eff}}(r_*) + \Tanh^{-1} T_2^{\text{eff}}(r_*)\right],
\end{equation}
where we have defined ``effective" tensions which depend on $r_*$,
\begin{equation}
T_i^{\text{eff}}(r_*) = \frac{T_i \alpha}{\sqrt{f_\alpha(r_*) - T_i^2 r_*^2}}.
\end{equation}
Note that we may also use the corner Einstein equation $\Theta(r_*) = M$, where the intersection angle is given in terms of $r_*$ in \eqref{intangle}, to write the effective tensions as functions of $M$:
\begin{equation}
T_i^{\text{eff}}(M) = T_i\sqrt{1-\frac{(1-T_i^2)\sin^2 M}{\left(T_j - T_i \cos M\right)^2}},\ \ j \neq i.
\end{equation}
In terms of the effective tensions, the Wald entropy term is the same as in the disconnected case. So, the remainder of the calculation is the same as above, and we have that the contribution to the BCFT partition function from the cut-and-glued configuration is
\begin{equation}
e^{-I_{\text{on-shell}}^{\text{ren}}} =  \left(\frac{1+T_1^{\text{eff}}}{1-T_1^{\text{eff}}}\right)^{\frac{c}{12}\alpha}\left(\frac{1+T_2^{\text{eff}}}{1-T_2^{\text{eff}}}\right)^{\frac{c}{12}\alpha}\,\tilde{q}^{-\frac{c}{24}\alpha^{2}}.
\end{equation}
Naively, we may think this is dual to a closed-string state $\vert\Delta^{\text{cl}}\rangle$ with the same dimension as the disconnected configuration \eqref{disconnecteddiskdimension}. However, the product of overlaps,
\begin{equation}
    \langle T_i \lvert \Delta^{\text{cl}} \rangle = \left(\frac{1+T_i^{\text{eff}}(M)}{1-T_i^{\text{eff}}(M)}\right)^{\frac{c}{12}\alpha},
\end{equation}
does not factorize in general for any value of the intersection mass strictly between $0$ and $\pi$, and so the existence of these cut-and-glued configurations presents a factorization puzzle \cite{Maldacena:2004rf}. We thus posit that they compute ensemble-averaged overlaps, as we discuss in Section \ref{sec:discussion}.

Now we consider the second case in \eqref{intdiskcases}. In this case, the conical line defect is hidden behind the EOW branes. As a result, there is no Wald entropy term because the Cauchy slice does not have a extra boundary in the interior (see Figure \ref{figs:disksliceintneg}). Thus, the on-shell action is simply proportional to the ADM mass,
\begin{equation}
I^{\text{ren}}_{\text{on-shell}} = 2\pi\,M_{\text{ADM}}^{\text{ren}}.
\end{equation}
The ADM mass is again \eqref{nonintdiskbraneadm}, and after again fixing the modulus \eqref{diskbranesmodulus}, we find that
\begin{equation}
	I^{\text{ren}}_{\text{on-shell}} = \frac{2\pi}{\kappa}\,\alpha^{2}\,\frac{\pi W}{\beta},
	\label{nonunitaryaction}
\end{equation}
and the corresponding contribution to the BCFT partition function is
\begin{equation}
  e^{-I^{\text{ren}}_{\text{on-shell}}} = \tilde{q}^{\frac{c}{24}\alpha^{2}}.
\end{equation}
The action \eqref{nonunitaryaction} has the opposite sign compared to \eqref{actionnonintdiskbranes}, so it gives the dimension
\begin{equation}
    \Delta^{\text{cl}} = \frac{c}{12}\,(1+\alpha^2) > \frac{c}{12},\label{dimDiskMergeBad}
\end{equation}
which is a heavy state above the black hole threshold. The overlaps are simply $\langle T_i \lvert \Delta^{\text{cl}} \rangle = 1$.

We conclude that the result \eqref{dimDiskMergeBad} is actually unphysical for the merging disk branes. Specifically, for fixed tensions $T_{1,2}$ and fixed defect masses $m,M$, we may use the condition that the branes intersect at $r = r_*$ to write $\alpha$ as a function of $r_*$ and the modulus $\frac{W}{\beta}$. We then use \eqref{intangle} to say that $\alpha$ depends only on the modulus. It follows from \eqref{dimDiskMergeBad} that the state dual to an intersecting-disk-brane configuration and supported by particular bulk parameters has \textit{modulus-dependent} dimension. This contradicts the fact that such data is supposed to be input for the BCFT.

While this seems problematic, we are able to avoid a puzzle surrounding these states. First, note that these intersecting-disk-brane configurations represent states above the black hole threshold, and so they are exponentially suppressed in the closed-string expansion of the BCFT partition function. We may ask if they are actually describing open-string contributions. Indeed, we find that an analytic continuation $\alpha \rightarrow i\alpha$ flips the sign of the on-shell action \eqref{nonunitaryaction}, and if we then $S$-transform $\tilde{q}\rightarrow q$, it matches with the action of an annulus brane configuration computed in the next section. This mapping is related to the analytic continuation of disk branes to strip branes \eqref{analyticontinuation}, as we further discuss in Section \ref{sec:discussion}.

\subsection{On-Shell Actions of Annulus-Brane Configurations}\label{sec:actionADM}

\begin{figure}[t]
\subfloat[\label{figs:stripsliceneg}]
{\includegraphics{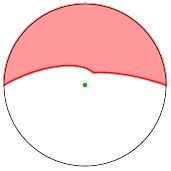}}\hspace{2cm}
\subfloat[\label{figs:stripslicepos}]
{\includegraphics{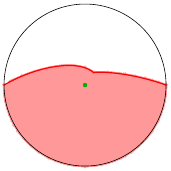}}
\caption{Slicings of annulus brane configurations used in the computation of their Euclidean on-shell actions. We identify $u$ with the angular ($\tau$) direction in \eqref{conicalglobalads}. In (b), the conical defect is integrated over in the slice, but it does not actually contribute to the on-shell action because it is orthogonal to each Cauchy slice.}
\label{figs:stripslice}
\end{figure}

Now we now compute Euclidean on-shell actions of the annulus brane configurations presented in Section \ref{subsec:braneconfs}. These configurations enjoy translation symmetry in the periodic $\tau$-direction. Hence this time, we foliate the conical global AdS$_3$ geometry \eqref{conicalglobalads} by constant-$\tau$ slices so that
\begin{equation}
    N = \sqrt{f_\alpha(r)},\quad u^a = \frac{1}{\sqrt{f_\alpha(r)}}\,\delta^a_\phi,\quad \xi^a = \delta^a_\tau.
\end{equation}
This foliation is shown in Figure \ref{figs:stripslice}. Note that there is no horizon in this foliation, so the Wald entropy will always be zero. Furthermore, the conical defect, when present in the bulk, entirely runs along the foliation direction $\tau$ and so does not contribute---see Appendix \ref{app:admmassproof} for details on this point. Thus, the Euclidean on-shell action of the annulus-brane configurations is entirely accounted for by the ADM mass.

\subsubsection*{Single Smooth Annulus Brane}

We first consider a single smooth annulus brane of tension $ T $ with the conical line defect being either behind or in front of the brane (depending on the sign of the brane tension). For positive and negative tensions, respectively, the representative Cauchy slices which we consider are
\begin{equation}
\begin{split}
\Sigma_{T > 0} &= \left\{\tau = 0,\ \ \phi_0 < \phi < \phi_0 + \frac{\pi}{\alpha},\ \ r > p(\phi;-T,\phi_0)\right\}^c,\\
\Sigma_{T < 0} &= \left\{\tau = 0,\ \ \phi_0 < \phi < \phi_0 + \frac{\pi}{\alpha},\ \ r > p(\phi;T,\phi_0)\right\},
\end{split}
\end{equation}
where the brane embedding is \eqref{stripbranepfunction}. The $^c$ on the set defining $\Sigma_{T>0}$ indicates that the positive-tension Cauchy slice is the complement of a bulk region bounded by a negative-tension brane.

In both cases, the on-shell action is entirely determined by the ADM mass and, recalling \eqref{tau0WidthVar}, takes the form
\begin{equation}
I_{\text{on-shell}}^{\text{ren}} = \tau_0\, M_{\text{ADM}}^{\text{ren}} = \Delta\phi \frac{\beta}{W} M_{\text{ADM}}^{\text{ren}},\label{actionADMStrip}
\end{equation}
where $\Delta\phi$ is a function of $\alpha$ that depends on the sign of the tension \eqref{Twidthsingle}. For now, however, we keep $\Delta\phi$ generic. By employing the cutoff surface $r = \Lambda$ and plugging into \eqref{ADMmassonshell3d}, we find that
\begin{equation}
M_{\text{ADM}}^{\text{ren}} = -\lim_{\Lambda \to \infty}\frac{1}{\kappa}\int_{\phi_0}^{\phi_0 + \Delta\phi} d\phi\,\Lambda \sqrt{f_\alpha(\Lambda)}\left(\frac{\sqrt{f_\alpha(\Lambda)}}{\Lambda} - 1\right) = - \Delta\phi\frac{\alpha^2}{2\kappa}.
\label{admgeneral}
\end{equation}
The Euclidean on-shell action and, consequently, the contribution to the partition function are
\begin{equation}
I_{\text{on-shell}}^{\text{ren}} = - \frac{(\alpha\Delta\phi)^2}{2\kappa}\frac{\beta}{W} \implies e^{-I_{\text{on-shell}}^{\text{ren}}} = q^{-\frac{c}{24}\left(\frac{\alpha\Delta\phi}{\pi}\right)^2},
\label{singlesmoothannulus}
\end{equation}
Thus, this configuration is dual to an open-string eigenstate $\lvert \Delta^{\text{op}}\rangle$ with dimension
\begin{equation}
\Delta^{\text{op}} = \frac{c}{12}\left[1 - \left(\frac{\alpha\Delta\phi}{\pi}\right)^2\right] = \begin{cases}
0,&\text{if}\ T < 0,\vspace{0.1cm}\\
\dfrac{c}{3}\,\alpha(1-\alpha),&\text{if}\ T > 0,
\end{cases}
\end{equation}
where we have utilized the equations \eqref{Twidthsingle} for the angular width in both cases. As a sanity check, observe that the configurations for all tensions corresponding to $\alpha = 1$, which are the original connected configurations in \cite{Takayanagi:2011zk}, simply describe open-string vacuum states $\Delta^{\text{op}} = 0$. Additionally, the ADM mass above for the positive-tension case is
\begin{equation}
M_{\text{ADM}}^{\text{ren}}|_{T > 0} = -\frac{c}{24}\alpha\,(2\alpha - 1),
\end{equation}
which is consistent with the literature \cite{Kawamoto:2022etl,Bianchi:2022ulu,Kusuki:2022wns,Kusuki:2022ozk}.\footnote{Note that matching the ADM mass across different coordinate systems is subtle because it does not transform as a scalar, but rather as a 1-form. We have checked that we match with \cite{Kusuki:2022ozk} by performing the appropriate coordinate transformation.}

For $\alpha < 1$, the negative-tension configurations, for which the bulk geometries do not contain the conical defect, correspond to vacuum states. However, the positive-tension configurations, for which the bulk geometries include the conical defect, describe excited states in the open-string channel. Furthermore, because of the requirement that the brane does not self-intersect, these states have a maximal dimension,
\begin{equation}
\left.\Delta^{\text{op}}\right|_{T > 0} < \frac{c}{12},
\end{equation}
which notably coincides with the black hole threshold. Thus, the open-string excited states constructed here fill the subthreshold window $\left(0,\frac{c}{12}\right)$.

\subsubsection*{Intersecting Annulus Branes}

\begin{figure}[t]
\subfloat[\label{figs:rstarplot}]
{\begin{tikzpicture}
\node[inner sep=0pt] (img1) at (0,0)
    {\includegraphics{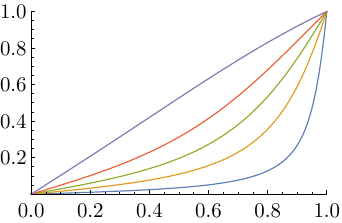}};
    \node at (-2.5,2.1) {$\rho_*$};
    \node at (3,-1.4) {$\alpha$};
    \node[rotate=34] at (-0.1,0.5) {$T = 0.9$};
    \node at (2.9,-0.7) {$T = 0.1$};
\end{tikzpicture}}
\hspace{1cm}
\subfloat[\label{figs:angleplot}]
{\begin{tikzpicture}
\node[inner sep=0pt] (img4) at (0,0)
    {\includegraphics{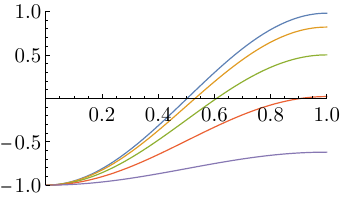}};
    \node at (-2.3,1.9) {$\cos{\Theta}$};
        \node at (3,0) {$\alpha$};
        \node[rotate=40] at (1,1.1) {$T = 0.1$};
    \node at (1.9,-1.3) {$T = 0.9$};
\end{tikzpicture}}
\caption{The intersection depth in terms of (a) the Poincar\'e disk radial coordinate \eqref{poincaredisk} and (b) the intersection angle of two annulus branes with tensions $T \equiv T_1 = -T_2$ as a function of the conical defect parameter $\alpha$.}
\label{figs:rstarcosthetaplots}
\end{figure}

Consider now intersecting annulus branes in the defect AdS$ _3 $ geometry. Having the defect outside of the bulk (and not ``hidden" by the EOW branes) requires at least one of the brane tensions to be negative (Figure \ref{figs:stripsliceneg}). Meanwhile for two positive tension branes, the defect will be visible (corresponding to the complement of the shaded region in Figure \ref{figs:stripslicepos}).

The Cauchy slice in this case is the part of a constant-$ \tau $ surface bounded by two annulus branes. However, recall that we are taking $\Delta\phi = \pi$ for these configurations, which implies that $\tau_0 = \pi\beta/W$. The ADM mass and resulting on-shell action are the same as in the smooth brane case \eqref{admgeneral}, and after setting $\Delta\phi = \pi$ we get\footnote{If we do not fix the interval length at $\pi$, then the on-shell action would still only depend on the product $\alpha\Delta\phi$, which itself is fixed by the intersection mass $M$.}
\begin{equation}
M_{\text{ADM}}^{\text{ren}} =  -\frac{\pi}{2\kappa}\,\alpha^{2}.
\end{equation}
To reiterate, this is the only contribution to the Euclidean on-shell action, even if the defect is present in the bulk. Thus, we plug into \eqref{actionADMStrip} to write
\begin{equation}
I^{\text{ren}}_{\text{on-shell}} = \tau_0\, M^{\text{ren}}_{\text{ADM}} = -\frac{\pi^2}{2\kappa}\,\alpha^{2}\,\frac{\beta}{W}
\label{diskbranesonshell}
\end{equation}
The corresponding contribution to the BCFT partition function is
\begin{equation}
  e^{-I^{\text{ren}}_{\text{on-shell}}} = q^{-\frac{c}{24}\alpha^{2}}.
  \label{stripbranecontribution}
\end{equation}
From the dependence on $q$, it follows that this geometry is dual to an eigenstate $\lvert \Delta^{\text{op}}\rangle$ of the open-string Hamiltonian $H^{\text{op}}$ with dimension (noting that the spin vanishes)
\begin{equation}
\Delta^{\text{op}} = \frac{c}{12}\,(1-\alpha^2).
    \label{stripbranedimension}
\end{equation}
The conical defect parameter $\alpha = \alpha(M)$ and hence $\Delta^{\text{op}}$ are fixed by the brane intersection mass $M$ through the corner Einstein equation $\Theta(\alpha) = M$.\footnote{This is still true even if we do not fix $\Delta\phi$, since we would ultimately get that the action is a function of the product $\alpha\Delta\phi$ which can then be traded for the intersection angle $\Theta$ as discussed in Section \ref{subsec:braneconfs}.}

\begin{figure}[t]
\subfloat[\label{figs:closebdy}]
{\includegraphics{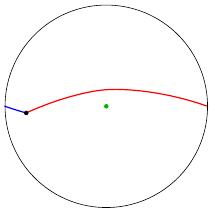}}
\hspace{2cm}
\subfloat[\label{figs:farbdy}]
{\includegraphics{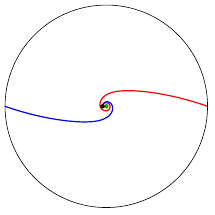}}
\caption{Two examples of intersecting annulus brane configurations. In (a) $\alpha = 0.95$, so the intersection is close to the conformal boundary and the configuration is almost like a single smooth brane. In (b) $\alpha = 0.2$, so the intersection point is close to the center. The configuration (a) is dual to a BCC operator with dimension $\Delta_{\text{bcc}} \approx 0$, and the configuration (b) is dual to $\Delta_{\text{bcc}} \approx \frac{c}{12}$.}
\label{figs:rstarcostheta}
\end{figure}

We now analyze the possible range of dimensions obtained from this model. For simplicity, we consider the special states for which the branes have opposite but equal tensions $T \equiv T_1 = -T_2$, as in the configuration of Figure \ref{figs:intstripbranes}. In this case, the intersection depth \eqref{striprs} is simply
\begin{equation}
    r_* =  \frac{\lvert T\lvert\,\alpha}{\sqrt{1-T^2}}\,\sec{\left(\frac{\pi \alpha}{2}\right)},
\end{equation}
and the intersection angle \eqref{stripTheta} becomes
\begin{equation}
    \cos{\Theta} = -T^2-(1-T^2)\cos{(\pi\alpha)}.
\end{equation}
These have been plotted as a function of $\alpha\in [0,1]$ in Figure \ref{figs:rstarcosthetaplots} for different values of the tension $T$. We find that for the range $\cos{\Theta} \in (-1,1-2T^2)$, we have that $\alpha \in (0,1)$. Thus by tuning $M$, we can engineer the range
\begin{equation}
    \Delta_{\text{bcc}}\in \left(0,\frac{c}{12}\right).
\end{equation}
Observe that the intersection point runs to the conformal boundary $r_* \rightarrow\infty$ when $\alpha \rightarrow 1$. This means that the configuration dual to a BCC operator of dimension $\Delta_{\text{bcc}} \rightarrow 0$ is arbitrarily close to a single smooth strip brane (see Figure \ref{figs:closebdy}). In the opposite limit $r_* \rightarrow 0$, the dimension $\Delta_{\text{bcc}} \rightarrow \frac{c}{12}$ and the configuration is a spiral (see Figure \ref{figs:farbdy}). We also find that the irreducible representation corresponding to the BCC operator \eqref{stripbranedimension} has degeneracy $\log{\mathcal{N}_{AB}^{\,\text{bcc}}} = O(c^0)$.

\section{Conclusions and Discussion}\label{sec:discussion} 

In this work, we consider the holographic model of Euclidean BCFTs in terms of EOW branes and extended it by the addition of scalar fields, point particles, and non-smooth brane intersections. The extended model allows for the description of primary operator exchanges in the closed-string channel (through scalar fields and point particles) and boundary-condition-changing operators in the open-string channel (through brane intersections). We also find novel wormhole configurations in our analysis.

To obtain dimensions of the dual operators, we compute Euclidean on-shell actions of the corresponding brane configurations. This requires taking into account Gibbons--Hawking--York terms on boundaries and Hayward terms on corners, and one also needs matter content at corners to support intersections. In 3-dimensional gravity however, there is no backreaction and the only effect of the Einstein equations is to fix the branes' extrinsic curvatures, their intersection angles, and deficit angles of conical defects. To simplify the computations, we derive a general formula that reduces the on-shell action to a boundary integral involving the Wald entropy and the ADM mass along the lines of \cite{wald_black_1993,iyer_properties_1994}. In doing so, we provide a generalization of the ADM mass formula for geometries with intersecting branes. In this process, we also see explicitly that actions of point particles that source conical line defects do not contribute directly to the on-shell actions of our bulk configurations. Lastly, we show that, in 3-dimensional gravity, the brane and brane-intersection matter content do not contribute to the ADM mass.

The general theme with heavy states in 3-dimensional gravity is that they are always dual to geometries containing a line defect supported by a point particle. For example, excited states are specifically dual to geometries containing a conical line defect of strength $\alpha$ supported by a point particle of mass $m$, with the relationship \eqref{pointparticlemass} between $\alpha$ and $m$ being determined from the equations of motion. Furthermore, the strength $\alpha$ of the conical defect is related to the scaling dimension of the dual state via,
\begin{equation}
\Delta^{\text{cl}} = \frac{c}{12}(1-\alpha^2),\ \ \Delta^{\text{op}} = \frac{c}{3}\alpha(1-\alpha),
\end{equation}
with excited closed-string states being well-defined for $\alpha \in (0,1)$ and excited open-string states being well-defined for $\alpha \in \left(\frac{1}{2},1\right)$. Both equations follows from the computations of on-shell actions. In both channels, one is able to engineer the full range of $\alpha$ by tuning $m$ such that the unitarity window $\Delta^{\text{cl,op}} \in \left(0,\frac{c}{12}\right)$ below the black hole threshold is filled.

Additionally, we consider geometries for which the line defect is a corner shared between the branes and is supported by a point particle of mass $M$. In the open-string channel, these are the only configurations compatible with BCFT on a cylinder and with two boundary conditions of different boundary entropies, and so we identify them as BCC operators. We specifically consider a one-parameter family of configurations subject to a stability condition \eqref{stabCond} relating $M$ the point particle mass on the conical line defect $m$.\footnote{This stability condition is only relevant for intersecting configurations in which we have a point particle in the bulk geometry.} These masses may be tuned together to control $\alpha$, which in turn controls the dimension of the dual operator,
\begin{equation}
\Delta_{\text{bcc}} = \frac{c}{12}(1-\alpha^2).
\end{equation}
Here we may also engineer the full range $\alpha \in (0,1)$ such that the unitarity window below the black hole threshold is filled.

\subsubsection*{The Wormhole Configuration}

We construct a family of intersecting disk-brane configurations obtained by cutting and gluing two disk branes whose tensions satisfy $T_1+T_2 > 0$ and which are supported by a matter stress tensor at the brane intersection. The resulting geometry is a type of a Euclidean wormhole where the ``future" and ``past" boundary states of the CFT are connected by the wormhole throat (see Figure \ref{figs:diskBraneCutGlueTorus}). In other words, cutting the geometry along the BCFT circle does not produce two disconnected components. Hence, the bulk geometry does not prepare a pure state in the closed-string Hilbert space (which would always be expected from the CFT side), but instead an entangled state similar to a thermofield double state. These types of wormholes are known as bra-ket wormholes \cite{chen_bra-ket_2020}, but our wormhole is different in the sense that its throat is bounded by two non-smoothly intersecting EOW branes.

We find that, for $\alpha = 1$, the configuration is dual to a closed-string state with the same scaling dimension as the vacuum state and whose boundary entropies depend on both tensions of the intersecting branes and the intersection mass.\footnote{The discussion readily generalizes to excited states corresponding to bulk configurations with a conical line defect $\alpha < 1$ running through the wormhole throat.} This is different from the usual geometry dual to the vacuum state---corresponding to two disconnected branes---where each boundary entropy is fixed by a single brane tension. Nonetheless, we find that the $\alpha = 1$ wormhole term in the partition function is still always subleading to that of the disconnected configuration. This is most readily seen by noting that the two configurations have the same ADM mass, and so the ratio of their partition functions is a function of the difference of their Wald entropies:
\begin{equation}
\frac{Z_{\text{wormhole}}}{Z_{\text{disconnected}}} = \exp\left(S_{\text{W}}^{\text{wormhole}} - S_{\text{W}}^{\text{disconnected}}\right).\label{wormholeSuppression}
\end{equation}
The identification used to construct the wormhole geometry also ensures that the horizon used to compute Wald entropy is ``shorter" than in the disconnected case, and so the ratio \eqref{wormholeSuppression} must be less than unity (and can in fact be small, depending on the brane tensions).

In line with the notion that gravity can compute ensemble averages of field-theoretic observables \cite{Saad:2019lba}, we propose that this $\alpha = 1$ wormhole configuration can be understood using ensemble averaging. The main reason for this is that the wormhole turns a pure state to an entangled state. That the wormhole configuration computes an ensemble-averaged quantity would be in spite of these states being below the black hole threshold \cite{Schlenker:2022dyo}, with the reason being that our theory includes defects. Note that analogous wormholes are also found in \cite{Chandra:2022bqq}.

Specifically, the wormhole should give a connected contribution to the averaged product of two overlaps which does not factorize:
\begin{equation}
    \overline{\langle A\lvert 0 \rangle \langle 0\lvert B\rangle} = \overline{\langle A\lvert 0 \rangle}\,\overline{\langle 0\lvert B\rangle} + \text{(wormhole)} + \ldots,\label{avgPart}
\end{equation}
where the leading factorized contribution is the standard disconnected disk-brane configuration. Such averaged products arise in the averaged partition function $\overline{Z_{AB}} $ when expanded in the closed-string channel. This equation indicates a factorization puzzle for partition functions because a vacuum overlap $ \langle A\lvert 0 \rangle $ is equal to the disk partition function $Z_A^{\text{disk}}$ with the boundary condition $A$ at the boundary of the disk.

We propose that the averaging in \eqref{avgPart} is performed over all boundary states $\ket{A}$ and $\ket{B}$ that respectively correspond to branes of fixed tensions $T_A$ and $T_B$, a position also taken by \cite{Kusuki:2022wns}. This ensures that $\overline{\langle A\lvert 0\rangle} = \langle T_A\lvert 0\rangle $ and $\overline{\langle 0\lvert B\rangle} = \langle 0\lvert T_B\rangle$, i.e. that these ``averaged" boundary-theory overlaps are computed by specific bulk configurations. Another way to make sense of the connected contributions in \eqref{avgPart} would be to employ the Coleman--Giddings--Strominger mechanism \cite{Giddings:1987cg,Coleman:1988cy,Giddings:1988cx} of integrating out the wormhole configuration and interpreting the effective action as a theory with random couplings, an approach which has recently been used to understand semiclassical 3-dimensional gravity without branes \cite{Chandra:2022bqq}. We leave further, more refined exploration of our bra-ket wormholes to future work.

\subsubsection*{Analytic Continuation of Brane Configurations} 

We show that two disk branes whose tensions satisfy $T_1 + T_2 \leq 0$, will always intersect inevitably in the open-string limit $\frac{W}{\beta} \rightarrow 0$. However, to support such intersections for all moduli $\frac{W}{\beta}$ requires the intersection mass term to depend on the modulus $M = M\left(\frac{W}{\beta}\right)$. This contradicts the lore that all bulk parameters are fixed by BCFT data (scaling dimensions and OPE coefficients) and are hence independent of the modulus. Fortunately, we also find that these configurations are dual to closed-string states that are above the black hole threshold $\Delta^{\text{cl}} = \frac{c}{12}\,(1+\alpha^2)\geq \frac{c}{12}$, so they would be exponentially suppressed in the closed-string limit $\frac{W}{\beta} \rightarrow \infty$. However, in the open-string limit $\frac{W}{\beta} \rightarrow 0$, their contributions may be parametrically large.

We observe that the on-shell action of this configuration is related to the on-shell action of an intersecting annulus brane configuration by simultaneous analytic continuation $\alpha \rightarrow i\alpha$ and $S$-transformation $\tilde{q}\rightarrow q$. The actions are related by
\begin{equation}
    e^{-I_{\text{disks}}} = \tilde{q}^{\frac{c}{24}(1+\alpha^2)-\frac{c}{24}} \rightarrow q^{\frac{c}{24}(1-\alpha^2)-\frac{c}{24}} = e^{-I_{\text{annuli}}}
\end{equation}
The transformed configuration thus has the interpretation as a sub-threshold open-string state of dimension $\Delta^{\text{op}} = \frac{c}{12}\,(1-\alpha^2)$. This relation between the actions might be explained by the relation between disk and strip brane embeddings themselves; they are related by the analytic continuation $\alpha \rightarrow i\alpha$ and subsequent coordinate transformation described in \eqref{analyticontinuation}.

In a more realistic holographic model, it is reasonable to expect that disk branes start to repel each other in the open-string limit $\frac{W}{\beta} \rightarrow 0$. This would mean that there will always be a non-zero geodesic distance between them, and so the problematic saddles would not even constitute mathematically consistent solutions in the first place.

\subsubsection*{Multi-Intersecting Configurations}

We focus on brane configurations that contain two branes intersecting non-smoothly at a single corner. In principle, one could also have multi-intersecting configurations consisting of multiple EOW brane components: two of the brane components are anchored to the conformal boundary while rest of the components are anchored between intersections (Figure \ref{figs:doubleintersection}). Multi-intersecting configurations require different corner terms at each intersection to support the configuration.

The methods of Section \ref{sec:adm} can be used to compute on-shell actions of such annulus-brane configurations easily. In such cases, the action is only sensitive to the ADM mass, which is an integral at the conformal boundary. Thus, any $\Delta\phi = \pi$ configuration involving multi-intersecting annulus branes has the same on-shell action as that of two intersecting annulus branes \eqref{stripbranecontribution} and hence is dual to an open-string state whose dimension has the same dependence on $\alpha$. The difference, however, lies in how the intersection masses determine the BCC scaling dimensions, since we would have that $\Delta^{\text{op}} = \Delta^{\text{op}}(M_1,M_2,\ldots)$. These configurations may describe a different possible range of BCC scaling dimensions than the configurations consisting of only two branes.

\subsubsection*{Future Directions}

In this paper, we focus only on 2-dimensional BCFTs that are dual to 3-dimensional Einstein gravity. This means that branes and intersections do not backreact on the geometry and that their actions do not directly contribute to the on-shell actions. A natural extension is to consider higher-dimensional setups where such backreaction effects becomes relevant. It would be interesting to construct intersecting EOW branes configurations in higher dimensions.

In our scalar model, we only study a probe limit of the full theory in which gravitational backreaction is suppressed. It would thus be interesting to study scalar-field interactions in backreacted backgrounds. Additionally, we only consider the classical saddle-point contribution to the bulk path integral without any quantum corrections. Hence, we do not produce the full Virasoro character containing contributions of all the descendants of the corresponding dual eigenstate $\lvert \Delta_\Phi \rangle$. The full character of the brane configurations would be reproduced by a 1-loop bulk computation along the lines of \cite{giombi_one-loop_2008,suzuki_one-loop_2022}.

One could also try to reproduce a unitary BCFT partition function by summing over all intersecting brane geometries with point particles, similarly to how \cite{Benjamin:2020mfz} addresses the non-unitarity of pure 3-dimensional gravity \cite{Maloney:2007ud,Keller:2014xba,Benjamin:2019stq}. It is an interesting question to understand if 3-dimensional gravity with EOW branes has a CFT dual.

We would like to understand how this bottom-up model is related to more refined top-down realizations of BCFT. A setup similar to our model is the tensionless limit of string theory on $\text{AdS}_3 \times S^3\times \mathbb{T}^4$ studied in \cite{Gaberdiel:2021kkp}. They also consider branes of disk and strip topology to which open strings can attach. These are D-branes, and they correspond to boundary conditions in a symmetric product orbifold theory (also studied by \cite{Belin:2021nck}). Concretely, our model of a scalar-field exchange between disk branes might be possible to understand as a point-particle limit of the closed-string exchange in \cite{Gaberdiel:2021kkp}.

\acknowledgments 

We thank Elena C\'aceres, Andreas Karch, Andrew Svesko, Manus Visser, David Wakeham, Chris Waddell, and Petar Simidzija for useful discussions during the completion of this work. JK was supported by the Osk. Huttunen Foundation. SS was supported by National Science Foundation (NSF) Grants PHY-1914679 and PHY-2112725.

\vfill
\pagebreak
\vfill

\begin{appendices}

\section{Derivation of the Corner Einstein Equation}\label{app:cornereinstein}

Consider the Einstein-Hilbert action with Gibbons--Hawking--York terms on all EOW branes $\mathcal{Q}$ and Hayward terms on all corners $\mathcal{C}$. In addition, we allow matter degrees of freedom in the bulk, on the branes, and on the corners. Schematically, the action is
\begin{equation}
I = -\frac{1}{2\kappa}\int_{\mathcal{M}} \sqrt{g}\,(R-2\Lambda-\mathcal{L}_\mathcal{M})- \frac{1}{\kappa}\int_{\mathcal{Q}} \sqrt{h}\,(K-\mathcal{L}_\mathcal{Q})- \frac{1}{\kappa}\int_{\mathcal{C}} \sqrt{\sigma}\,(\Theta-\mathcal{L}_\mathcal{C}).
\label{actioncornerderivation}
\end{equation}
The variational problem consists of keeping the coordinates and embeddings of the boundaries fixed while varying the inverse metric. The extremum of the action then fixes the metric $\mathcal{M}$. Defining the bulk, boundary, and corner stress tensors as
\begin{equation}
T_{ab}^{\mathcal{M}} = -\frac{2}{\sqrt{g}}\frac{\partial(\sqrt{g}\,\mathcal{L}_\mathcal{M})}{\partial g^{ab}},\ \ T_{ab}^{\mathcal{Q}} = -\frac{2}{\sqrt{h}}\frac{\partial(\sqrt{h}\,\mathcal{L}_\mathcal{Q})}{\partial h^{ab}},\ \ T_{ab}^{\mathcal{C}} = -\frac{2}{\sqrt{\sigma}}\frac{\partial(\sqrt{\sigma}\,\mathcal{L}_\mathcal{C})}{\partial \sigma^{ab}},
\end{equation}
the variation of \eqref{actioncornerderivation} with two intersecting branes $\mathcal{Q}_{s}$ ($s = 1,2$) sharing a corner $\mathcal{C} \equiv \mathcal{Q}_1 \cap \mathcal{Q}_2$ becomes
\begin{equation}
\begin{split}
\delta I =\ &-\frac{1}{2\kappa}\int_{\mathcal{M}} \sqrt{g}\,\left(G_{ab} + \Lambda g_{ab}+\frac{1}{2} T^{\mathcal{M}}_{ab}\right)\delta g^{ab}\\
&\ \ -\sum_{s=1}^2\frac{1}{2\kappa}\int_{\mathcal{Q}_s} \sqrt{h_s}\,\left(K_{sab}-K_s h_{sab}+ T^{\mathcal{Q}_s}_{ab}\right)\delta h^{ab}_s
- \sum_{s=1}^2\frac{1}{2\kappa}\int_{\mathcal{Q}_s} \sqrt{h_s}\,D_{sa}\deltabar U^{a}_s\\
&\ \ +\frac{1}{2\kappa}\int_{\mathcal{C}}  \sqrt{\sigma}\,\left(\Theta\,\sigma_{ab}- T^{\mathcal{C}}_{ab}\right)\delta \sigma^{ab}-\frac{1}{\kappa}\int_{\mathcal{C}} \sqrt{\sigma}\,\delta \Theta,
\end{split}
\label{fullvariation}
\end{equation}
where $ D_{s}^a $ is the covariant derivative compatible the induced metric of $\mathcal{Q}_s$ and
\begin{equation}
\deltabar U_s^{a} = n_{sb}\,\delta g^{ab} + 2g^{ab}\,\delta n_{sb}.
\end{equation}
Here $n^a_s$ is the outward-pointing unit normal vector of $\mathcal{Q}_s$. Note that the $s$ is \textit{not} a spatial index but rather a label corresponding to the brane.

The first two lines of \eqref{fullvariation} come from the variation of the Einstein-Hilbert and Gibbons--Hawking--York terms \cite{chakraborty_novel_2017,jiang_surface_2019}. Additionally, the total derivative term $D_a \deltabar U^a$ sums with the variation of the Hayward term. We deduce that the bulk, boundary, and corner Einstein equations are
\begin{align}
&G_{ab} + \Lambda g_{ab}+\frac{1}{2}T^{\mathcal{M}}_{ab}=0,\label{einsteinBulk}\\
&K_{sab}-K_s h_{sab}+T^{\mathcal{Q}_s}_{ab}=0,\label{einsteinBdry}\\
&\Theta\,\sigma_{ab}-T^{\mathcal{C}}_{ab} = 0.\label{einsteinCorner}
\end{align}
Upon imposing these, the variation \eqref{fullvariation} should vanish. However, when we do so, we are left with
\begin{equation}
\delta I = -\frac{1}{2\kappa}\int_{\mathcal{C}} \sqrt{\sigma}\left(2\delta \Theta+t_{1a}\deltabar U^{a}_1+t_{2a}\deltabar U^{a}_2\right).\label{remaining}
\end{equation}
where $t_s^a $ a unit tangent vector of $\mathcal{Q}_s$ positioned as an outward-pointing unit normal vector of $\partial \mathcal{Q}_s = \mathcal{C}$---see Figure \ref{figs:cornerCancel}.

\input{figs/cornerCancel}

So, for the variational principle to be consistent with \eqref{einsteinBulk}--\eqref{einsteinCorner}, the remaining corner term \eqref{remaining} has to vanish. This is indeed the case without imposing extra conditions on the variation. When the embeddings of the boundaries are kept fixed by \eqref{einsteinBdry}, we have the identity (see \cite{jiang_surface_2019})
\begin{equation}
\delta n_{sa} = \delta \omega_s\, n_{sa}, \ \ \delta \omega_s = -\frac{1}{2} n_{sa}n_{sb}\, \delta g^{ab}.
\label{deltana}
\end{equation}
It follows that
\begin{equation}
t_{sa}\, \deltabar U^{a}_s = t_{sa}n_{sb}\, \delta g^{ab}
\label{rdeltabarU}
\end{equation}
where we have used $ t_{sa}n^{a}_s = 0 $. Meanwhile, the two sets of orthonormal unit vectors are related as
\begin{equation}
\begin{split}
n_{2}^{a} &= n_{1}^{a}\,\cos{\Theta}+t_{1}^{a}\,\sin{\Theta},\\
t_{2}^{a} &= n_{1}^{a}\,\sin{\Theta}-t_{1}^{a}\,\cos{\Theta}.
\end{split}
\label{vectorrotation}
\end{equation}
so that $ \cos{\Theta} = n_1\cdot n_2 $. We may write the variation of $\Theta$ as
\begin{equation}
\delta\Theta = -\delta(g^{ab} n_{1a} n_{2b}) \csc{\Theta}  = (\delta\omega_1 - \delta \omega_2) \cot{\Theta}- t_{1a}n_{1b}\, \delta g^{ab}.\label{deltatheta}
\end{equation}
where we have used \eqref{deltana} and \eqref{vectorrotation} to perform various simplifications. Note that the definition of $\Theta$ is symmetric under the exchange $ 1\leftrightarrow 2 $, and adding the exchanged equation to \eqref{deltatheta} yields
\begin{equation}
2\,\delta\Theta = -(t_{1a}n_{1b}+t_{2a}n_{2b})\,\delta g^{ab}.
\end{equation}
By combining this with \eqref{rdeltabarU}, we see that the variation \eqref{remaining} indeed vanishes.

\section{Foliating AdS$_{d+1}$ into AdS$_d$}\label{app:foliations} 

As discussed in \cite{Aharony:2003qf}, families of Karch--Randall branes may be viewed as foliations of bulk AdS$_{d+1}$ space into AdS$_d$ slices of constant extrinsic curvature.\footnote{Implicit in the definition of a ``foliation" is that the individual AdS$_{d}$ ``leaves" are smooth and topologically connected. We will not explicitly discuss non-smoothness or disconnectedness in this appendix, but note that foliations are still relevant to realizing such characteristics.} In this appendix, we will describe the foliations which realize a variety of topologies in the boundary field theory---namely half-spaces, disks, strips, and annuli.

Foliations yielding the first two topologies are standard in the AdS/BCFT literature \cite{Takayanagi:2011zk,Fujita:2011fp} and can be done rather easily in Euclidean pure AdS obtained from Wick rotation of Lorentzian pure AdS. The strip, while less standard, can also be done in pure AdS by compactifying the half-space foliation. However, to obtain a manifestly annular boundary topology in general dimension, it is more natural foliate Euclidean AdS-Schwarzschild \cite{Cooper:2018cmb} obtained from Wick rotation of Lorentzian AdS-Schwarzschild.

While the solutions constructed from the foliations described in this section are those of pure Euclidean gravity with an RS term,\footnote{Note that we will set the bulk AdS radius $\ell = 1$ for convenience.}
\begin{equation}
I = -\frac{1}{2\kappa}\int_{\mathcal{M}} d^{d+1}X \sqrt{g}\,\left[R + \frac{d(d-1)}{\ell^2} \right] - \frac{1}{\kappa}\int_{\mathcal{Q}} d^d\hat{x} \sqrt{h}\,(K-T),
\end{equation}
they are also be relevant in the probe limit of the Einstein + scalar theory considered in the main text.

We note that much of the discussion below is in general $d$, but specification to $d = 2$---the case of interest in this paper---is readily apparent. Interestingly, there is something special which happens precisely when $d = 2$; the annular foliation of the Euclidean BTZ metric is actually an identification of the strip foliation of the pure AdS$_3$ metric. This is because the pure AdS$_3$ and Euclidean BTZ metrics are locally equivalent up to a coordinate transformation.

\subsection{Half-Spaces and Disks in Pure AdS}\label{sec:planarSpherical}

We first review the standard foliations which provide either half-space or a disk on the conformal boundary. We start with the following Euclidean metric ansatz for the full equations of motion,
\begin{equation}
ds^2 = \frac{1}{f(\vartheta)^2}\left(d\vartheta^2 + ds_{\text{AdS}_d}^2\right),\ \ \vartheta \in (0,\pi).\label{foliation}
\end{equation}
The coordinate $\vartheta$ labels a particular Euclidean AdS$_{d}$ slice, and backreaction manifests through the freedom of $f(\vartheta)$.\footnote{Keeping $f(\vartheta)$ generic would be interesting to study effects subleading to the probe limit of matter.} For pure AdS$_{d+1}$, the warp factor is
\begin{equation}
f(\vartheta) = \sin\vartheta.\label{emptyf}
\end{equation}
A simple foliation is one where we slice the bulk into planes,
\begin{equation}
ds_{\text{AdS}_d}^2 = \frac{1}{w^2}(dw^2 + d\vec{x}^2),\ \ w \in (0,\infty).\label{adsSlicePlanar}
\end{equation}
Here, $\vec{x} \in \mathbb{R}^{d-1}$ parameterizes the transverse flat directions. Each constant-$\vartheta$ slice is a Poincar\'e patch of Euclidean AdS$_d$ on an upper half-plane. They share a conformal boundary at $w \to 0$. We depict the planar foliation of AdS$_{d+1}$ in Figure \ref{figs:adsFoliationPlanar}.

\input{figs/adsFoliationsPlanarSpherical}

An alternate foliation is the one for which each constant-$\vartheta$ slice is Euclidean AdS$_d$ in a disk,
\begin{equation}
ds_{\text{AdS}_d}^2 = \frac{1}{w^2}\left[dw^2 + \left(1 - \frac{w^2}{4\mathcal{R}^2}\right)^2 \mathcal{R}^2 d\Omega_{d-1}^2\right],\ \ w \in (0,2\mathcal{R}],\label{adsSlice}
\end{equation}
Here, $\mathcal{R} > 0$ is a free parameter\footnote{Technically $\mathcal{R}$ has length-dimension $1$, so the appropriate \textit{dimensionless} free parameter is $\mathcal{R}/\ell$.} labeling the particular foliation we choose, while $d\Omega_{d-1}$ is the line element of the transverse $(d-1)$-sphere. Geometrically, $\mathcal{R}$ is the radius of the $(d-1)$-sphere on the boundary (up to the $w^{-1}$ boundary divergence) at which the constant-$\vartheta$ slices intersect. We depict this disk foliation in Figure \ref{figs:adsFoliationSpherical}.

Each individual leaf of each of these foliations is a valid KR brane which solves the embedding equation
\begin{equation}
K_{ab} - (K-T)\,h_{ab} = 0,\label{embed}
\end{equation}
with the tension $T$ of the leaf $\vartheta = \theta$ being
\begin{equation}
T = -(d-1)\cos\theta.\label{tensionVal}
\end{equation}
Thus, the induced AdS$_d$ curvature radius $\bar{\ell}$ is related to the tension by
\begin{equation}
\bar{\ell}^2 = \frac{1}{\sin^2\vartheta} = \frac{1}{1-T^2/(d-1)^2}.\label{curvatureInduced}
\end{equation}
So, we restrict the bulk coordinates to $\vartheta \in (0,\theta]$ to obtain AdS$_{d+1}$ space with an EOW KR brane. This ``excision" of the subspace $\vartheta > \theta$ yields a holographic BCFT state on either half-space (for the planar foliation---see Figure \ref{figs:bcftDualsPlane}) or a disk of radius $\mathcal{R}$ (for the disk foliation---see Figure \ref{figs:bcftDualsSphere}). Note that $\theta > \frac{\pi}{2}$ characterizes positive tension branes, $\theta < \frac{\pi}{2}$ describes negative tension branes, and $\theta = \frac{\pi}{2}$ is the zero tension brane.

\input{figs/bcftDualsPlaneSphere}

As an aside, observe that by taking $\mathcal{R} \to \infty$ in \eqref{adsSlice}, we obtain \eqref{adsSlicePlanar}. Thus, the planar foliation is actually a limit of disk foliations. So for now, we will focus on providing details about the disk.

\subsubsection*{Disk Branes in Poincar\'e Coordinates}

Now consider pure AdS in Poincar\'e coordinates,
\begin{equation}
ds^2 = \frac{dz^2 + dy^2 + d\vec{x}^2}{z^2}.\label{poincareMet}
\end{equation}
where $z > 0$ and $(y,\vec{x}) \in \mathbb{R}^d$. We find this coordinate system to be more convenient in the main text for studying a scalar field between two disconnected disk EOW branes. To use it, we require the disk foliation of Poincar\'e space which yields a disk of radius $\mathcal{R}$ on the boundary. We may apply a conformal mapping to the disk brane $\vartheta = \theta$ in the foliation coordinates \eqref{foliation} to write \cite{Berenstein:1998ij,Fujita:2011fp}
\begin{equation}
(z + \mathcal{R} \cot\theta)^2 + y^2 + \vec{x}^2 = \mathcal{R}^2 \csc^2\theta.\label{sphereBrane}
\end{equation}
It is useful to write this in terms of brane tension $T$, instead. By doing so, we get
\begin{equation}
\left[z - \frac{\mathcal{R} T}{(d-1)\sqrt{1-T^2/(d-1)^2}}\right]^2 + y^2 + \vec{x}^2 = \frac{\mathcal{R}^2}{1-T^2/(d-1)^2}.
\end{equation}
The symmetry of these branes suggest working in an alternate set of spherical coordinates in the subspace $(y,\vec{x}) \in \mathbb{R}^{d}$,
\begin{equation}
(y,\vec{x}) \to (\varrho,\psi_1,...,\psi_{d-2},\phi),\label{planetospherePoincare}
\end{equation}
where $\varrho$ is a radial coordinate and $\psi_1,...,\psi_{d-2},\phi$ parameterize a $(d-1)$-sphere; we take $\phi$ to be the $2\pi$-periodic angle. The Poincar\'e metric takes the form
\begin{equation}
ds^2 = \frac{dz^2 + d\varrho^2 + \varrho^2 d\Omega_{d-1}^2}{z^2}.\label{sphericalPoincare}
\end{equation}
Thus the disk branes are rotation-invariant,
\begin{equation}
\left[z - \frac{\mathcal{R} T}{(d-1)\sqrt{1-T^2/(d-1)^2}}\right]^2 + \varrho^2 = \frac{\mathcal{R}^2}{1-T^2/(d-1)^2}.\label{poincareBranes}
\end{equation}

\subsubsection*{Disk Branes in Global Coordinates}

Another embedding we use in the main text is that of the disk branes in global coordinates. This coordinate chart is how one explicitly realizes Euclidean BCFT states on the cylinder. The embedding is described in \cite{Cooper:2018cmb}; we review their construction here.

We start with a generic spherically symmetric metric which fills the $\mathbb{R} \times S^{d-1}$ cylinder, so the bulk is topologically $\mathbb{R} \times D^d$ (where $D^d$ denotes a $d$-dimensional disk). Schematically, the metric in $(\tau,r,\psi_1,...,\psi_{d-2},\phi)$ coordinates is
\begin{equation}
ds^2 = f(r)d\tau^2 + \frac{dr^2}{f(r)} + r^2 d\Omega_{d-1}^2.\label{adsCylinder}
\end{equation}
where $\tau \in \mathbb{R}$ and $r > 0$, with $r = \infty$ being the conformal boundary. Pure AdS corresponds to
\begin{equation}
f(r) = r^2 + 1.
\end{equation}
To get the disk branes in these coordinates, we want a foliation into AdS$_{d}$ slices in balls anchored to the boundary at some $\tau = \tau_0$. From the embedding equation \eqref{embed}, we may find that the spherically symmetric branes are described by the equation of motion \cite{Cooper:2018cmb}\footnote{This is technically only one branch describing the leaves of the foliation, and we may also consider the same expression but without a minus sign. The induced geometry would still be AdS$_d$, but the embedding equation \eqref{embed} would imply that such a brane has tension $-T$.}
\begin{equation}
\frac{d\tau}{dr} = -\frac{T r}{(d-1)f(r)}\frac{1}{\sqrt{f(r)-T^2 r^2/(d-1)^2}}.\label{dtdr}
\end{equation}
By integrating (noting that $\tau(\infty) = \tau_0$), we have that the branes (for any $d$) are
\begin{equation}
\begin{split}
\tau(r) - \tau_0
&= \int_r^{\infty} d\hat{r}\, \frac{T \hat{r}}{(d-1)f(\hat{r})}\frac{1}{\sqrt{f(\hat{r})-T^2 \hat{r}^2/(d-1)^2}}\\
&= \text{Tanh}^{-1}\left[\frac{T}{\sqrt{(d-1)^2 (r^2 + 1) - r^2 T^2}}\right],\label{globalBranes}
\end{split}
\end{equation}
where $T$ is the tension.\footnote{Note that \cite{Cooper:2018cmb} does not have the factors of $(d-1)$. This is because they incorporate it into the RS action.} As before, these KR branes have an induced AdS$_{d}$ geometry with the radius given by \eqref{curvatureInduced}. The foliation is depicted in Figure \ref{figs:adsGFoliationSpherical}, and the bulk geometry with one of these leaves as an EOW brane is depicted in Figure \ref{figs:bcftDualGSphere}.

\input{figs/adsGFoliationBCFTdualSphere}

The above foliation is obtained independently of the previous constructions. So, as a sanity check, we should ensure that these disk branes are equivalent to those of the Poincar\'e patch. Specifically, we first note that there exists a coordinate transformation from \eqref{sphericalPoincare} to \eqref{adsCylinder} with $f(r) = r^2 + 1$,
\begin{equation}
\tau = \frac{1}{2}\log\left(z^2 + \varrho^2\right),\ \ r = \frac{\varrho}{z}.\label{poincaretoglobal}
\end{equation}
Applying this coordinate transformation to \eqref{globalBranes} produces \eqref{poincareBranes} so long as the free parameters $\tau_0$ and $\mathcal{R}$ are related by
\begin{equation}
\mathcal{R} = e^{\tau_0}.
\end{equation}

\subsection{Strips in Pure AdS}\label{sec:strip}

We now discuss the foliation of pure AdS into $d$-dimensional strips, by which we mean $\mathbb{R} \times D^{d-1}$. As it turns out, this topology comes about when mapping the half-space foliation in Poincar\'e coordinates to the global AdS cylinder. In other words, we may think about the strip foliation as a compactification of the planar foliation on each $D^{d-1}$ slice.

First, note that the half-space foliation in Poincar\'e coordinates \eqref{poincareMet} is described by
\begin{equation}
\frac{y}{z} = \cot\theta = -\frac{T}{(d-1)\sqrt{1 - T^2/(d-1)^2}},\label{planarPoincare}
\end{equation}
In order to transform into global coordinates, we use \eqref{planetospherePoincare}. Concretely, without loss of generality we may specify the transformation such that
\begin{equation}
y = \varrho\sin\phi\,\prod_{i=1}^{d-2} \sin\psi_i,
\end{equation}
where we take $\{\psi_i\}$ to be angles such that $\psi_i \in [0,\pi]$ for $i < d-2$ and $\phi \sim \phi + 2\pi$. By then using \eqref{poincaretoglobal}, we rewrite \eqref{planarPoincare} as
\begin{equation}
r\sin\phi \prod_{i=1}^{d-2} \sin\psi_i = -\frac{T}{(d-1)\sqrt{1 - T^2/(d-1)^2}}.\label{stripEmbed}
\end{equation}
Within the pure AdS cylinder \eqref{adsCylinder}, we note that this foliation is $\tau$-independent and, on each $\tau$-slice, has the topology of a $(d-1)$-disk. Thus a single leaf indeed has topology $\mathbb{R} \times D^{d-1}$---that of the $d$-dimensional strip.

\input{figs/adsGFoliationBCFTdualStrip}

To get a clearer idea of how this foliation looks and why we refer to its leaves as strips, we restrict to the $(r,\phi)$ subspace. On this disk, note that the $r \to \infty$ limit only makes sense in \eqref{stripEmbed} if $\psi_i = 0$ or $\pi$. As such, each leaf of the foliation projects onto a curve which anchors to the conformal boundary at antipodal points (Figure \ref{figs:adsGFoliationStrip}). By taking one of these leaves as an EOW brane, we are left with a configuration of the sort depicted in Figure \ref{figs:bcftDualGStrip}.

\subsection{Annuli in Euclidean AdS-Schwarzschild}\label{sec:connectedBrane}

We now describe how to construct the smooth, connected branes which yield annular boundary topologies (equivalent to a finite cylinder). These branes arise as a natural foliation of the Euclidean AdS-Schwarzschild geometry.

The boundary topology of Euclidean AdS-Schwarzschild is a toroid of the form $S^{1} \times S^{d-1}$, but it may also be treated as a finite cylinder with $(d-1)$-spherical cross-sections and the ends identified. The bulk metric filling this toroid is spherically symmetric and thus of the form \eqref{adsCylinder}, which we now decorate with tildes to avoid confusion with the previous metric:
\begin{equation}
ds^2 = \tilde{f}(\tilde{r})d\tilde{\tau}^2 + \frac{d\tilde{r}^2}{\tilde{f}(\tilde{r})} + \tilde{r}^2 d\tilde{\Omega}_{d-1}^2.\label{tildedAdSCylinder}
\end{equation}
However, this time we have that
\begin{equation}
\tilde{f}(\tilde{r}) = \tilde{r}^2 + 1 - \frac{r_h^{d-2}}{\tilde{r}^{d-2}} \left(r_h^2 + 1\right).
\end{equation}
This metric with range $\tilde{r} > r_h$ covers the whole Euclidean manifold. Furthermore, in this geometry, $2\pi$-periodic angle $\tilde{\phi}$ parameterizes the non-contractible circle of the solid toroid. Meanwhile, $\tilde{\tau}$ parameterizes a contractible circle direction. We usually relate this periodicity in $\tilde{\tau}$ to the inverse temperature,
\begin{equation}
\beta_h =\frac{4 \pi r_{h}}{(d-2) + d r_{h}^{2}},
\end{equation}
so we may take the domain of $\tilde{\tau}$ to be $\left[-\frac{\beta_h}{2},\frac{\beta_h}{2}\right]$.\footnote{Note that this $\beta_h$ is unrelated to the $\beta$ defined in our review of BCFT in Section \ref{sec:BCFT}. Instead, it will be related to $W$---i.e. the width of the boundary cylinder.}

Topologically, we want a smooth ``annular" foliation of the AdS-Schwarzschild geometry; such a foliation is depicted in Figure \ref{figs:schFoliation}. To do this, without loss of generality we may consider solutions that are symmetric about $\tilde{\tau} = 0$ and so have $\left.\frac{d\tilde{r}}{d\tilde{\tau}}\right|_{\tilde{\tau} = 0} = 0$.

\input{figs/schFoliationBCFTdual2}

The equation of motion \eqref{dtdr} is branched at this throat. Instead of just taking one branch as we did before, we must use both branches to write the full solution, with the signs such that $\frac{d\tilde{\tau}}{d\tilde{r}} > 0$ in the $\tilde{\tau} > 0$ branch and $\frac{d\tilde{\tau}}{d\tilde{r}} < 0$ in the $\tilde{\tau} < 0$ branch.
\begin{equation}
\frac{d\tilde{\tau}}{d\tilde{r}} = \text{sgn}(\tilde{\tau}) \frac{|T| \tilde{r}}{(d-1)\tilde{f}(\tilde{r})}\frac{1}{\sqrt{\tilde{f}(\tilde{r}) - T^2 \tilde{r}^2/(d-1)^2}}.\label{eomBH}
\end{equation}
Each solution is labeled by the radius of the throat $\tilde{r}(\tilde{\tau} = 0) = r_0$ which is constrained by the equation of motion
\begin{equation}
\tilde{f}(r_0) = \left(\frac{T r_0}{d-1}\right)^2.\label{throatCondition}
\end{equation}
By integrating each branch of \eqref{eomBH} from the throat, we may obtain the two halves of the embedding. For example, the $\tilde{\tau} > 0$ branch follows
\begin{equation}
\tilde{\tau}(\tilde{r})=\int_{r_{0}}^{\tilde{r}} d \hat{r} \frac{|T| \hat{r}}{\tilde{f}(\hat{r}) \sqrt{\tilde{f}(\hat{r})-T^{2} \hat{r}^{2}/(d-1)^2}},\label{upperembed}
\end{equation}
while the $\tilde{\tau} < 0$ branch is the same expression times a minus sign.

Given some embedding of this form, we may treat it as an EOW KR brane satisfying \eqref{embed} and excise part of the bulk to get a state on an annulus. Since the topology is toroidal, there is a choice in which part of the bulk we excise (see Figure \ref{figs:annularDual}). This ambiguity manifests in the sign of the tension \cite{Cooper:2018cmb}.

For the branes to have positive tension, we keep the part of the bulk which includes the horizon $\tilde{r} = r_h$---geometrically represented as the central axis of the cylinder or the central cycle of the torus. The resulting half-length of the boundary interval $\tau_0^+$ is then
\begin{equation}
\tau_0^+
= \frac{2\pi r_h}{(d-2) + dr_h^2} - \int_{r_0}^{\infty} d\tilde{r} \frac{T\tilde{r}}{\tilde{f}(\tilde{r})\sqrt{\tilde{f}(\tilde{r})-T^2 \tilde{r}^2/(d-1)^2}}.
\end{equation}
For the branes to have negative tension however, we excise the part of the bulk which includes the horizon, in which case the half-length of the boundary interval $\tau_0^-$ is
\begin{equation}
\tau_0^-
= -\int_{r_0}^{\infty} d\tilde{r} \frac{T\tilde{r}}{\tilde{f}(\tilde{r})\sqrt{\tilde{f}(\tilde{r})-T^2 \tilde{r}^2/(d-1)^2}}.
\end{equation}
Unlike the strip foliation in pure AdS, the annular foliation is not generically antipodal on the $(\tilde{r},\tilde{\tau})$ disk. However, there is an exception---the case of $d = 2$. This is not a coincidence; when $d = 2$, the annular foliations is precisely a quotient of the strip foliation.

\subsubsection*{Euclidean BTZ}

Much of the above expressions become simpler in the Euclidean BTZ metric obtained by setting $d = 2$---the case of primary interest in this paper. Here, the function $\tilde{f}(\tilde{r})$ is simply quadratic while $\beta_h$ has a simple relationship with $r_h$,
\begin{equation}
\tilde{f}(\tilde{r}) = \tilde{r}^2 - r_h^2,\ \ \beta_h = \frac{2\pi}{r_h}.
\end{equation}
Additionally, the constraint \eqref{throatCondition} which determines the throat radius significantly simplifies to
\begin{equation}
\frac{r_0}{r_h} = \frac{1}{\sqrt{1-T^2}}.
\end{equation}
As such, we may solve the integrals expressions in the embeddings above in closed form. Doing so leads us to the following expression for the embedding (covering both the $\tilde{\tau} > 0$ and $\tilde{\tau} < 0$ branches):
\begin{equation}
\tan^2(r_h \tilde{\tau}) = \frac{\tilde{r}^2(1-T^2) - r_h^2}{T^2 r_h^2}.\label{embAnnular}
\end{equation}
This allows us to see that the half-length of the boundary interval is completely tension-independent, simply being a quarter of a full $\tilde{\tau}$-cycle.
\begin{equation}
\tau_0^+ = \tau_0^- = \frac{\pi}{2r_h} = \frac{\beta_h}{4}.\label{halflength2d}
\end{equation}
When viewing the projection of these annular branes onto the $(\tau,r)$ disk, this condition means that they intersect the conformal boundary at antipodal points. Put another way, the width of the boundary cylinder is
\begin{equation}
W = \frac{\pi}{r_h} = \frac{\beta_h}{2}.\label{lengthCyl}
\end{equation}

\subsection{Equivalence of Strip and Annular Foliations in $d = 2$}\label{app:stripAnnEq}

We now demonstrate that, in $d = 2$, the strip foliation of pure AdS is in truth the same as the annular foliation in Euclidean BTZ up to a global topological identification. We exploit this fact in the main text to perform all of our calculations using the pure AdS metric.

We start with pure AdS$_3$ and its associated strip brane \eqref{stripEmbed},
\begin{equation}
ds^2 = (r^2 + 1)d\tau^2 + \frac{dr^2}{r^2 + 1} + r^2 d\phi^2,\ \ r(\phi) = -\frac{T}{\sqrt{1-T^2}}\csc\phi.
\end{equation}
This matches with our conventions in the main text \eqref{stripbranepfunction} (taking $\alpha = 1$ and $\phi_0 = 0$). Furthermore, we take this opportunity to perform the identification
\begin{equation}
\tau \sim \tau + 2\pi r_h,
\end{equation}
for some $r_h > 0$. We then apply the coordinate transformation
\begin{equation}
\tau = r_h \tilde{\phi},\ \ r = \frac{1}{r_h}\sqrt{\tilde{r}^2 - r_h^2},\ \ \phi = r_h \left(\tilde{\tau} + \frac{\pi}{2r_h}\right).
\end{equation}
The metric then becomes
\begin{equation}
ds^2 = (\tilde{r}^2 - r_h^2)d\tilde{\tau}^2 + \frac{d\tilde{r}^2}{\tilde{r}^2 - r_h^2} + \tilde{r}^2 d\tilde{\phi}^2,
\end{equation}
with $\tilde{r} > r_h$, $\tilde{\tau} \sim \tilde{\tau} +  \frac{2\pi}{r_h}$, and $\tilde{\phi} \sim \tilde{\phi} + 2\pi$. This is precisely the Euclidean BTZ metric filling the torus and with inverse temperature $\beta_h = \frac{2\pi}{r_h}$. Furthermore, by applying this very coordinate transformation to the strip brane, we have that
\begin{equation}
\sec^2\left(r_h \tilde{\tau}\right) = \frac{(\tilde{r}^2 - r_h^2)(1-T^2)}{T^2 r_h^2},
\end{equation}
By applying trigonometric identities, we may rewrite this as
\begin{equation}
\tan^2(r_h \tilde{\tau}) = \sec^2(r_h\tilde{\tau}) - 1 = \frac{\tilde{r}^2(1-T^2) - r_h^2}{T^2 r_h^2}.
\end{equation}
This precisely matches with \eqref{embAnnular}, so we indeed have that the strip and annular foliations in $d = 2$ are the same up to a global identification of the former. Thus in the main text, we may restrict ourselves to studying the disk and strip branes in pure AdS.

\section{Well-Definedness of Cutting and Gluing Disk Branes}\label{app:proofNoCaseii}

In this appendix, we show that the cutting-and-gluing procedure for creating an intersecting-disk-brane configuration in the bulk is only valid when
\begin{equation}
T_1 + T_2 > 0.\label{tensionBdApp}
\end{equation}
Recall that the disk branes are embedded in the bulk \eqref{conicalglobalads} as $\tau = -F_1(r)$ and $\tau = F_2(r)$, where
\begin{equation}
F_i(r) \equiv F(r;T_i,\tau_i) = \tau_i + \frac{1}{\alpha}\Tanh^{-1}\left(\frac{T_i \alpha}{\sqrt{f_\alpha(r) - T_i^2 r^2}}\right).
\end{equation}
We cut the first brane at $\tau = -\tau_{1*}$ and the second at $\tau = \tau_{2*}$, with both defined such that
\begin{equation}
\tau_{i*} = F_i(r_*),\label{tausS}
\end{equation}
for the same $r_* > 0$. Thus we may glue both branes along the circle $r = r_*$. As discussed in Section \ref{subsec:braneconfs}, this is only a consistent procedure if
\begin{equation}
\tau_{1*} + \tau_{2*} > \tau_1 + \tau_2.\label{tauineqAp}
\end{equation}
We show that this implies \eqref{tensionBdApp}. In doing so, we first use \eqref{tausS} to write
\begin{equation}
\tau_{1*} + \tau_{2*} = \tau_1 + \tau_2 + \frac{1}{2\alpha}\log\left[\left(\frac{\sqrt{f_\alpha(r_*) - T_1^2 r_*^2} + T_1 r_*}{\sqrt{f_\alpha(r_*) - T_1^2 r_*^2} - T_1 r_*}\right)\left(\frac{\sqrt{f_\alpha(r_*) - T_2^2 r_*^2} + T_2 r_*}{\sqrt{f_\alpha(r_*) - T_2^2 r_*^2} - T_2 r_*}\right)\right].\label{tauequalities}
\end{equation}
Now, we consider just the product inside of the log which we denote as $\Pi$. Note that
\begin{equation}
\left.\Pi\right|_{T_1 + T_2 = 0} = 1.
\end{equation}
By keeping $T_1$ fixed, we may write
\begin{equation}
\pdv{\Pi}{T_2} = \frac{2 f_\alpha(r_*) r_*}{\sqrt{f_\alpha(r_*) - T_2^2 r_*^2}\left(f_\alpha(r_*) - 2T_2 r_*\sqrt{f_\alpha(r_*) - T_2^2 r_*^2}\right)}\left(\frac{\sqrt{f_\alpha(r_*) - T_1^2 r_*^2} + T_1 r_*}{\sqrt{f_\alpha(r_*) - T_1^2 r_*^2} - T_1 r_*}\right).
\end{equation}
The first factor can be shown to be positive by noting that $\left[f_\alpha(r_*) - 2T_2^2 r_*^2\right]^2 > 0$. The second factor is equal to $e^{2\alpha(\tau_{1*} - \tau_1)}$ and so is also positive. We thus deduce that
\begin{equation}
\left.\Pi\right|_{T_1 + T_2 > 0} > 1,\ \ \ \  \left.\Pi\right|_{T_1 + T_2 \leq 0} \leq 1.
\end{equation}
It follows that the log term in \eqref{tauequalities} is positive precisely when \eqref{tensionBdApp} is satisfied. Thus, \eqref{tauineqAp} holds when $T_1 + T_2 > 0$ but is broken otherwise, and so we can only cut and glue disk branes when \eqref{tensionBdApp} is satisfied.

\section{Euclidean On-Shell Action as a Boundary Integral}\label{app:admmassproof}

In this appendix, we prove the boundary version of the ADM mass formula \eqref{ADMmasswithbranes}, which takes into account branes and their intersections in the bulk and serves as a more tractable formula for the on-shell action via \eqref{definingeq}. All of our notation is defined in Section \ref{sec:adm}. To start, we recall the formula here for convenience:
\begin{align}
M_{\text{ADM}}^{\text{ren}} =\ &-\frac{1}{\kappa}\int_{\mathcal{B}\cap \Sigma}\sqrt{\hat{\gamma}}\,\bigl(u_{a}r_{b}\, \nabla^{[a}\xi^{b]}+ N\,(\mathcal{K}-\mathcal{L}_\mathcal{B})\bigr)\nonumber\\
&\ \ -\frac{1}{\kappa}\int_{\mathcal{Q}\cap \Sigma}\sqrt{\hat{h}}\,\bigl(u_{a}n_{b}\, \nabla^{[a}\xi^{b]} + N\,(K-\mathcal{L}_{\mathcal{Q}})\bigr)-\frac{1}{\kappa}\int_{\mathcal{C}\cap \Sigma}N\sqrt{\hat{\sigma}}\,(\Theta-\mathcal{L}_{\mathcal{C}}).\label{ADMmasswithbranesappendix}
\end{align}
We prove \eqref{ADMmasswithbranesappendix} by showing that it satisfies the equation \eqref{definingeq}. This amounts to an explicit version of the calculation done in covariant phase space formalism \cite{wald_black_1993} but with intersecting EOW branes present. Notably, we explicitly show that conical defects that orthogonally run through the Cauchy slices $\Sigma$ are not included in the computation of the Euclidean on-shell action when the ADM mass has the expression \eqref{ADMmasswithbranesappendix}.

We can rewrite the first term as
\begin{equation}
\begin{split}
-\frac{1}{\kappa}\int_{\mathcal{B}\cap \Sigma}\sqrt{\hat{\gamma}}\,u_{a}r_{b}\, \nabla^{[a}\xi^{b]} =\ &-\frac{1}{\kappa}\int_{\Sigma}\sqrt{\hat{g}}\,u_a\nabla_{b}\, \nabla^{[a}\xi^{b]}\\
&\ \ + \frac{1}{\kappa}\int_{\mathcal{Q}\cap\Sigma} \sqrt{\hat{h}}\,u_{a}n_{b}\, \nabla^{[a}\xi^{b]}+\frac{1}{\kappa}\int_{\mathcal{H}} \sqrt{\hat{\sigma}_{\mathcal{H}}}\,u_an_b^{\mathcal{H}}\, \nabla^{[a}\xi^{b]},
\label{noetherchargeapp}
\end{split}
\end{equation}
because the boundary $ \partial \Sigma $ of the Cauchy slice consists of the branes $ \mathcal{Q}\cap \Sigma $ and the horizon $ \mathcal{H} $.\footnote{Stokes' theorem does not involve contributions from corners in $\partial\Sigma$.}  Here, $n_a^{\mathcal{H}}$ is the outward-pointing unit normal vector of the horizon. Now note that, from the Killing vector identity $ \nabla_{a}\nabla_{b}\,\xi_{c} = \tensor{R}{_c_b_a^d}\xi_{d} $, it follows that\footnote{In our conventions, the Riemann tensor is defined as $[\nabla_a,\nabla_b]v_c = \tensor{R}{_a_b_c^d} v_d$ for any vector field $v^a$.}
\begin{equation}
\nabla_{b}\nabla^{[a}\xi^{b]} = -\nabla_{b} \nabla^{b}\xi^{a} =  \tensor{R}{^a_b}\xi^{b},\label{ricciKilling}
\end{equation}
where the first equality is due to $ \nabla^{(a}\xi^{b)} = 0 $, and then we use $ R_{ab} = g^{cd}R_{acbd} $. Furthermore, the last term in \eqref{noetherchargeapp} is
\begin{equation}
\frac{1}{\kappa}\int_{\mathcal{H}} \sqrt{\hat{\sigma}_{\mathcal{H}}}\,u_an_b^{\mathcal{H}}\, \nabla^{[a}\xi^{b]} = \frac{\kappa_{\text{s}}}{2\pi}\,S_{\text{W}},
\label{waldentropyexplicitly}
\end{equation}
where $S_{\text{W}}$ is the Wald entropy \eqref{definingeq} and $\kappa_{\text{s}}>0$ is the surface gravity of $\mathcal{H}$. This comes from the fact that Killing vectors satisfy the general identity $\left.\nabla_{[a}\xi_{b]}\lvert_{\mathcal{H}} = \pm 2 \kappa_{\text{s}} u_{[a}n_{b]}^{\mathcal{H}}\right|_{\mathcal{H}}$. In other words, $\nabla_{[a}\xi_{b]}$ reduces to the binormal of the horizon times surface gravity. By applying \eqref{ricciKilling}--\eqref{waldentropyexplicitly} to \eqref{noetherchargeapp} and recalling that $\xi^a = N u^a$, we get
\begin{equation}
-\frac{1}{\kappa}\int_{\mathcal{B}\cap \Sigma}\sqrt{\hat{\gamma}}\,u_{a}r_{b}\, \nabla^{[a}\xi^{b]} = -\frac{1}{\kappa}\int_{\Sigma}N\sqrt{\hat{g}}\,R_{ab}\,u^{a}u^{b} + \frac{1}{\kappa}\int_{\mathcal{Q}\cap\Sigma} \sqrt{\hat{h}}\,u_{a}n_{b}\, \nabla^{[a}\xi^{b]}+\frac{\kappa_{\text{s}}}{2\pi}S_{\text{W}}.
\end{equation}
Substituting the above expression to \eqref{ADMmasswithbranesappendix}, the Noether charge terms on the branes cancel and we are left with
\begin{align}
M_{\text{ADM}}^{\text{ren}}-\frac{\kappa_{\text{s}}}{2\pi}S_{\text{W}}
=\ &-\frac{1}{\kappa}\int_{\Sigma}N\sqrt{\hat{g}}\,R_{ab}\,u^a u^{b}-\frac{1}{\kappa}\int_{\mathcal{Q}\cap \Sigma}N\sqrt{\hat{h}}\, (K-\mathcal{L}_{\mathcal{Q}})-\frac{1}{\kappa}\int_{\mathcal{C}\cap\Sigma} N\sqrt{\hat{\sigma}}\,(\Theta-\mathcal{L}_{\mathcal{C}})\nonumber\\
&\ \ -\frac{1}{\kappa}\int_{\mathcal{B}\cap \Sigma}N\sqrt{\hat{\gamma}}\, (\mathcal{K}-\mathcal{L}_{\mathcal{B}}).
\label{ADMalmost}
\end{align}
Now, we consider the bulk term over $\Sigma$. The possible contribution to the Ricci tensor of a codimension-2 conical defect $\mathcal{D}$ that non-trivially intersects with $\Sigma$ is given by \cite{fursaev_description_1995}
\begin{equation}
\int_\Sigma N\sqrt{\hat{g}}\,R_{ab} = \int_{\Sigma\backslash (\mathcal{D} \cap \Sigma)} N\sqrt{\hat{g}}\,R_{ab}  + 2\pi\,(1-\alpha)\int_{\mathcal{D}\cap\Sigma} N\sqrt{\hat{g}}\,(n_{1a}^{\mathcal{D}}n_{1b}^{\mathcal{D}} + n_{2a}^{\mathcal{D}}n_{2b}^{\mathcal{D}}),
\end{equation}
where $n_{ia}^{\mathcal{D}}$ are two mutually orthogonal unit vectors spanning the 2-dimensional subspace of $\Sigma$ transverse to $\mathcal{D} \cap \Sigma$. Since we only consider such conical defects tangent to the Killing symmetry,\footnote{Specifically, this covers the conical defect for the case of strip branes. In the case of disk branes, the defect does not intersect each slice ($\mathcal{D} \cap\Sigma = \varnothing$) and is actually described by the horizon $\mathcal{H}$.} we have that $u^an_{ia}^{\mathcal{D}} = 0$ and thus get (on-shell)
\begin{equation}
\frac{1}{\kappa}\int_\Sigma N \sqrt{g}R_{ab}u^a u^b = \frac{1}{\kappa}\int_{\Sigma\backslash (\mathcal{D}\cap\Sigma)} \sqrt{g}R_{ab}u^a u^b = \frac{1}{2\kappa} \int_{\Sigma\backslash (\mathcal{D}\cap\Sigma)} \sqrt{g}(R-2\Lambda),
\end{equation}
where we have used the identity\footnote{Recall that the Einstein tensor is $ G_{ab} = R_{ab}-\frac{1}{2}Rg_{ab} $.}
\begin{equation}
R_{ab}u^{a}u^{b} = (G_{ab}+\Lambda g_{ab})u^{a}u^{b}+\frac{1}{2}(R-2\Lambda)
\label{Rtautauidentity}
\end{equation}
and the fact that $G_{ab}+\Lambda g_{ab} = 0$ on-shell and away from the defect.

Lastly, since $ \xi^{a} $ is a Killing vector for all $ u \in [0,u_0]$ labeling the Cauchy slices \eqref{admMet}, we can use
\begin{equation}
\int_{\mathcal{M}} \sqrt{g} = \int_{0}^{u_0} du\,N\int_{\Sigma} \sqrt{\hat{g}} = u_0\int_{\Sigma} N\sqrt{\hat{g}}
\end{equation}
to replace integrals over $\Sigma$ and its subspaces by full-space integrals divided by $u_0$. Upon doing so, \eqref{ADMalmost} becomes
\begin{equation}
I^{\text{ren}}_{\text{on-shell}} = u_0\,\bigg(M_{\text{ADM}}^{\text{ren}} - \frac{\kappa_{\text{s}}}{2\pi}\,S_{\text{W}}\biggr),
\end{equation}
where
\begin{equation}
\begin{split}
I^{\text{ren}}_{\text{on-shell}}
=\ &-\frac{1}{2\kappa}\int_{\mathcal{M}\backslash \mathcal{D}}\sqrt{g}(R-2\Lambda)-\frac{1}{\kappa}\int_{\mathcal{Q}}\sqrt{h} (K-\mathcal{L}_{\mathcal{Q}})\\
&\ \ -\frac{1}{\kappa}\int_{\mathcal{C}}\sqrt{\sigma}(\Theta-\mathcal{L}_{\mathcal{C}})-\frac{1}{\kappa}\int_{\mathcal{B}}\sqrt{\gamma} (\mathcal{K}-\mathcal{L}_{\mathcal{B}})
\end{split}
\end{equation}
is the (renormalized) Euclidean on-shell action which \textit{does not} contain any contribution from the conical defects but does contain Gibbons--Hawking--York and Hayward terms.

\section{On-Shell Action of Non-Intersecting Disk Branes}\label{app:gravActionexplicit}

In this appendix, we compute the Euclidean on-shell action of two non-intersecting disk branes embedded in the conical geometry \eqref{conicalglobalads} (Figure \ref{figs:nonintDiskBranes}). The calculation is also done in Section \ref{sec:actiondiskconfs}, but here we do it explicitly rather than by employing the ADM mass.

\input{figs/nonintDiskBranes}

First, note that the regulated region bounded by the branes is $ -F_1(r) < \tau < F_2(r) $ with $0<r<\Lambda$ and $F_i(r) = F(r;T_i,\tau_i)$ (see \eqref{diskbrane}). The outward-pointing unit normal vectors of the branes are
\begin{equation}
n_{1a} = -\Omega^{-1}\,\partial_a(\tau + F_1(r)), \quad n_{2a} = \Omega^{-1}\,\partial_a(\tau - F_2(r)),
\end{equation}
where $\Omega$ is a normalization factor. Explicitly,
\begin{equation}
n_{1a} = \biggl( -\sqrt{f_{\alpha}(r)-T^{2}_1r^{2}},\frac{T_1r}{f_{\alpha}(r)},0 \biggr), \quad n_{2a} = \biggl( \sqrt{f_{\alpha}(r)-T^{2}_2r^{2}},\frac{T_2r}{f_{\alpha}(r)},0 \biggr).\label{normsBranesDisks}
\end{equation}
The regularized Einstein-Hilbert part of the total action is
\begin{align}
I_{\mathcal{M}}
&= \frac{4\pi}{\kappa} \int_0^\Lambda r \left[F_1(r) + F_2(r)\right]\nonumber\\
&= \frac{2\pi}{\kappa}\,\Lambda^{2}\,[F_1(\Lambda)+F_2(\Lambda)]-\frac{2\pi}{\kappa}\int_{0}^{\Lambda}dr\,r^{2}\,[F'_1(r)+F'_2(r)],\label{einsteinHilbDisks}
\end{align}
where we have integrated by parts. The brane term in the action is
\begin{equation}
I_{\mathcal{Q}_1}+I_{\mathcal{Q}_2} = \frac{2\pi}{\kappa}\int_{0}^{\Lambda}dr\,(r^{2} + \alpha^{2})\,[F'_1(r) + F'_2(r)],\label{branesDisks}
\end{equation}
where we have used $\sqrt{h_i} = -f_\alpha(r) F'_i(r)/T_i$ and $K_i= 2T_i$. Summing \eqref{einsteinHilbDisks} and \eqref{branesDisks}, we get
\begin{align}
&I_{\mathcal{M}} + I_{\mathcal{Q}_1} + I_{\mathcal{Q}_2}\nonumber\\
&=\frac{2\pi}{\kappa}\,\Lambda^{2}\,[F_1(\Lambda)+F_2(\Lambda)]+\frac{2\pi\alpha^{2}}{\kappa}\int_{0}^{\Lambda}dr\,[F'_1(r)+F'_2(r)]\nonumber\\
&= \frac{2\pi}{\kappa}\,(\Lambda^{2}+\alpha^{2})\,[F_1(\Lambda)+F_2(\Lambda)]-\frac{2\pi\alpha^{2}}{\kappa}\,\biggl[\tau_1+\tau_2+\frac{1}{\alpha}\,(\Tanh^{-1}{T_1}+\Tanh^{-1}{T_2})\biggr],
\end{align}
where we have evaluated the integral explicitly. The last step is to add the Gibbons--Hawking--York and Hayward terms associated with the cutoff surface $\mathcal{B}_\Lambda$ defined as $r = \Lambda$. To do so, note that the outward-pointing unit normal vector field of the cutoff surface is
\begin{equation}
r_{a} = \frac{1}{\sqrt{f_\alpha(\Lambda)}}(0,1,0),\label{normsCutoffLambda}
\end{equation}
and the trace of the extrinsic curvature is
\begin{equation}
\mathcal{K} = \frac{2f_\alpha(\Lambda)+\Lambda f_\alpha'(\Lambda)}{2\Lambda\sqrt{f_\alpha(\Lambda)}}.
\end{equation}
Noting that $ \sqrt{\gamma} = \Lambda\sqrt{f_\alpha(\Lambda)} $, the Gibbons--Hawking--York term at the conformal boundary becomes 
\begin{equation}
I_{\mathcal{B}_\Lambda} = -\frac{2\pi}{\kappa}\,(2\Lambda^{2}+\alpha^{2})\,[F_1(\Lambda)+F_2(\Lambda)]
\end{equation}
By computing the inner products of \eqref{normsBranesDisks} and \eqref{normsCutoffLambda}, we compute the intersection angles appearing in the Hayward terms as:
\begin{equation}
\Theta_{1}^{\Lambda} = \text{Cos}^{-1}\left(\frac{T_1 \Lambda}{\sqrt{f_\alpha(\Lambda)}}\right),\quad \Theta_{2}^{\Lambda} = \text{Cos}^{-1}\left(\frac{T_2 \Lambda}{\sqrt{f_\alpha(\Lambda)}}\right).
\end{equation}
The corner terms are thus
\begin{equation}
I_{\mathcal{B}_\Lambda\cap \mathcal{Q}_1}+I_{\mathcal{B}_\Lambda\cap \mathcal{Q}_2} =-\frac{2\pi}{\kappa}\,\Lambda\,\biggl[\text{Cos}^{-1}{\biggl(\frac{T_1\Lambda}{\sqrt{\Lambda^{2}+\alpha^{2}}}\biggr)}+\text{Cos}^{-1}{\biggl(\frac{T_2\Lambda}{\sqrt{\Lambda^{2}+\alpha^{2}}}\biggr)}\biggr].
\end{equation}
So, summing everything together and expanding around large $\Lambda$, the total regularized action is
\begin{equation}
\begin{split}
&I_{\mathcal{M}} + I_{\mathcal{Q}_1} + I_{\mathcal{Q}_2}+I_{\mathcal{B}_\Lambda\cap \mathcal{Q}_1}+I_{\mathcal{B}_\Lambda\cap \mathcal{Q}_2}+I_{\mathcal{B}_\Lambda}\\
&= -\frac{2\pi}{\kappa}\biggl[(\tau_1+\tau_2)\,\Lambda^{2} +\biggl(\text{Cos}^{-1} T_1+\text{Cos}^{-1} T_2+\frac{T_1}{\sqrt{1-T_1^{2}}}+\frac{T_2}{\sqrt{1-T_2^{2}}}\biggr)\,\Lambda\biggr.\\
&\hspace{5cm}\biggl.+\alpha^2\, (\tau_1+\tau_2)+\alpha\,(\Tanh^{-1}{T_1}+\Tanh^{-1}{T_2})+\mathcal{O}\biggl(\frac{1}{\Lambda}\biggr) \biggr]
\end{split}
\end{equation}
To cancel the divergences, we add the following counterterms at the cutoff boundary and corners:
\begin{align}
I_{\text{ct}} &= \frac{a_1}{\kappa}\int_{\mathcal{B}_\Lambda}\sqrt{\gamma}+\frac{b_1}{\kappa}\int_{\mathcal{B}_\Lambda\cap \mathcal{Q}_1} \sqrt{\sigma_1}+\frac{b_2}{\kappa}\int_{\mathcal{B}_\Lambda\cap \mathcal{Q}_2} \sqrt{\sigma_2}\\
&= \frac{2\pi}{\kappa}\biggl[a_1\,(\tau_1+\tau_2)\,\Lambda^{2} +\biggl( b_1+b_2+\frac{a_1T_1}{\sqrt{1-T_1^{2}}}+\frac{a_1T_2}{\sqrt{1-T_2^{2}}}\biggr)\,\Lambda+\frac{1}{2}\, a_1\alpha^{2}\,(\tau_1+\tau_2)+\mathcal{O}\biggl(\frac{1}{\Lambda}\biggr)\biggr],\nonumber
\end{align}
The divergences are cancelled by setting
\begin{equation}
a_1 = 1, \quad b_1 = \text{Cos}^{-1} T_1, \quad b_2 = \text{Cos}^{-1} T_2,
\end{equation}
and the renormalized on-shell action is
\begin{equation}
I_{\text{on-shell}}^{\text{ren}} = -\frac{2\pi}{\kappa}\Bigl[\frac{1}{2}\,\alpha^{2}\,(\tau_1+\tau_2)+\alpha\,(\Tanh^{-1}{T_1}+\Tanh^{-1}{T_2})\Bigr],
\label{twodiskbraneaction}
\end{equation}
which agrees with the result \eqref{nonintdiskbraneaction} of the computation in the main text.
\vfill

\end{appendices}

\bibliographystyle{jhep}
\bibliography{references.bib}
\end{document}

%% file: figs/cylinderDuals.tex
\begin{figure}
\subfloat[Disconnected\label{figs:disconnected}]
{\begin{tikzpicture}
\draw[-] (-1,0) to (-1,2);
\draw[-] (1,0) to (1,2);

\draw[-,thick,red] (-1,0) arc (180:360:1 and 0.5);
\draw[-,thick,red,dashed] (-1,0) arc (180:0:1 and 0.5);
\draw[-,thick,red,fill=red!20] (0,2) circle (1 and 0.5);

\node at (1.2,0) {$A$};
\node at (1.2,2) {$B$};

\node[white] at (2,0) {$A$};
\node[white] at (2,2) {$B$};
\node[white] at (-2,0) {$A$};
\node[white] at (-2,2) {$B$};

\draw[-,draw=none,fill=red!40] (-1,0) arc (180:0:1 and 0.75) arc (0:-180:1 and 0.5);
\draw[-,dashed] (-1,1) arc (180:0:1 and 0.5);

\draw[-,thick,red] (-1,0) arc (180:360:1 and 0.5);
\draw[-,thick,red] (0,2) circle (1 and 0.5);
\draw[-,thick,red,dashed] (-1,0) arc (180:0:1 and 0.5);

\draw[-,thick,red] (-1,0) arc (180:0:1 and 0.75);
\draw[-,thick,red,fill=red!40] (-1,2) arc (180:360:1 and 0.75) arc (360:180:1 and 0.5);

\draw[-] (-1,1) arc (180:360:1 and 0.5);

\end{tikzpicture}}\qquad
\subfloat[Smooth and connected\label{figs:connectedSmooth}]
{\begin{tikzpicture}
\draw[-] (-1,0) to (-1,2);
\draw[-] (1,0) to (1,2);

\draw[-,thick,red] (-1,0) arc (180:360:1 and 0.5);
\draw[-,thick,red,dashed] (-1,0) arc (180:0:1 and 0.5);
\draw[-,thick,red,fill=red!20] (0,2) circle (1 and 0.5);

\node at (1.2,0) {$A$};
\node at (1.2,2) {$B$};

\node[white] at (2,0) {$A$};
\node[white] at (2,2) {$B$};
\node[white] at (-2,0) {$A$};
\node[white] at (-2,2) {$B$};

\draw[-,dashed] (-1,1) arc (180:0:1 and 0.5);

\draw[-,dashed,red,thick,fill=red!40] (0,0) circle (1 and 0.5);

\draw[-,thick,red,fill=red!40] (-1,0) arc (-90:90:0.5 and 1) to (1,2) arc (90:270:0.5 and 1) arc (0:-180:1 and 0.5);


\draw[-,thick,red] (-1,0) arc (180:360:1 and 0.5);
\draw[-,thick,red,fill=red!20] (0,2) circle (1 and 0.5);

\draw[-,thick,red,dashed] (-1,2) arc (90:0:0.5 and 1);
\draw[-,thick,red,dashed] (1,2) arc (90:180:0.5 and 1);

\draw[-,thick,red] (-0.5,1) arc (180:360:0.5 and 0.15);
\draw[-,thick,red,dashed] (-0.5,1) arc (180:0:0.5 and 0.15);

\draw[-] (-1,1) arc (180:360:1 and 0.5);

\end{tikzpicture}}\qquad
\subfloat[Non-smooth and connected\label{figs:connectedNonSmooth}]
{\begin{tikzpicture}
\draw[-] (-1,0) to (-1,2);
\draw[-] (1,0) to (1,2);

\draw[-,thick,red] (-1,0) arc (180:360:1 and 0.5);
\draw[-,thick,red,dashed] (-1,0) arc (180:0:1 and 0.5);
\draw[-,thick,red,fill=red!20] (0,2) circle (1 and 0.5);

\node at (1.2,0) {$A$};
\node at (1.2,2) {$B$};

\node[white] at (2,0) {$A$};
\node[white] at (2,2) {$B$};
\node[white] at (-2,0) {$A$};
\node[white] at (-2,2) {$B$};

\draw[-,dashed] (-1,1) arc (180:0:1 and 0.5);

\draw[-,thick,red] (-1,0) arc (180:0:1 and 1.25);
\draw[-,thick,red,fill=red!40] (-1,2) arc (180:360:1 and 1.25) arc (360:180:1 and 0.5);
\draw[-,draw=none,fill=red!40] (-1,0) arc (180:0:1 and 1.25) arc (0:-180:1 and 0.5);

\draw[-,red,thick] (-1,0) arc (180:126.25:1 and 1.25);
\draw[-,red,thick] (1,0) arc (0:180-126.25:1 and 1.25);
\draw[-,red,thick] (-1,0) arc (180:360:1 and 0.5);

\draw[-,red,thick] (-1,0) arc (180:126.25:1 and 1.25);
\draw[-,blue,thick,dashed] (-0.5913,1.008) arc (180:0:0.59 and 0.15);
\draw[-,blue,thick] (-0.5913,1.008) arc (180:360:0.59 and 0.15);

\draw[-,thick,red] (-1,0) arc (180:360:1 and 0.5);
\draw[-,thick,red] (0,2) circle (1 and 0.5);
\draw[-,thick,red,dashed] (-1,0) arc (180:0:1 and 0.5);

\draw[-] (-1,1) arc (180:360:1 and 0.5);
\end{tikzpicture}}
\caption{Possible brane configurations for the holographic dual to a BCFT state on a cylinder. $A$ and $B$ represent possible conformal boundary conditions at each connected component of the boundary. Figure \ref{figs:connectedNonSmooth} in particular represents a non-smooth merging of two different branes.}
\label{figs:cylinderDuals}
\end{figure}

%% file: figs/openClosedDuality.tex
\begin{figure}
\subfloat[Open-string channel\label{figs:openChannel}]
{\begin{tikzpicture}
\draw[-,draw=none,fill=red!25] (-1,0) arc (180:360:1 and 0.5) to (1,2) arc (0:-180:1 and 0.5) to (-1,0);
\draw[-,thick,red] (-1,0) arc (180:360:1 and 0.5);
\draw[-,thick,red,dashed] (-1,0) arc (180:0:1 and 0.5);
\draw[-,thick,red,fill=red!10] (0,2) circle (1 and 0.5);

\node at (1.2,0) {$A$};
\node at (1.2,2) {$B$};

\draw[-,thick,black!50] (0,-0.5) to (0,1.5);
\draw[-,thick,black!50] (-0.4,-0.45) to (-0.4,1.55);
\draw[-,thick,black!50] (-0.75,-0.325) to (-0.75,1.67);
\draw[-,thick,black!50] (-1,0) to (-1,2);
\draw[-,thick,black!50] (0.4,-0.45) to (0.4,1.55);
\draw[-,thick,black!50] (0.75,-0.325) to (0.75,1.67);
\draw[-,thick,black!50] (1,0) to (1,2);

\draw[-,dashed,thick,black!50] (-0.6,0.4) to (-0.6,1.615);
\draw[-,thick,black!50] (-0.6,1.615) to (-0.6,2.4);
\draw[-,dashed,thick,black!50] (-0.2,0.47) to (-0.2,1.5);
\draw[-,thick,black!50] (-0.2,1.5) to (-0.2,2.5);
\draw[-,dashed,thick,black!50] (0.2,0.47) to (0.2,1.5);
\draw[-,thick,black!50] (0.2,1.5) to (0.2,2.5);
\draw[-,dashed,thick,black!50] (0.6,0.4) to (0.6,1.615);
\draw[-,thick,black!50] (0.6,1.615) to (0.6,2.4);

\draw[-,thick,red] (-1,0) arc (180:360:1 and 0.5);
\draw[-,thick,red] (0,2) circle (1 and 0.5);



\node at (-1.5,1) {};
\node at (1.5,1) {};
\end{tikzpicture}}\qquad\qquad
\subfloat[Closed-string channel\label{figs:closedChannel}]
{\begin{tikzpicture}
\draw[-,draw=none,fill=red!25] (-1,0) arc (180:360:1 and 0.5) to (1,2) arc (0:-180:1 and 0.5) to (-1,0);
\draw[-,thick,red] (-1,0) arc (180:360:1 and 0.5);
\draw[-,thick,red,dashed] (-1,0) arc (180:0:1 and 0.5);
\draw[-,thick,red,fill=red!10] (0,2) circle (1 and 0.5);

\node at (1.3,0) {$\ket{A}$};
\node at (1.3,2) {$\ket{B}$};

\draw[-,thick,black!50] (-1,0.5) arc (180:360:1 and 0.5);
\draw[-,thick,dashed,black!50] (-1,0.5) arc (180:0:1 and 0.5);

\draw[-,thick,black!50] (-1,1) arc (180:360:1 and 0.5);
\draw[-,thick,dashed,black!50] (-1,1) arc (180:0:1 and 0.5);

\draw[-,thick,black!50] (-1,1.5) arc (180:360:1 and 0.5);
\draw[-,thick,black!50] (0,2) arc (90:30:1 and 0.5);
\draw[-,thick,black!50] (0,2) arc (90:150:1 and 0.5);
\draw[-,thick,dashed,black!50] (-1,1.5) arc (180:0:1 and 0.5);

\draw[-,thick,red] (-1,0) arc (180:360:1 and 0.5);
\draw[-,thick,red] (0,2) circle (1 and 0.5);



\node at (-1.5,1) {};
\node at (1.5,1) {};

\end{tikzpicture}}
\caption{The slicings of the cylinder corresponding to the (a) open-string and (b) closed-string channels. In (a), the cylinder is sliced into intervals along the periodic coordinate, and so the partition function is thermal. In (b), the boundary conditions represent initial and final states belonging to a CFT quantized on the circle.}
\label{figs:openClosedDuality}
\end{figure}

%% file: figs/hOpenClosedPlane.tex
\begin{figure}
\centering
\subfloat[Open-string Hamiltonian\label{figs:hopenPlane}]{
\begin{tikzpicture}[scale=0.95]
\node at (-2,0) {};
\draw[-,draw=none,fill=red!25] (-1,0) arc (180:360:1 and 0.5) to (1,2) arc (0:-180:1 and 0.5) to (-1,0);
\draw[-,thick,red] (-1,0) arc (180:360:1 and 0.5);
\draw[-,thick,red,dashed] (-1,0) arc (180:0:1 and 0.5);
\draw[-,thick,red,fill=red!10] (0,2) circle (1 and 0.5);

\draw[-] (-1,0) to (-1,2);
\draw[-] (1,0) to (1,2);

\draw[-,thick,black!50] (-0.5,-0.45) to (-0.5,1.55);
\draw[-,thick,black!50] (0.5,-0.45) to (0.5,1.55);

\draw[-,thick,red] (-1,0) arc (180:360:1 and 0.5);
\draw[-,thick,red] (0,2) circle (1 and 0.5);

\draw[->,thick] (0,0.75) arc (270:300:1 and 0.5);
\draw[-,thick] (0,0.75) arc (270:240:1 and 0.5);

\node at (0,1) {$H^{\text{op}}$};

\draw[->,very thick] (1.5,1) to (3.5,1);
\node at (2.5,1.3) {Unwrap};
\node at (2.5,0.7) {into strip};

\draw[draw=none,fill=red!25] (4,2) to (7,2) to (7,0) to (4,0) to (4,2);
\draw[<->,thick,red] (4,2) to (7,2);
\draw[<->,thick,red] (4,0) to (7,0);

\draw[-,thick,black!50] (5.5-0.5,0) to (5.5-0.5,2);
\draw[-,thick,black!50] (5.5+0.5,0) to (5.5+0.5,2);

\draw[->,thick] (5,1) to (6,1);
\node at (5.5,1.25) {$H^{\text{op}}$};

\draw[->,very thick] (7.5,1) to (9.5,1);
\node at (8.5,1.3) {Conformal};
\node at (8.5,0.7) {transformation};


\draw[draw=none,fill=red!25] (10,0+0.25) to (10,1.5+0.25) to (13,1.5+0.25) to (13,0+0.25) to (10,0+0.25);
\draw[<->,thick,red] (10,0+0.25) to (13,0+0.25);
\node at (11.5,0+0.25) {$\bullet$};

\draw[-,thick,black!50] (11.875,0+0.25) arc (0:180:0.375);
\draw[-,thick,black!50] (12.75,0+0.25) arc (0:180:1.25);
\draw[->,very thick] (11.5+0.375/1.414,0.375/1.414+0.25) to (11.5+1.25/1.414,1.25/1.414+0.25);
\node at (11.5+0.807,0.45+0.25) {$\mathcal{L}_0$};
\end{tikzpicture}
}\\

\subfloat[Closed-string Hamiltonian\label{figs:hclosedPlane}]{
\begin{tikzpicture}[scale=0.95]
\node at (-2,0) {};
\draw[-,draw=none,fill=red!25] (-1,0) arc (180:360:1 and 0.5) to (1,2) arc (0:-180:1 and 0.5) to (-1,0);
\draw[-,thick,red] (-1,0) arc (180:360:1 and 0.5);
\draw[-,thick,red,dashed] (-1,0) arc (180:0:1 and 0.5);
\draw[-,thick,red,fill=red!10] (0,2) circle (1 and 0.5);

\draw[-] (-1,0) to (-1,2);
\draw[-] (1,0) to (1,2);

\draw[-,thick,black!50] (-1,0.5) arc (180:360:1 and 0.5);
\draw[-,thick,dashed,black!50] (-1,0.5) arc (180:0:1 and 0.5);

\draw[-,thick,black!50] (-1,1.5) arc (180:360:1 and 0.5);
\draw[-,thick,black!50] (0,2) arc (90:30:1 and 0.5);
\draw[-,thick,black!50] (0,2) arc (90:150:1 and 0.5);
\draw[-,thick,dashed,black!50] (-1,1.5) arc (180:0:1 and 0.5);

\draw[-,thick,red] (-1,0) arc (180:360:1 and 0.5);
\draw[-,thick,red] (0,2) circle (1 and 0.5);

\draw[->,thick] (0,0) to (0,1);

\draw[draw=none,fill=red!25] (0.5,0.4) to (0.5,0.6) to (0.1,0.6) to (0.1,0.4) -- cycle;
\node at (0.35,0.5) {$H^{\text{cl}}$};

\draw[->,very thick] (1.5,1) to (3.5,1);
\node at (2.5,1.3) {Conformal};
\node at (2.5,0.7) {transformation};

\draw[-,thick,red,fill=red!25] (5.5,1) circle (1.5);
\draw[-,thick,red,fill=white] (5.5,1) circle (0.25);
\draw[-,thick,black!50] (5.5,1) circle (0.65);
\draw[-,thick,black!50] (5.5,1) circle (1.1);

\draw[->,thick] (5.5+0.65/1.414,1+0.65/1.414) to (5.5+1.1/1.414,1+1.1/1.414);

\node at (6.75,1.4) {$\mathcal{L}_0 + \bar{\mathcal{L}}_0$};
\end{tikzpicture}
}
\caption{The conformal transformations by which we see that the open-string Hamiltonian and closed-string Hamiltonian respectively correspond to (a) dilatation on the upper half-plane generated by $\mathcal{L}_0$ and (b) dilatation on the full plane generated by $\mathcal{L}_0 + \bar{\mathcal{L}}_0$. The additional $-c$ term in \eqref{openHam} and \eqref{closedHam} comes from the Schwarzian derivatives of these conformal transformations.}
\label{figs:hOpenClosedPlane}
\end{figure}

%% file: figs/diskstripbranes.tex
\begin{figure}
\centering
\subfloat[Disk brane in conical AdS\label{figs:diskbrane}]{
\begin{tikzpicture}
\node[white] at (-1.7,0) {$\tau = \tau_0$};
\node[white] at (1.7,0) {$\tau$};

\node[white] at (1.7,0) {$\tau = \tau_0$};
\node[white] at (-1.7,0) {$\tau = \tau_0$};

\draw[->,very thick] (1.3,-1) to (1.3,0);
\node at (1.5,-0.5) {$\tau$};

\draw[-,very thick] (0,-2.25) arc (270:210:1 and 0.25);
\draw[->,very thick] (0,-2.25) arc (270:330:1 and 0.25);
\node at (0.75,-2.5) {$\phi$};

\draw[-] (-1,-1.75) to (-1,0.5);
\draw[-,black!20] (-1,0.5) to (-1,1.75);
\draw[-] (1,-1.75) to (1,0.5);
\draw[-,black!20] (1,0.5) to (1,1.75);

\draw[-] (-1,-1.75) arc (180:360:1 and 0.25);
\draw[-,dashed] (-1,-1.75) arc (180:0:1 and 0.25);

\draw[-,draw=none,fill=red!40] (-1,0.5) .. controls (-0.4,2) and (0.4,2) .. (1,0.5) arc (0:-180:1 and 0.25);

\draw[-,draw=none,fill=red!40] (0,0.5) circle (1 and 0.25);

\draw[-,red,thick,dashed] (-1,0.5) arc (180:0:1 and 0.25);

\draw[-,very thick,black!30!green] (0,-1.75) to (0,0.25);
\draw[-,dashed,very thick,black!30!green] (0,0.25) to (0,1.65);

\draw[-,red!75] (0,1.625) .. controls (-0.4,1.25) and (-0.5,0.4) .. (-0.5,0.275);
\draw[-,red!75] (0,1.15-0.25) arc (270:180:0.67 and 0.25);
\draw[-,red!75] (0,1.15-0.25) arc (270:360:0.67 and 0.25);

\draw[-,red,thick] (-1,0.5) arc (180:360:1 and 0.25);

\end{tikzpicture}}\qquad
\subfloat[Strip brane in conical AdS\label{figs:stripbrane}]{
\begin{tikzpicture}
\node[white] at (-1.7,0) {$\tau = \tau_0$};
\node[white] at (1.7,0) {$\tau$};

\node[white] at (1.7,0) {$\tau = \tau_0$};
\node[white] at (-1.7,0) {$\tau = \tau_0$};

\draw[->,very thick] (1.3,-1) to (1.3,0);
\node at (1.5,-0.5) {$\tau$};

\draw[-,very thick] (0,-2.25) arc (270:210:1 and 0.25);
\draw[->,very thick] (0,-2.25) arc (270:330:1 and 0.25);
\node at (0.75,-2.5) {$\phi$};

\draw[-,black!20] (-1,-1.75) to (-1,1.75);
\draw[-] (1,-1.75) to (1,1.75);

\draw[-,draw=none,fill=red!40] (-0.2,-2) to[bend left] (0.2,-1.5) to (0.2,2) to[bend right] (-0.2,1.5) to (-0.2,-2);


\draw[-] (1,-1.75) arc (0:-101.5:1 and 0.25);
\draw[-,dashed] (1,-1.75) arc (0:78.5:1 and 0.25);
\draw[-,black!20] (-1,-1.75) arc (180:270-11.5:1 and 0.25);
\draw[-,dashed,black!20] (-1,-1.75) arc (180:78.5:1 and 0.25);

\draw[-,red!75] (-0.2,-2+1.75) to[bend left] (0.2,-1.5+1.75);

\draw[-,red,thick,dashed] (0.2,-1.5) to (0.2,1.5);
\draw[-,red,thick] (0.2,1.5) to (0.2,2);

\draw[-,very thick,black!30!green] (0,-1.75) to (0,1.75);

\draw[-] (1,1.75) arc (0:-101.5:1 and 0.25);
\draw[-] (1,1.75) arc (0:78.5:1 and 0.25);
\draw[-,black!20] (-1,1.75) arc (180:270-11.5:1 and 0.25);
\draw[-,black!20] (-1,1.75) arc (180:78.5:1 and 0.25);

\draw[-,red,thick] (-0.2,-2) to (-0.2,1.5);





\end{tikzpicture}}
\caption{(a) A disk brane and (b) a strip brane in conical AdS. The conical defect at $r = 0$ is represented by the green line. The disk brane is a $\phi$-symmetric slice of the bulk, while the strip brane is a $\tau$-symmetric slice of the bulk.}
\label{figs:diskstripbranes}
\end{figure}

%% file: figs/diskBraneCutGlueTorus.tex
\begin{figure}
\centering
\begin{tikzpicture}
\draw[-] (0,1) arc (90:-90:2 and 1);
\draw[-,black!20] (0,1) arc (90:270:2 and 1);
\draw[-,black!20] (0,-0.1) arc (-90:-180:0.4 and 0.2);
\draw[-] (0,-0.1) arc (-90:0:0.4 and 0.2);
\draw[-,black!20] (0,0.1) arc (90:160:0.4 and 0.1);
\draw[-] (0,0.1) arc (90:20:0.4 and 0.1);

\draw[-,draw=none,fill=red!40] (0,1) .. controls (-1.75,1) and (-1.25,-0.25) .. (-1,-0.25) .. controls (-0.75,-0.25) and (-1,0.2) .. (0,0.1);
\draw[-,draw=none,fill=red!40] (0,-1) .. controls (-1.75,-1) and (-1.25,0.25) .. (-1,0.25) .. controls (-0.75,0.25) and (-1,-0.2) .. (0,-0.1);

\draw[-,draw=none,fill=red!20] (0,0.55) circle (0.225 and 0.45);
\draw[-,draw=none,fill=red!20] (0,-0.55) circle (0.225 and 0.45);

\draw[-,dashed,red,thick] (0,-1) arc (-90:90:0.225 and 0.45);
\draw[-,dashed,red,thick] (0,1) arc (90:-90:0.225 and 0.45);

\draw[-,red,thick] (0,1) .. controls (-1.75,1) and (-1.25,-0.25) .. (-1,-0.25) .. controls (-0.75,-0.25) and (-1,0.2) .. (0,0.1);

\draw[-,red,thick] (0,-1) .. controls (-1.75,-1) and (-1.25,0.25) .. (-1,0.25) .. controls (-0.75,0.25) and (-1,-0.2) .. (0,-0.1);


\draw[-,draw=none,fill=red!40] (-0.74,0.4) to (-1.22,0.4) to (-1.22,-0.4) to (-0.74,-0.4) -- cycle;

\draw[-,blue,thick,dashed] (-1.96/2,0) circle (0.24 and 0.12);

\draw[-,very thick,dashed,black!30!green] (0,0) circle (1 and 0.5);
\draw[-,very thick,black!30!green] (1,0) arc (0:102.5:1 and 0.5);
\draw[-,very thick,black!30!green] (1,0) arc (0:-102.5:1 and 0.5);

\draw[-,blue,thick] (-1.96/2-0.24,0) arc (180:0:0.24 and 0.12);

\draw[-,red,thick] (0,-1) arc (-90:-270:0.225 and 0.45);
\draw[-,red,thick] (0,1) arc (-270:-90:0.225 and 0.45);
\end{tikzpicture}
\caption{The configuration obtained by cutting and gluing two disk branes together at a disk of radius $r = r_*$ (shown in blue). This may be interpreted as the embedding of two disk branes in a torus such that they intersect without ``hiding" the defect.}
\label{figs:diskBraneCutGlueTorus}
\end{figure}

%% file: figs/intersectAnnulus.tex
\begin{figure}
\centering
\begin{tikzpicture}
\draw[-,red,thick,fill=red!30] (0,0) circle (2 and 1);
\draw[-,thick,blue] (0,-0.2) circle (1 and 0.4);
\draw[-,red,thick,fill=white] (0,0.125) circle (0.5 and 0.25);
\draw[-,red,thick] (0,0.125) circle (0.5 and 0.25);

\draw[-,red!75] (-0.5,0.125) to[bend left] (-1,-0.25);
\draw[-,red!75] (-2,0) .. controls (-1.8,-0.5) and (-1,-0.75)..  (-1,-0.25);

\draw[-,red!75] (0.5,0.125) to[bend right] (1,-0.25);
\draw[-,red!75] (2,0) .. controls (1.8,-0.5) and (1,-0.75)..  (1,-0.25);

\draw[-,red!75] (-0.2,-0.1) to[bend left] (-0.5,-0.55);
\draw[-,red!75] (-0.75,-0.925) .. controls (-0.5,-0.9) and (-0.4,-0.55) ..  (-0.5,-0.55);

\draw[-,red!75] (0.2,-0.1) to[bend right] (0.5,-0.55);
\draw[-,red!75] (0.75,-0.925) .. controls (0.5,-0.9) and (0.4,-0.55) ..  (0.5,-0.55);

\draw[-,red!75] (0.75,0.925) ..controls (0.65,0.5)  and (0.5,0.25).. (0.325,0.325);
\draw[-,red!75] (-0.75,0.925) ..controls (-0.65,0.5)  and (-0.5,0.25).. (-0.325,0.325);

\draw[-,thick,blue] (0,-0.2) circle (1 and 0.4);
\draw[-,red,thick,fill=white] (0,0.125) circle (0.5 and 0.25);

\draw[-,draw=none,fill=red!40] (-0.5,0.14) .. controls (-0.25,0.25) and (0.25,0.25) .. (0.5,0.14) arc (0:180:0.5 and 0.25);
\draw[-,red,thick] (0,0.125) circle (0.5 and 0.25);

\draw[-,very thick,black!30!green] (0,-0.125) circle (1.45 and 1.25/2);

\end{tikzpicture}
\caption{The configuration obtained by making two annular branes intersect at a finite radius $r = r_*$ (shown in blue). We also show the conical defect (in green). Both the corner and the conical are $\tau$-cycles.}
\label{figs:intersectAnnulus}
\end{figure}

%% file: figs/disconnectedAnnulusDual.tex
\begin{figure}
\centering
\begin{tikzpicture}[scale=1]
\draw[-,black!30] (-3,0) to (4,0) to (5,1) to (-2,1) to (-3,0);

\draw[-,draw=none,fill=red!10] (-0.005,0.5) arc (245:-65:2.4);
\draw[-,draw=none,fill=red!10] (1,0.525) circle (1.05 and 0.3);

\draw[-,red,thick] (-0.05+0.005,0.51) arc (190:350:1.05 and 0.3);
\draw[-,red,dashed,thick] (-0.05+0.005,0.51) arc (190:-10:1.05 and 0.3);

\draw[-,draw=none,fill=red!40] (0.65,0.5) arc (245:-65:0.825);
\draw[-,draw=none,fill=red!40] (1,0.525) circle (0.35 and 0.1);

\node at (1,2.07) {$\bullet$};
\node at (1,5.08) {$\bullet$};
\draw[-,very thick] (1,2.07) to (1,5.08);

\draw[-,red!75] (1,2.07) to[bend right] (0.85,0.44);
\draw[-,red!75] (0.18,1.25) arc (180:360:0.82 and 0.15);

\draw[-,red!95] (1,5.08) to[bend right] (0.65,0.3);
\draw[-,red!95] (-1.377,2.5) arc (180:360:2.39 and 0.45);

\draw[-,red,thick] (0.65+0.005,0.51) arc (190:350:0.35 and 0.1);
\draw[-,red,dashed,thick] (0.65+0.005,0.51) arc (190:-10:0.35 and 0.1);

\draw[->] (3.5,0.525) to (3.2,0.225);
\draw[->] (3.5,0.525) to (3.924,0.525);
\draw[->] (3.5,0.525) to (3.5,0.949);

\node at (3,0.225) {$x$};
\node at (4.1,0.625) {$y$};
\node at (3.3,1.049) {$z$};

\node at (1.9,1.8) {$\mathcal{Q}_1$};
\node at (3,4.4) {$\mathcal{Q}_2$};

\end{tikzpicture}\qquad
\begin{tikzpicture}[scale=0.72]
\draw[->,black!70] (-3,0) to (3,0);
\node at (3,-0.5) {$\sqrt{y^2 + x^2}$};
\draw[->,black!70] (0,0) to (0,6.25);
\node at (-0.3,6.35) {$z$};

\draw[-] (-2.55/12-0.25,0) arc (180:145:0.25);
\node at (-2.55/12-0.5,0.3) {$\theta_1$};

\draw[-] (2.55/2-0.2,0) arc (180:33:0.2);
\node at (2.55/2+0.1,0.45) {$\theta_2$};

\draw[-,very thick,red] (2.55/2,0) arc (-65:245:3);
\draw[-,very thick,red] (2.55/12,0) arc (-65:245:0.5);

\draw[-,very thick,red!40] (2.55/2,0) to (2.55/12,0);
\draw[-,very thick,red!40] (-2.55/2,0) to (-2.55/12,0);

\node[red] at (2.55/2,0) {$\bullet$};
\node[red] at (-2.55/2,0) {$\bullet$};
\node[red] at (2.55/12,0) {$\bullet$};
\node[red] at (-2.55/12,0) {$\bullet$};

\node at (0,0.95) {$\bullet$};
\node at (0,5.7) {$\bullet$};
\draw[-,very thick] (0,0.95) to (0,5.7);

\node at (0.65,1) {$\mathcal{Q}_1$};
\node at (2.55,4.9) {$\mathcal{Q}_2$};

\end{tikzpicture}
\caption{On the left, we show the region $ \mathcal{M} $ bounded by two spherical branes $ \mathcal{Q}_1 $ and $ \mathcal{Q}_2 $. On the right, we take a transverse slice to better depict the angles $\theta_1$ and $\theta_2$ that $\mathcal{Q}_1$ and $\mathcal{Q}_2$ respectively make with the conformal boundary. The minimal geodesic between the two branes, which is ultimately the main object of interest in this section, is shown in black.}
\label{figs:disconnectedAnnulusDual}
\end{figure}

%% file: figs/admSlices.tex
\begin{figure}
\centering
\subfloat[$\phi$-foliation\label{figs:phifoliation}]{\begin{tikzpicture}

\node[white] at (-1.85,0) {$\tau = \tau_0$};
\node[white] at (1.85,0) {$\tau$};

\node[white] at (1.85,0) {$\tau = \tau_0$};
\node[white] at (-1.85,0) {$\tau = \tau_0$};


\draw[-,very thick] (0,-2.25) arc (270:210:1 and 0.25);
\draw[->,very thick] (0,-2.25) arc (270:330:1 and 0.25);
\node at (0,-2.5) {$u = \phi$};

\draw[-,black] (-1,-1.75) to (-1,1.75);
\draw[-] (1,-1.75) to (1,1.75);

\draw[-,draw=none,fill=red!40] (0,1.75) to (1,1.75) to (1,-1.75) to (0,-1.75) to (0,1.75);
\draw[-,red,thick] (1,1.75) to (1,-1.75);

\draw[-,draw=none,fill=red!40] (0,1.75) to (-0.5,1.53349) to (-0.5,-1.96651) to (0,-1.75) to (0,1.75);
\draw[-,red,thick] (-0.5,1.53349) to (-0.5,-1.96651);

\draw[-] (1,-1.75) arc (0:-180:1 and 0.25);
\draw[-,dashed] (1,-1.75) arc (0:180:1 and 0.25);
\draw[-] (0,1.75) circle (1 and 0.25);

\draw[-,very thick,black!30!green] (0,-1.75) to (0,1.75);

\draw[-] (1,1.75) arc (0:-101.5:1 and 0.25);
\draw[-] (1,1.75) arc (0:78.5:1 and 0.25);

\end{tikzpicture}}\quad
\subfloat[$\tau$-foliation\label{figs:taufoliation}]{\begin{tikzpicture}
\node[white] at (-1.85,0) {$\tau = \tau_0$};
\node[white] at (1.85,0) {$\tau$};

\node[white] at (1.85,0) {$\tau = \tau_0$};
\node[white] at (-1.85,0) {$\tau = \tau_0$};

\draw[->,very thick] (1.3,-1) to (1.3,0);
\node at (1.85,-0.5) {$u = \tau$};

\draw[-,very thick,white] (0,-2.25) arc (270:210:1 and 0.25);
\draw[->,very thick,white] (0,-2.25) arc (270:330:1 and 0.25);
\node[white] at (0,-2.5) {$u = \phi$};

\draw[-,red,thick,fill=red!40] (0,-0.875) circle (1 and 0.25);
\draw[-,red,thick,fill=red!40] (0,0.875) circle (1 and 0.25);
\draw[-,red,thick,fill=red!40] (0,0) circle (1 and 0.25);

\draw[-,black] (-1,-1.75) to (-1,1.75);
\draw[-] (1,-1.75) to (1,1.75);

\draw[-] (1,-1.75) arc (0:-180:1 and 0.25);
\draw[-,dashed] (1,-1.75) arc (0:180:1 and 0.25);
\draw[-] (0,1.75) circle (1 and 0.25);

\draw[-,very thick,black!30!green] (0,-1.75) to (0,-1.125);
\draw[-,thick,red] (0,-1.125) arc (-90:-130:1 and 0.25);
\draw[-,thick,red] (0,-1.125) arc (-90:-50:1 and 0.25);
\draw[-,thick,red] (0,-0.25) arc (-90:-50:1 and 0.25);
\draw[-,thick,red] (0,-0.25) arc (-90:-130:1 and 0.25);
\draw[-,thick,red] (0,0.625) arc (-90:-130:1 and 0.25);
\draw[-,thick,red] (0,0.625) arc (-90:-50:1 and 0.25);

\draw[-,very thick,black!30!green] (0,-0.875) to (0,-0.275);
\draw[-,very thick,black!30!green] (0,0) to (0,0.6);
\draw[-,very thick,black!30!green] (0,1.75) to (0,0.875);

\draw[-] (0,1.75) circle (1 and 0.25);

\end{tikzpicture}}
\caption{Two ADM foliations of conical AdS$_3$ \eqref{conicalglobalads}. In (a), we take the foliation coordinate $u$ to be $\phi$, which is periodic. Additionally, the $u$-cycle shrinks to $0$ at $r = 0$, which is a horizon that acts as a boundary for each Cauchy slice $\Sigma$. In (b), we take the foliation coordinate to be $\tau$ and we may make it periodic (resulting in a decomposition of the torus into disks). For this foliation, the slices do not contain horizons acting as boundaries in the bulk, and the conical line defect runs orthogonally to the slices. These are the decompositions relevant to our particular configurations.}
\label{figs:admSlices}
\end{figure}

%% file: figs/cornerCancel.tex
\begin{figure}
\centering
\begin{tikzpicture}

\draw[-] (1,1.25) arc (90:37:0.25);
\node at (0.85,0.75) {$\mathcal{C}$};
\node at (1.2,1.4) {$\Theta$};

\draw[-,thick,red] (-1,1) to (1,1);
\draw[-,thick,blue] (2,-0.5) to (1,1);

\draw[->,red] (1,1) to (1.75,1);
\node[red] at (1.8,0.7) {$t_1^a$};
\draw[->,red] (1,1) to (1,1.75);
\node[red] at (1,2) {$n_1^a$};
\draw[->,blue] (1,1) to (0.584,1.624);
\node[blue] at (0.3,1.6) {$t_2^a$};
\draw[->,blue] (1,1) to (1.624,1.416);
\node[blue] at (1.9,1.4) {$n_2^a$};

\node[black] at (1,1) {$\bullet$};

\node[red] at (-0.75,1.3) {$\mathcal{Q}_1$};
\node[blue] at (2.3,-0.25) {$\mathcal{Q}_2$};
\end{tikzpicture}
\caption{The normal and tangent vectors for two branes $\mathcal{Q}_1$ and $\mathcal{Q}_2$ located at their shared corner $\mathcal{C}$. Note that for either brane, the corresponding normal and tangent vectors are orthogonal. The normal vectors of the two branes form an angle $\Theta$.}
\label{figs:cornerCancel}
\end{figure}
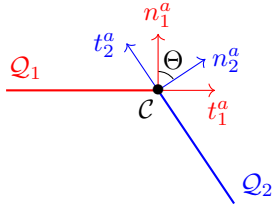

%% file: figs/adsFoliationsPlanarSpherical.tex
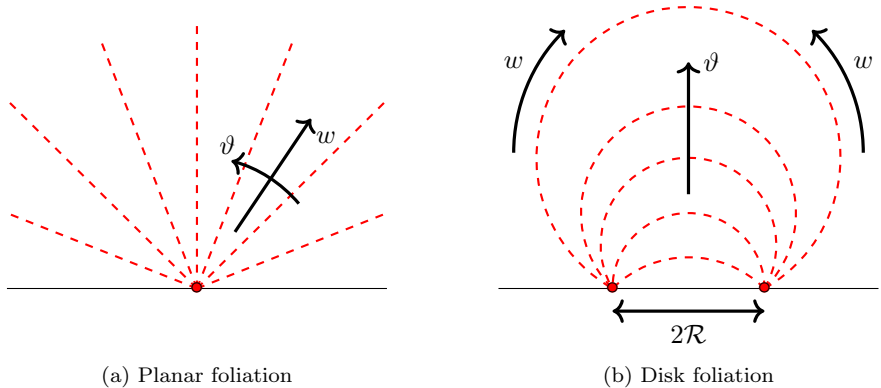
\begin{figure}
\subfloat[Planar foliation\label{figs:adsFoliationPlanar}]
{\begin{tikzpicture}
\draw[-] (-2.5,0) to (2.5,0);

\draw[-,thick,red,dashed] (0,0) to (2.5,1);
\draw[-,thick,red,dashed] (0,0) to (2.5,2.5);
\draw[-,thick,red,dashed] (0,0) to (1.25,3.25);
\draw[-,thick,red,dashed] (0,0) to (0,3.5);
\draw[-,thick,red,dashed] (0,0) to (-1.25,3.25);
\draw[-,thick,red,dashed] (0,0) to (-2.5,2.5);
\draw[-,thick,red,dashed] (0,0) to (-2.5,1);

\node[red] at (0,0) {$\bullet$};
\node at (0,0) {$\circ$};

\draw[->,very thick] (1.75*0.766,1.75*0.643) arc (40:75:1.75);
\node at (0.4,1.9) {$\vartheta$};

\draw[->,very thick] (0.5,0.75) to (1.5,2.25);
\node at (1.7,2) {$w$};

\draw[<->,draw=none] (1,-0.3) to (-1,-0.3);
\node[white] at (0,-0.6) {$2\mathcal{R}$};
\end{tikzpicture}}\qquad\qquad
\subfloat[Disk foliation\label{figs:adsFoliationSpherical}]
{\begin{tikzpicture}
\draw[-] (-2.5,0) to (2.5,0);

\draw[-,red,thick,dashed] (1,0) arc (-60:240:2);
\draw[-,red,thick,dashed] (1,0) arc (-45:225:1.414);
\draw[-,red,thick,dashed] (1,0) arc (-30:210:1.155);
\draw[-,red,thick,dashed] (1,0) arc (0:180:1);
\draw[-,red,thick,dashed] (1,0) arc (45:135:1.414);

\node[red] at (-1,0) {$\bullet$};
\node at (-1,0) {$\circ$};
\node[red] at (1,0) {$\bullet$};
\node at (1,0) {$\circ$};

\draw[->,very thick] (0,1.25) to (0,3);
\node at (0.3,3) {$\vartheta$};

\draw[->,very thick] (2.3,1.8) arc (0:45:2.3);
\draw[->,very thick] (-2.3,1.8) arc (180:135:2.3);
\node at (2.3,3) {$w$};
\node at (-2.3,3) {$w$};

\draw[<->,very thick] (1,-0.3) to (-1,-0.3);
\node at (0,-0.6) {$2\mathcal{R}$};
\end{tikzpicture}}
\caption{Transverse slices of (a) the planar foliation and (b) the disk foliation (labeled by $\mathcal{R}$ which represents the radius of the $(d-1)$-sphere on the boundary) of AdS$_{d+1}$. $\vartheta$ labels different AdS$_d$ slices, starting at $0$ and ending at $\pi$. In the planar foliation, $w = 0$ is the conformal boundary of each individual slice, while $w = \infty$ is the Poincar\'e horizon of each slice. In the disk foliation, $w = 0$ is still the conformal boundary of each slice, but it reaches a finite maximum value of $w = 2\mathcal{R}$ at the ``middle."}
\label{figs:bcftDualsPlaneSphere}
\end{figure}

%% file: figs/bcftDualsPlaneSphere.tex
\begin{figure}
\subfloat[Half-space configuration\label{figs:bcftDualsPlane}]
{\begin{tikzpicture}
\draw[-] (0,0) to (2,0) to (3,1) to (1,1) to (0,0);
\draw[-,black!20] (0,0) to (-2,0) to (-1,1) to (1,1) to (0,0);
\draw[-,draw=none,fill=red!40] (0,0) to (1,1) to (-0.5,2.5) to (-1.5,1.5) to (0,0);
\draw[-,thick,red] (0,0) to (1,1);

\end{tikzpicture}}\qquad\qquad
\subfloat[Disk configuration\label{figs:bcftDualsSphere}]
{\begin{tikzpicture}[scale=1.2]
\draw[-,black!30] (-1,0) to (2,0) to (3,1) to (0,1) to (-1,0);

\draw[-,draw=none,fill=red!40] (0.65,0.5) arc (245:-65:0.825);

\draw[-,draw=none,fill=red!40] (1,0.525) circle (0.35 and 0.1);

\draw[-,red!75] (1,2.07) to[bend right] (0.85,0.44);
\draw[-,red!75] (0.18,1.25) arc (180:360:0.82 and 0.15);

\draw[-,red,thick] (0.65+0.005,0.51) arc (190:350:0.35 and 0.1);
\draw[-,red,dashed,thick] (0.65+0.005,0.51) arc (190:-10:0.35 and 0.1);

\end{tikzpicture}}
\caption{The bulk configuration obtained when selecting a leaf of (a) the planar foliation or (b) the disk foliation as an end-of-the-world KR brane (in red). For the half-space configuration, we excise the part of the bulk ``behind" the brane. For the disk configuration, we excise the part of the bulk ``outside" of the brane.}
\label{figs:bcftDualsPlaneSphere3d}
\end{figure}
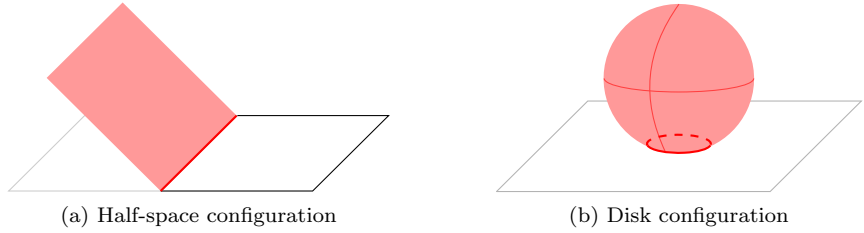

%% file: figs/adsGFoliationBCFTdualSphere.tex
\begin{figure}
\centering
\subfloat[Disk foliation of global\label{figs:adsGFoliationSpherical}]{
\begin{tikzpicture}
\node[white] at (-1.7,0) {$\tau = \tau_0$};
\node[white] at (1.7,0) {$\tau$};

\draw[-] (-1,1.75) to (-1,-1.75);
\draw[-] (1,-1.75) to (1,1.75);

\draw[-,thick,red,dashed] (-1,0) to (1,0);

\draw[-,thick,red,dashed] (-1,0) .. controls (-0.4,1) and (0.4,1) .. (1,0);
\draw[-,thick,red,dashed] (-1,0) .. controls (-0.6,2) and (0.6,2) .. (1,0);

\draw[-,thick,red,dashed] (-1,0) .. controls (-0.4,-1) and (0.4,-1) .. (1,0);
\draw[-,thick,red,dashed] (-1,0) .. controls (-0.6,-2) and (0.6,-2) .. (1,0);

\node[red] at (-1,0) {$\bullet$};
\node at (-1,0) {$\circ$};
\node[red] at (1,0) {$\bullet$};
\node at (1,0) {$\circ$};

\node at (1.7,0) {$\tau = \tau_0$};
\node[white] at (-1.7,0) {$\tau = \tau_0$};

\draw[->,very thick] (-1.5,-0.5) to (-1.5,0.5);
\node at (-1.8,0.5) {$\tau$};

\draw[->,very thick] (0,0.25) to (0.5,0.25);
\draw[->,very thick] (0,0.25) to (-0.5,0.25);
\draw[-,very thick] (0,0.35) to (0,0.15);

\node at (0,0.45) {$r$};

\end{tikzpicture}}\qquad
\subfloat[Disk brane in global\label{figs:bcftDualGSphere}]{
\begin{tikzpicture}
\node[white] at (-1.7,0) {$\tau = \tau_0$};
\node[white] at (1.7,0) {$\tau$};

\node[white] at (1.7,0) {$\tau = \tau_0$};
\node[white] at (-1.7,0) {$\tau = \tau_0$};

\draw[-] (-1,-1.75) to (-1,0.5);
\draw[-,black!20] (-1,0.5) to (-1,1.75);
\draw[-] (1,-1.75) to (1,0.5);
\draw[-,black!20] (1,0.5) to (1,1.75);

\draw[-] (-1,-1.75) arc (180:360:1 and 0.25);
\draw[-,dashed] (-1,-1.75) arc (180:0:1 and 0.25);

\draw[-,draw=none,fill=red!40] (-1,0.5) .. controls (-0.4,2) and (0.4,2) .. (1,0.5) arc (0:-180:1 and 0.25);

\draw[-,draw=none,fill=red!40] (0,0.5) circle (1 and 0.25);

\draw[-,red!75] (0,1.625) .. controls (-0.4,1.25) and (-0.5,0.4) .. (-0.5,0.275);
\draw[-,red!75] (0,1.15-0.25) arc (270:180:0.67 and 0.25);
\draw[-,red!75] (0,1.15-0.25) arc (270:360:0.67 and 0.25);

\draw[-,red,thick] (-1,0.5) arc (180:360:1 and 0.25);
\draw[-,red,thick,dashed] (-1,0.5) arc (180:0:1 and 0.25);

\draw[white,->,very thick] (-1.5,-0.5) to (-1.5,0.5);
\node[white] at (-1.8,0.5) {$\tau$};
\draw[->,very thick] (1.5,-0.5) to (1.5,0.5);
\node at (1.8,0.5) {$\tau$};

\draw[-,very thick] (0,-1) arc (-90:-155:1 and 0.25);
\draw[->,very thick] (0,-1) arc (-90:-25:1 and 0.25);
\node at (0.8,-1.25) {$\phi$};

\end{tikzpicture}}
\caption{(a) A transverse slice of the disk foliation of Euclidean pure AdS$_{d+1}$ in global coordinates. $\tau$ increases upward while $r$ increases outward from the central axis. (b) The bulk configuration obtained by selecting a leaf of the disk foliation as an end-of-the-world KR brane (in red), with all transverse spherical coordinates fixed except for the periodic angle $\phi$. We excise the part of the bulk ``above" the brane.}
\end{figure}
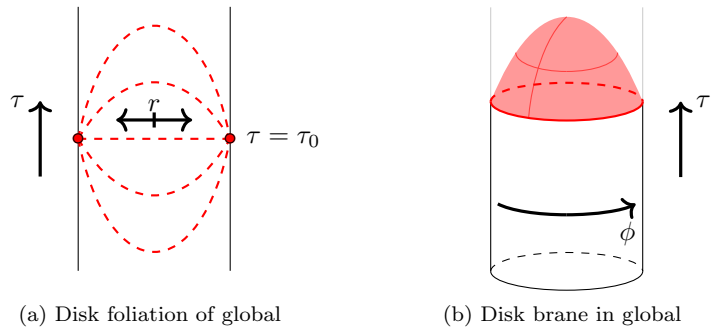

%% file: figs/adsGFoliationBCFTdualStrip.tex
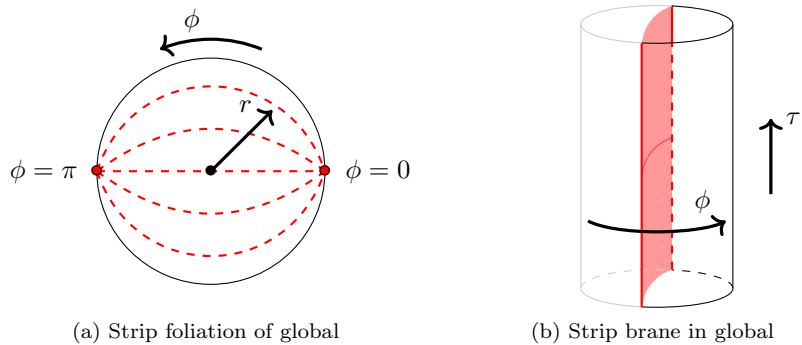
\begin{figure}
\centering
\subfloat[Strip foliation of global\label{figs:adsGFoliationStrip}]{
\begin{tikzpicture}

\draw[-] (0,0) circle (1.5);

\draw[-,thick,red,dashed] (-1.5,0) to (1.5,0);
\draw[-,thick,red,dashed] (-1.5,0) .. controls (-0.5,0.75) and (0.5,0.75) .. (1.5,0);
\draw[-,thick,red,dashed] (-1.5,0) .. controls (-1,1.5) and (1,1.5) .. (1.5,0);
\draw[-,thick,red,dashed] (-1.5,0) .. controls (-0.5,-0.75) and (0.5,-0.75) .. (1.5,0);
\draw[-,thick,red,dashed] (-1.5,0) .. controls (-1,-1.5) and (1,-1.5) .. (1.5,0);

\node[red] at (-1.5,0) {$\bullet$};
\node at (-1.5,0) {$\circ$};
\node[red] at (1.5,0) {$\bullet$};
\node at (1.5,0) {$\circ$};

\draw[->,very thick] (0,1.75) arc (0+90:22.5+90:1.75);
\draw[-,very thick] (0,1.75) arc (0+90:90-22.5:1.75);
\node at (-0.25,2) {$\phi$};

\draw[->,very thick,white] (0,-1.75) arc (270:270+22.5:1.75);
\draw[-,very thick,white] (0,-1.75) arc (270:270-22.5:1.75);

\node at (2.2,0) {$\phi = 0$};
\node at (-2.2,0) {$\phi = \pi$};

\draw[->,very thick] (0,0) to (1.414/1.75,1.414/1.75);
\node at (1.414/1.75-0.35,1.414/1.75+0.05) {$r$};
\node at (0,0) {$\bullet$};

\end{tikzpicture}}\qquad
\subfloat[Strip brane in global\label{figs:bcftDualGStrip}]{
\begin{tikzpicture}
\node[white] at (-1.7,0) {$\tau = \tau_0$};
\node[white] at (1.7,0) {$\tau$};

\node[white] at (1.7,0) {$\tau = \tau_0$};
\node[white] at (-1.7,0) {$\tau = \tau_0$};

\draw[-,black!20] (-1,-1.75) to (-1,1.75);
\draw[-] (1,-1.75) to (1,1.75);

\draw[-,draw=none,fill=red!40] (-0.2,-2) to[bend left] (0.2,-1.5) to (0.2,2) to[bend right] (-0.2,1.5) to (-0.2,-2);


\draw[-] (1,-1.75) arc (0:-101.5:1 and 0.25);
\draw[-,dashed] (1,-1.75) arc (0:78.5:1 and 0.25);
\draw[-,black!20] (-1,-1.75) arc (180:270-11.5:1 and 0.25);
\draw[-,dashed,black!20] (-1,-1.75) arc (180:78.5:1 and 0.25);

\draw[-,red!75] (-0.2,-2+1.75) to[bend left] (0.2,-1.5+1.75);

\draw[-,red,thick,dashed] (0.2,-1.5) to (0.2,1.5);
\draw[-,red,thick] (0.2,1.5) to (0.2,2);

\draw[-] (1,1.75) arc (0:-101.5:1 and 0.25);
\draw[-] (1,1.75) arc (0:78.5:1 and 0.25);
\draw[-,black!20] (-1,1.75) arc (180:270-11.5:1 and 0.25);
\draw[-,black!20] (-1,1.75) arc (180:78.5:1 and 0.25);

\draw[-,red,thick] (-0.2,-2) to (-0.2,1.5);





\draw[white,->,very thick] (-1.5,-0.5) to (-1.5,0.5);
\node[white] at (-1.8,0.5) {$\tau$};
\draw[->,very thick] (1.5,-0.5) to (1.5,0.5);
\node at (1.8,0.5) {$\tau$};

\draw[-,very thick] (0,-1) arc (-90:-155:1 and 0.25);
\draw[->,very thick] (0,-1) arc (-90:-25:1 and 0.25);
\node at (0.6,-0.6) {$\phi$};

\end{tikzpicture}}
\caption{(a) The strip foliation of Euclidean pure AdS$_{d+1}$ in global coordinates, projected onto the $(r,\phi)$ disk. The leaves anchor to antipodal points on the conformal boundary. (b) The bulk configuration obtained by selecting a leaf of the strip foliation as an end-of-the-world KR brane (in red), with the transverse $(\psi_1,...,\psi_{d-2})$ coordinates fixed. We excise the part of the bulk to the ``left" the brane.}
\end{figure}

%% file: figs/schFoliationBCFTdual2.tex
\begin{figure}
\centering
\subfloat[Annular foliation of AdS-Schwarzschild\label{figs:schFoliation}]{\begin{tikzpicture}[scale=0.9]


\draw[-] (1,-2.5) to (1,2.5);

\draw[-] (1,2.5) to (1.2,2.5);
\node at (1.765+0.05+0.08,2.5) {$\tilde{\tau} = \dfrac{\beta_h}{2}$};
\draw[-] (1,-2.5) to (1.2,-2.5);
\node at (1.9+0.05+0.085,-2.5) {$\tilde{\tau} = -\dfrac{\beta_h}{2}$};


\node at (3.5,0) {};
\node at (-2.5,0) {};

\draw[-,thick] (-0.5,2.5) to (1,2.5);
\draw[->,thick] (-0.5,2.5) to (0,2.5);
\draw[->,thick] (-0.5,2.5) to (0.5,2.5);
\draw[-,thick] (-0.5,-2.5) to (1,-2.5);
\draw[->,thick] (-0.5,-2.5) to (0,-2.5);
\draw[->,thick] (-0.5,-2.5) to (0.5,-2.5);

\draw[-,thick,red,dashed] (1,1.5) .. controls (0.75,0.2) and (0.75,-0.2) .. (1,-1.5);

\draw[-,thick,red,dashed] (1,1.5) .. controls (0.15,0.5) and (0.15,-0.5) .. (1,-1.5);

\draw[-,thick,red,dashed] (1,1.5) .. controls (-0.5,1.25) and (-0.5,-1.25) .. (1,-1.5);

\node[red] at (1,1.5) {$\bullet$};
\node at (1,1.5) {$\circ$};
\node[red] at (1,-1.5) {$\bullet$};
\node at (1,-1.5) {$\circ$};

\draw[->,very thick] (1.5,-0.5) to (1.5,0.5);
\node at (1.8,0.5) {$\tilde{\tau}$};
\draw[->,draw=none,very thick] (1.5,-0.5) to (1.5,0.5);

\draw[-,very thick,dashed] (-0.5,2.5) to (-0.5,-2.5);

\draw[->,very thick] (-0.5,0) to (0,0);
\draw[-,very thick] (-0.5,0.1) to (-0.5,-0.1);

\node at (-0.25,0.3) {$\tilde{r}$};

\node[black!30!green] at (-0.5,0) {$\bullet$};

\end{tikzpicture}}\qquad\qquad
\subfloat[Annular brane in AdS-Schwarzschild\label{figs:annularDual}]{\begin{tikzpicture}[scale=1.125]
\node at (0,-2) {};

\draw[-,red,thick,fill=red!30] (0,0) circle (2 and 1);
\draw[-,red,thick,fill=white] (0,0.125) circle (0.5 and 0.25);
\draw[-,red,thick] (0,0.125) circle (0.5 and 0.25);

\draw[-,red!75] (-0.5,0.125) .. controls (-0.7,-0.4) and (-1.8,-0.4) .. (-2,0);
\draw[-,red!75] (0.5,0.125) .. controls (0.7,-0.4) and (1.8,-0.4) .. (2,0);

\draw[-,red!75] (-0.2,-0.1) to[bend left] (-0.75,-0.925);
\draw[-,red!75] (0.2,-0.1) to[bend right] (0.75,-0.925);

\draw[-,red!75] (-0.75,0.925) ..controls (-0.7,0.5)  and (-0.5,0.25).. (-0.25,0.35);
\draw[-,red!75] (0.75,0.925) ..controls (0.7,0.5)  and (0.5,0.25).. (0.25,0.35);

\draw[-,draw=none,fill=red!40] (-0.5,0.14) .. controls (-0.25,0.25) and (0.25,0.25) .. (0.5,0.14) arc (0:180:0.5 and 0.25);
\draw[-,red,thick] (0,0.125) circle (0.5 and 0.25);

\draw[-,very thick,black!30!green] (0,0) circle (1.25 and 0.625);

\node at (0,-2.35) {};
\node at (-2.25,0) {};
\node at (2.25,0) {};

\draw[<-,very thick] (0,-1.25) arc (-90:-210:0.4 and 0.8);
\node[] at (0.2,-1.3) {$\tilde{\tau}$};

\draw[->,very thick] (1.5,0) arc (0:45:1.5 and 0.75);
\draw[-,very thick] (1.5,0) arc (0:-45:1.5 and 0.75);
\node[] at (1.7,0) {$\tilde{\phi}$};
\end{tikzpicture}}
\caption{(a) A slice of the annular foliation of Euclidean AdS-Schwarzschild with the spherical directions fixed and the center point of the $\tilde{\tau}$-cycle blown-up to a vertical line. $\tilde{\tau}$ increases upward while $\tilde{r}$ increases outward. The ends $\tilde{\tau} = \pm \beta_h/2$ are identified. (b) The bulk configuration obtained by selecting a leaf of the annular foliation as an end-of-the-world brane, with the transverse $(\tilde{\psi}_1,...,\tilde{\psi}_{d-2})$ coordinates fixed. The central green line is the non-contractible horizon. Each given leaf partitions the bulk into a piece containing the horizon and a piece not containing the horizon. Keeping the horizon corresponds to $T > 0$, and excising the horizon corresponds to $T < 0$.}
\label{figs:schFoliationBCFTdual}
\end{figure}
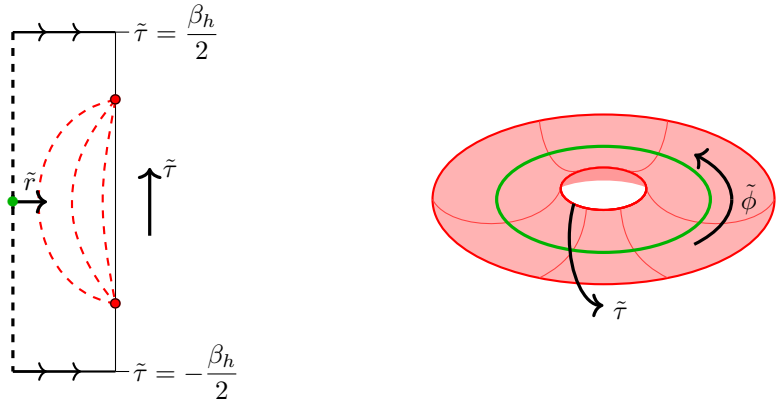

%% file: figs/nonintDiskBranes.tex
\begin{figure}
\centering
\begin{tikzpicture}
\node[white] at (-1.7,0) {$\tau = \tau_0$};
\node[white] at (1.7,0) {$\tau$};

\node[white] at (1.7,0) {$\tau = \tau_0$};
\node[white] at (-1.7,0) {$\tau = \tau_0$};

\draw[->,very thick] (1.3,-1) to (1.3,0);
\node at (1.5,-0.5) {$\tau$};

\draw[-,very thick] (0,-0.75) arc (270:210:1 and 0.25);
\draw[->,very thick] (0,-0.75) arc (270:330:1 and 0.25);
\node at (0.75,-1) {$\phi$};

\draw[-] (-1,-1.75) to (-1,0.5);
\draw[-,black!20] (-1,0.5) to (-1,1.75);
\draw[-,black!20] (-1,-0.5-1.25) to (-1,-1.75-1.25);
\draw[-] (1,-1.75) to (1,0.5);
\draw[-,black!20] (1,0.5) to (1,1.75);
\draw[-,black!20] (1,-0.5-1.25) to (1,-1.75-1.25);

\draw[-] (-1,-1.75) arc (180:360:1 and 0.25);
\draw[-,dashed] (-1,-1.75) arc (180:0:1 and 0.25);

\draw[-,draw=none,fill=red!40] (-1,0.5) .. controls (-0.4,2) and (0.4,2) .. (1,0.5) arc (0:-180:1 and 0.25);

\draw[-,draw=none,fill=red!40] (0,0.5) circle (1 and 0.25);

\draw[-,red,thick,dashed] (-1,0.5) arc (180:0:1 and 0.25);


\draw[-,draw=none,fill=red!40] (-1,-0.5-1.25) .. controls (-0.4,-2-1.25) and (0.4,-2-1.25) .. (1,-0.5-1.25) arc (0:-180:1 and 0.25);

\draw[-,draw=none,fill=red!25] (0,-0.5-1.25) circle (1 and 0.25);

\draw[-,red,thick,dashed] (-1,-0.5-1.25) arc (180:0:1 and 0.25);


\draw[-,very thick,black!30!green] (0,-2) to (0,0.25);
\draw[-,dashed,very thick,black!30!green] (0,0.25) to (0,1.65);
\draw[-,dashed,very thick,black!30!green] (0,-1.75) to (0,-2.875);

\draw[-,red!75] (0,1.625) .. controls (-0.4,1.25) and (-0.5,0.4) .. (-0.5,0.275);
\draw[-,red!75] (0,1.15-0.25) arc (270:180:0.67 and 0.25);
\draw[-,red!75] (0,1.15-0.25) arc (270:360:0.67 and 0.25);
\draw[-,red,thick] (-1,0.5) arc (180:360:1 and 0.25);

\draw[-,red!75] (0,-1.625-1.25) .. controls (-0.4,-1.25-1.4) and (-0.5,-0.5-1.4) .. (-0.5,-0.275-1.7);
\draw[-,red!75] (0,-1.15+0.25-1.7) arc (270:195:0.67 and 0.25);
\draw[-,red!75] (0,-1.15+0.25-1.7) arc (270:345:0.67 and 0.25);
\draw[-,red,thick] (-1,-0.5-1.25) arc (180:360:1 and 0.25);

\draw[->] (-0.67,1.15) to (-1,1.4);
\node at (-0.75,1.65) {$n_2^a$};
\draw[->] (-0.67,-1.15-1.25) to (-1,-1.4-1.25);
\node at (-0.75,-1.65-1.25) {$n_1^a$};

\node at (0.5,-1.75-1.25) {$T_1$};
\node at (0.5,1.75) {$T_2$};
\end{tikzpicture}
\caption{Two non-intersecting disk branes with tensions $T_1$ and $T_2$ embedded in conical AdS$_3$ \eqref{conicalglobalads}. They respectively have outward-pointing unit normal vectors $n_1^a$ and $n_2^a$.}
\label{figs:nonintDiskBranes}
\end{figure}
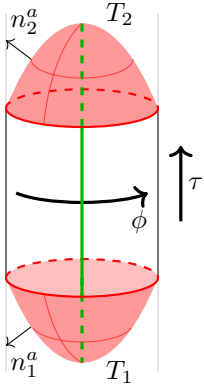